%% file: mainProject.tex
\documentclass[a4paper, 10pt, twocolumn, authoryear]{article}
\usepackage[left=1.6cm, right=1.6cm, top=3.0cm, bottom=3.0cm]{geometry}
\usepackage[affil-it]{authblk}

\usepackage{graphicx}
\usepackage{amssymb}

\usepackage{amsmath}				
\usepackage[nolist]{acronym}		
\usepackage{booktabs}
\usepackage[flushleft]{threeparttable}
\usepackage{tikz}
\usetikzlibrary{patterns,decorations.pathmorphing,arrows,intersections}
\usepackage{mathrsfs}
\usepackage{pgfplots}
\usepackage{subcaption}
\usepackage{adjustbox}			
\usepackage[]{placeins}

\usepackage{natbib}

\begin{document}
 
\title{Simulation of scour around arbitrary offshore foundations based on the Volume-of-Fluid method combined with a Bingham model}

\author[1]{Janek Meyer \thanks{}}
\author[2]{Kai Graf}
\author[3]{Thomas Slawig}
\affil[1]{Yacht Research Unit Kiel, R$\&$D-Centre University of Applied Sciences Kiel, Germany}
\affil[2]{Institute of Naval Architecture, University of Applied Sciences Kiel, Germany}
\affil[3]{Departement of Computer Science, Kiel University, Germany}

\renewcommand\Authands{ and }
\renewcommand\Affilfont{\itshape\small}

\date{5th December 2020}
%
\twocolumn[
  \begin{@twocolumnfalse}
    \maketitle
    \begin{abstract}
	\input{./chapters/abstract}
	\newline
	\newline
	\noindent\textit{Keywords:}
	\newline
	Scour, Sediment, Volume-of-Fluid, Bingham, Finite Volume Method, Internal Wall Function, OpenFOAM
    \end{abstract}
    \vspace{1cm}
  \end{@twocolumnfalse}
]
{
  \renewcommand{\thefootnote}%
  {\fnsymbol{footnote}}
  \footnotetext[1]{Corresponding author\\ Email address: \texttt{info@janekmeyer.de}}
}
%
%
  \input{./chapters/Introduction}

  \input{./chapters/Methods}
  \input{./chapters/Results}
  \input{./chapters/conclusion}
  \input{./chapters/acknowledgements}
  \FloatBarrier
\bibliographystyle{apalike}

\bibliography{Verzeichnis}

\end{document}

%% file: chapters/abstract.tex
\noindent This paper presents a method for the simulation of scour around arbitrary offshore structures.
It is based on the solution of the Reynolds-Averaged-Navier-Stokes equations implemented in the OpenFOAM framework.
The sediment is simulated with the help of a Bingham model,
which basically models a solid sediment behavior by introducing a very high viscosity.
The relative pressure used by the Bingham model is estimated with a new approach based on the solution of a Poisson equation.
The position of the sediment surface is calculated with the Volume-of-Fluid approach using a high-resolution scheme.
To keep the typical wall characteristics without demanding a fine grid, the common wall functions
are transferred to the domain internal sediment walls.
Furthermore, additional modifications are applied to model a solid sediment wall inside the solution domain.
The new internal wall function implementation is validated with a 2D test case. 
The results show a very good agreement to common wall functions and a significant improvement compared to its negligence.
Furthermore the solver is used to simulate the scour downstream of an apron and the scour around a vertical cylinder in current.
The results are compared to experiments presented in the literature and show good agreement.
The applicability onto arbitrary structures is demonstrated by applying the solver onto a vertical cylinder with a mudplate.
The current development state is able to resolve all important physical flow and scour
phenomena. The results also unveil that modeling of the suspension and the treatment of the internal wall need
additional attention.

%% file: chapters/Introduction.tex
\section{Introduction}
\label{Introduction} 
Scour is the erosion of sediment on the seabed near offshore structures like offshore wind power stations.
To avoid an excessive weakening of the structural strength the driving depth of respective offshore foundings is increased
and scour protection systems might be applied.
Furthermore the classification society may dictate the application of scour monitoring systems.
As all measures result in substantial extra costs it is of large interest to be able to predict the magnitude of the scour.
Currently most common prediction methods are based on very simple empirical equations,
which are restricted to be applied to simple obstacles like circular piles.
They may overestimate the scour many times.

In the last decades one has begun to develop simulation methods for the scour prediction.
In this context it has been shown that the long time scale of the scour development
in combination with the small time scale of the vortices shedded by the offshore structures
is challenging and leads to high computational costs.
Furthermore, the structures can be of complex shape (for example mudplates).
In addition the forces are not easy to model as they are often generated by a combination of a current and irregular waves.

The flow and scour around a vertical cylinder exposed to current were investigated experimentally and numerically by \cite{Roulund2005}.
The sediment shape is represented by a domain boundary.
The numerical model is based on the Exner equation and a grid morphing approach to calculate the sediment motion.
The numerical simulation captures all the main features and agrees reasonably well with the experiment.

The numerical simulation of the flow and scour around a vertical cylinder exposed to steady current
was investigated systematically by \cite{Baykal2015}.
The numerical model is also based on the Exner equation and a grid morphing approach.
It is shown that both, vortex shedding and the suspended load, needs to be resolved
to get the correct sediment shape downstream of the pile. 
This study was extended to waves and backfilling by \cite{Baykal2017}.
It is shown that the numerical model is applicable to scouring and backfilling under waves.
In advance, the vortices around a vertical wall-mounted cylinder exposed to waves were investigated by \cite{Sumer1997}.
Furthermore the backfilling of a scour hole around a pile in waves was investigated by \cite{Sumer2012}.
A first review of the applied methods has been given in \cite{Sumer2001} and second review has been given in \cite{Sumer2007}.
These methods are principally based on empirical bedload formulas.

Another model based on the Exner equation and a grid morphing approach was presented by \cite{Stahlmann2013}.
The model is used to simulate the scour around a tripod structure in waves.
Experimental results are also given for this scenario and a good agreement between the simulation and the experiment has been observed.

A new mesh deformation method was presented by \cite{Sattar2017}.
The mobile bed is modeled using the Exner equation.
As stated out, the model has overcome major limitations regarding mesh deformation and large amplitude bed movement.

A different modeling approach, which does not require conventional bedload and suspended load assumptions was described by \cite{Cheng2017}.
The method uses a multi-dimensional eulerian two-phase model with proper closure terms for the momentum exchange
to model the sediment transport and it does not require empirical bedload formulas.
The Volume-of-Fluid (VOF) method is used to describe the sediment shape.
Furthermore, a modified version of the k-$\epsilon$ turbulence model is applied.
The method has been applied to small-scaled problems like steady and oscillatory sheet flow
or scour downstream of an apron.
Additional enhancements to the model were presented by \cite{Chauchat2017}, where also a modified k-$\omega$ turbulence model is added.
In \cite{Nagel2017} and \cite{Nagel2020} this model has been used to simulate the scour around a vertical cylinder in current.
The presented results show a qualitatively good  agreement to the experiment on the upstream side of the pile.
However, on the downstream side a sediment accumulation is reported instead of an erosion.
Furthermore, the simulations require relatively long computational costs and 6,000 CPU hours were required for only 10 seconds of the simulation.

Another different method to simulate fluid-soil-body interaction was given by \cite{Volkner2015}.
The soil is modeled using a Bingham model and the sediment shape is described with the Volume-of-Fluid method.
The method has been applied onto a three-phase flow dam break test, a soil collapse test
and offshore groundings and jacking operations in current, waves and wind.
The work has its focus on the grounding and jacking operations and an application onto a scour problem is not given.
Similar Bingham-based approaches to simulate the soil were given by \cite{Ulrich2013}, \cite{Fourtakas2016} and \cite{Manenti2012}.
These works use the Smoothed-Particle-Hydrodynamics method to calculate the flow and to represent the shape of the sediment.
In \cite{Manenti2012} the Mohr-Coulomb Erosion Criterion is used to calculate the Bingham-viscosity.
The Bingham approach has been compared to the Shields Erosion Criterion by applying both onto a 2D flushing simulation.
The comparison to the flushing experiment shows, that both approaches reproduce the main features.
Nevertheless, \cite{Manenti2012} concludes that the Shields Erosion Criterion seems preferable for practical applications.
%
Furthermore, the Bingham model approach was used earlier by \cite{Liu1989} and \cite{Huang1997}.

Our main goal of this work is to develop a scour simulation method
which is applicable onto complex arbitrary shaped offshore structures.
Furthermore, the solution time should allow to use the method for industrial applications.
For the final method, the environmental forces acting onto the structures should result from current, waves or a combination of both.
Nevertheless, as the first developing step this paper will only consider currents.
All methods based on grid morphing are systematically excluded for an application on arbitrary complex shaped structures.
Whereas the Volume-of-Fluid method has a good potential for such cases and has been chosen for this work.

The sediment solver is implemented as an addition to our free-surface OpenFOAM-solver \citep{Meyer2016}.
This free-surface solver has been used in a variety of industrial and scientific projects (\cite{Graf2016} and \cite{Graf2017})
and is also able to calculate high quality wave simulations \citep{Meyer2017}.

The paper first describes the numerical model with all important details.
Afterward, the method is validated qualitatively against a wall-function test case,
scour downstream of an apron and scour around a vertical cylinder in current.
Furthermore, a first application on a complex shaped geometry is given
by investigating a vertical cylinder with an additional mudplate in current.

%% file: chapters/Methods.tex
\section{Methods}
\label{Methods} 
\subsection{Governing equations}
\label{Governing equations} 

For the calculation of the free-surface flow the incompressible unsteady Reynolds-averaged Navier-Stokes equations are solved
using the finite volume method.
The Volume-of-Fluid (VOF) method introduced by \cite{Hirt1981} is used for the calculation of the free-surface.
The momentum conservation equation,
the continuity equation and the conservation equation for the transport of the volume fraction $\alpha$ are defined as
\begin{equation}
\label{eq_momentum_p}
 \frac{\partial\rho \mathbf{u}}{\partial t}+\nabla \cdot (\rho \mathbf{u}\mathbf{u})
 -\nabla \cdot \mu_e \left(\nabla \mathbf{u} + (\nabla \mathbf{u})^T \right)
 =-\nabla p+\rho \mathbf{g}
\end{equation}
\begin{equation}
\label{eq_continuity}
 \nabla \cdot \mathbf{u} = 0
\end{equation}
\begin{equation}
\label{eq_vof_transport}
 \frac{\partial \alpha_i}{\partial t} +   \nabla \cdot ( \alpha_i \mathbf{u}) = 0
\end{equation}
with the volume fraction $\alpha_i$ for the $i$th phase,
the velocity vector $\mathbf{u}$, the pressure $p$, the gravity vector $\mathbf{g}$, the density $\rho$
and the effective dynamic viscosity $\mu_e = \mu_l + \mu_t$ consisting of the laminar dynamic viscosity $\mu_l$
and the dynamic eddy viscosity $\mu_t$.
The flow properties are then calculated by
\begin{equation}
\rho = \sum_i \rho_i \alpha_i \text{ ,}\hspace{0.5cm}
\mu = \sum_i \mu_i \alpha_i \hspace{0.25cm}\text{and}\hspace{0.25cm}
1 = \sum_i \alpha_i \hspace{0.2cm}.
\end{equation}
The free-sediment-surface is defined by the volume fraction $\alpha_s = 0.6$,
with $\alpha_s$ being the volume fraction of the sediment phase.
The dynamic eddy viscosity $\mu_t$ is calculated with a modified version of OpenFOAMs implementation
of the k-$\omega$-SST two equation model \citep{Menter2003}.
The original implementation in OpenFOAM excludes the density from space and time derivations
which is only a valid operation for one-phase flows with constant density.
A suitable formulation for the turbulent kinetic energy $k$ and the specific dissipation rate $\omega$
in a multi-phase flow is
\begin{equation}
  \label{eq_k}
 \frac{\partial \rho k}{\partial t}
 + \nabla \cdot \left( \rho \mathbf{u} k \right)
 - \nabla \cdot \left( \rho \left( \nu_l + \alpha_k \nu_t \right) \nabla k \right)
 =
 \rho P_k
 - \rho \beta^* k \omega
\end{equation}
and
\begin{equation}
\begin{split}
 \label{eq_omega}
 \frac{\partial \rho \omega}{\partial t}
 + \nabla \cdot \left( \rho \mathbf{u} \omega \right)
 - \nabla \cdot \left( \rho \left( \nu_l + \alpha_\omega \nu_t \right) \nabla \omega \right)\\
 =
 \rho \gamma S_2
 - \rho \beta \omega^2 
 + 2 \left(1 - F_1 \right) \rho
 \frac{\alpha_{\omega2}}{\omega}
 \left( \nabla k \cdot \nabla \omega \right)
 \end{split}
\end{equation}
with
\begin{equation}
\label{eq_S2}
 S_2 = \left( \nabla \mathbf{u} + \left( \nabla \mathbf{u} \right) ^T\right)
 :
 \left( \nabla \mathbf{u} + \left( \nabla \mathbf{u} \right) ^T\right)
\end{equation}
where $:$ stands for the double inner product,
\begin{equation}
\label{eq_G}
 G = \nu_t S_2
\end{equation}
and the production rate for the kinetic energy
\begin{equation}
\label{eq_Pk}
 P_k = \text{min}\left(G , 10.0 \beta^* \omega \right) \hspace{2mm}\text{.}
\end{equation}
The closure coefficients are
$\alpha_{\omega1}=0.5$,
$\alpha_{\omega2}=0.856$,
$\alpha_{k1}=0.85$,
$\alpha_{k2}=1.0$,
$\beta_1=0.075$,
$\beta_2=0.0828$,
$\gamma_1=\frac{5}{9}$,
$\gamma_2=0.44$,
and $\beta^*=0.09$.
The values for $\alpha_{\omega}$, $\alpha_k$, $\beta$ and $\gamma$ are than calculated with the following generalized blending function
\begin{equation}
 \phi_\text{blend} = F_1 \phi_1 + \left(1 - F_1 \right) \phi_2
\end{equation}
with the blending factor
\begin{equation}
 F_1 = 
 \text{tanh}
 \left(
 \left(
 \text{min}
      \left(
      \text{min}
      \left(
	    F^*,
	    \frac{4 \alpha_{\omega2} k}{\text{CD}_{k\omega}^+ y^2}
      \right),
      10
 \right)
 \right)^4
 \right)
\end{equation}
with the wall distance $y$,
\begin{equation}
 F^* = \text{max}\left(\frac{k^{0.5}}{\beta^* y \omega}, \frac{500 \nu_\text{l}}{y^2 \omega}\right) \hspace{2mm}\text{,}
\end{equation}
and
\begin{equation}
 \text{CD}_{k\omega}^+ = \text{max}\left(\text{CD}_{k\omega} ,10^-10 \right)
\end{equation}
with
\begin{equation}
 \text{CD}_{k\omega} = \frac{2 \alpha_{\omega2}}{\omega} \nabla k  \cdot \nabla \omega \hspace{2mm}\text{.}
\end{equation}
The blending value $F_2$ is calculated by
\begin{equation}
 F_2 = 
 \text{tanh}
 \left(
	\left(
	      \text{min}
	      \left(
		    F^*,
		    100
	      \right)
	\right)^2
 \right)\hspace{2mm}\text{.}
\end{equation}
Finally, the kinematic and the dynamic eddy viscosities are calculated by
\begin{equation}
  \label{eq_nut}
  \nu_t = \frac{a_1 k}{\text{max} \left( a_1 \omega, F_2 S_2^{0.5} \right)}
\end{equation}
and
\begin{equation}
  \label{eq_mut}
  \mu_t = \rho \nu_t \hspace{2mm}\text{.}
\end{equation}  
For all equations the finite volume method is used to discretize the spatial derivatives.
After integrating the equations over the volume, the resulting volume integrals are converted to surface integrals using the Gauss theorem.
This leads to a face based method, requiring interpolation schemes, which interpolate the cell-centered variables to the faces.
%
\subsection{Sediment simulation}
\label{Sediment simulation}
\subsubsection{Bingham model}
\label{Bingham model}

The sediment phase is modeled with an approach based on Bingham models \citep{Volkner2015}.
\cite{Volkner2015} used the method to simulate offshore grounding and jacking operations.
Treating the sediment as a Bingham plastic, allows to model the sediment as a non-Newtonian fluid
using a variable dynamic viscosity $\mu_\text{Bingham}$ in the diffusive term of the momentum equation \eqref{eq_momentum_p}.
This leads to the following modified momentum equation
\begin{equation}
\label{eq_momentum_Bingham}
\begin{split}
 \frac{\partial\rho \mathbf{u}}{\partial t}+\nabla \cdot (\rho \mathbf{u}\mathbf{u})
 -\nabla \cdot \left( \mu_e + \mu_\text{Bingham} \right) \left(\nabla \mathbf{u} + (\nabla \mathbf{u})^T \right)\\
 =-\nabla p+\rho \mathbf{g}
 \end{split}
\end{equation}
If the local shear stress in the sediment is below the sediment specific yield stress the sediment should behave as a solid.
The solid behavior is modeled by using a relatively high local dynamic Bingham viscosity.
If the local shear stress is higher than the yield stress the sediment begins to behave like a fluid.
This behavior is modeled by reducing the local dynamic Bingham viscosity.

For the calculation of the Bingham viscosity \cite{Volkner2015} distinguish between
the Bingham viscosity $\mu_\text{soil}$ for the soil and the Bingham viscosity $\mu_\text{susp}$ for the suspension.
Furthermore, we use a different interpolation scheme for the soil viscosity.
Considering the later applied finite volume method and Gauss theorem, we will write the momentum equation as
\begin{equation}
\label{eq_momentum_Bingham_final}
\begin{split}
 \frac{\partial\rho \mathbf{u}}{\partial t}+\nabla \cdot (\rho \mathbf{u}\mathbf{u})
 -\nabla \cdot \left( \mu_e + \mu_\text{susp} \right) \left(\nabla \mathbf{u} + (\nabla \mathbf{u})^T \right)\\
 -\nabla \cdot  \mu_\text{soil} \left(\nabla \mathbf{u} + (\nabla \mathbf{u})^T \right)
 =-\nabla p+\rho \mathbf{g}
 \end{split}
\end{equation}
using two diffusive terms emphasizing the two different interpolation schemes applied later.
To calculate the soil viscosity \cite{Volkner2015} use a combination of the von Mises and Mohr-Coulomb yield criterion
leading to the following equation for the dynamic soil viscosity $\mu_{s}$
 \begin{equation}
 \label{eq_mu_soil}
 \mu_\text{soil}^*
 =
 \frac{\tau_f}{\left(4 j \right)^{0.5}}
 \end{equation}
 Here, $\tau_f$ is the yield stress and $j$ is the second invariant of the strain rate tensor.
 The yield stress $\tau_f$ is estimated with the von Mises and Mohr-Coulomb yield criterion 
\begin{equation}
 \label{eq_tau_f}
 \tau_f
 =
 p_\text{rel} \text{ sin}\left( \phi \right)
 + c \text{ cos} \left( \phi \right)
\end{equation}
with the relative pressure $p_\text{rel}$, the cohesion $c$ and the internal friction angle $\phi$.
The calculation of the relative pressure is not straightforward and our own approach is described in section \ref{Relative pressure}.
The second invariant of the strain rate tensor is calculated by
\begin{equation}
\label{eq_j}
  j = 0.5
  \left( \nabla \mathbf{u} + \left( \nabla \mathbf{u} \right)^T	\right)
  :
  \left( \nabla \mathbf{u} + \left( \nabla \mathbf{u} \right)^T	\right)
\end{equation}

where $:$ is the double inner product of two tensors.
For the calculation of the suspension viscosity \cite{Volkner2015} give the following equation
\begin{equation}
 \mu_\text{susp}^* = \frac{\tau_w}{\left(4 j \right)^{0.5}}
\end{equation}
with 
\begin{equation}
 \tau_w = C_f \rho_G \mathbf{u} \cdot \mathbf{u}
 \hspace{0.2cm}.
\end{equation}
Here $C_f$ is the empirical friction coefficient set to 0.01 and $\rho_G$ is the density of the granular soil phase.

The estimated sediment viscosities are bounded between a lower and an upper bound $\mu_{s_\text{lower}}$ and $\mu_{s_\text{upper}}$.
For the lower and upper bounds, values of $25 \frac{Ns}{m^2}$ and $1500\frac{Ns}{m^2}$ were suggested by \cite{Volkner2015} .
These bounds were estimated by simulating a soil collapse test with different values.
This test does not reflect the main phenomenons of scouring
and our own simulations have shown that the lower limit should be decreased, see section \ref{Scour around a vertical pile}.
The upper bound is adopted as it stands.

For the numerical stability of the solution algorithm, it is common practice to avoid discontinuities of the final spatial characteristics of the viscosities.
Nevertheless, we decided to keep the discontinuities at the sediment surface to avoid an influence onto the result from a blending zone.
In comparison to previous simulations with a blending zone applied at the sediment surface we could not observe any differences in the solver stability.
Equations \eqref{eq_blend_mu_soil} and \eqref{eq_blend_mu_susp} show our final applied blending zones.

\begin{equation}
\label{eq_blend_mu_soil}
\mu_\text{soil} =
    \begin{cases} 
    0.0	& \textrm{ if } \alpha_{s_i} \le 0.6\\
    \mu_\text{soil}^* & \textrm{ if } 0.6 < \alpha_{s_i}\\
    \end{cases}
\end{equation}

\begin{equation}
\label{eq_blend_mu_susp}
\mu_\text{susp} =
    \begin{cases} 
    0.0	& \textrm{ if } \alpha_{s_i} \le 0.005\\
    \frac{\alpha_{s_i} - 0.005}{0.01 - 0.005} \mu_\text{susp}^* & \textrm{ if } 0.005 < \alpha_{s_i} \le 0.01\\
    \mu_\text{susp}^*	& \textrm{ if } 0.01 < \alpha_{s_i} \le 0.6\\  
    0.0	& \textrm{ if } 0.6 < \alpha_{s_i}\\
    \end{cases}
\end{equation}
With respect to \cite{Nnadi1992} and \cite{Ulrich2013} the suspension viscosity should be blended out below a sediment volume fraction of $0.3$.
As our current model does not include a proper generation of the suspension layer, we have decided to decrease this value to $0.01$.
This guarantees an influence of the suspension, although the suspension concentration might be low.
\subsubsection{Creeping}
\label{Creeping}
To suppress further creeping of the solid sediment \cite{Volkner2015} suggest to damp the velocities,
if the sediment viscosity reaches a specific percentage of the upper bound.
Our own way to damp the velocities is based on implicitly relaxing the final momentum equation to a target velocity of $0\frac{m}{s}$.
The local relaxation factor of the $i\text{th}$ cell is calculated by
%
%
\begin{equation}
r_i =
    \text{min} \left( 1.0, \text{max}\left(0.0, 1.0 - \frac{\mu_i - 0.7 \mu_{s_\text{upper}}}{\left(0.9 - 0.7 \right) \mu_{s_\text{upper}}} \right) \right)\\
    \hspace{2mm}.
\end{equation}
%
\subsubsection{Interpolation of the soil viscosity}
\label{Interpolation of the soil viscosity}
Integrating the momentum equation \eqref{eq_momentum_Bingham_final} over the volume and applying the Gauss theorem leads to a face based method,
requiring interpolation schemes to interpolate the cell-centered variables to the faces.
Using the typical linear interpolation for the sediment viscosity in the diffusive term might be the first choice.
But, due to the jump behavior of this viscosity such linear interpolation will lead to significant problems.

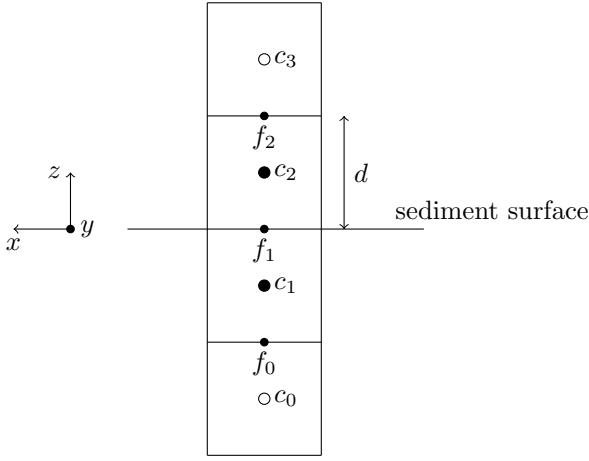
\begin{figure}
 \centering
    \begin{tikzpicture}[scale=1.5]
    
    \draw[black] (0.0,2.0) -- (1.0,2.0);
    \draw[black] (0.0,1.0) -- (1.0,1.0);
    \draw[black] (-0.7,0.0) -- (1.9,0.0);  
    \draw[black] (0.0,-1.0) -- (1.0,-1.0);
    \draw[black] (0.0,-2.0) -- (1.0,-2.0);
    
    \draw[black] (0.0,-2.0) -- (0.0,2.0);
    \draw[black] (1.0,-2.0) -- (1.0,2.0);
    
    \draw[black, <->] (1.2,0.0) -- (1.2,1.0);
    \draw[black, thick] (1.2,0.5) node[right]{$d$};
    
    \draw[black, thick] (2.5,0.0) node[above]{sediment surface};
    
    \node[draw,circle,inner sep=1.5pt] at (0.5,1.5) {}; \draw[black, thick] (0.5,1.5) node[right]{$c_{3}$};
    \node[draw,circle,inner sep=1.5pt,fill] at (0.5,0.5) {}; \draw[black, thick] (0.5,0.5) node[right]{$c_{2}$};
    \node[draw,circle,inner sep=1.5pt,fill] at (0.5,-0.5) {}; \draw[black, thick] (0.5,-0.5) node[right]{$c_{1}$};
    \node[draw,circle,inner sep=1.5pt] at (0.5,-1.5) {}; \draw[black, thick] (0.5,-1.5) node[right]{$c_{0}$};
    
    \node[draw,circle,inner sep=1.0pt,fill] at (0.5,1.0) {}; \draw[black, thick] (0.5,1.0) node[below]{$f_{2}$};
    \node[draw,circle,inner sep=1.0pt,fill] at (0.5,0.0) {}; \draw[black, thick] (0.5,0.0) node[below]{$f_{1}$};
    \node[draw,circle,inner sep=1.0pt,fill] at (0.5,-1.0) {}; \draw[black, thick] (0.5,-1.0) node[below]{$f_{0}$};
    
    \draw[->] (-1.2,-0.0) -- (-1.2,0.5) node[left] {$z$};
    \draw[->] (-1.2,-0.0) -- (-1.7,0.0) node[below] {$x$};
    \node[draw,circle,inner sep=1.0pt,fill] at (-1.2,0.0) {}; \draw[black, thick] (-1.2,0.0) node[right]{$y$};
%
    \end{tikzpicture}
 \caption{Cells divided by sediment-surface}
 \label{Cells devided by free surface}
\end{figure}

The following example should explain the problem in more detail.
Afterward, the solution of the problem is given.
Figure \ref{Cells devided by free surface} shows a uniform one-dimensional grid with four cells in the vertical z-direction.
The cells $c_0$ and $c_3$ act as fixed value boundary cells.
In the two other directions $x$ and $y$ one can assume cells acting as a zero gradient boundary condition.
The two upper cells $c_2$ and $c_3$ are filled with water and therefore have a sediment viscosity of zero.
The two lower cells $c_0$ and $c_1$ are filled with sediment acting as a solid
and therefore have a sediment viscosity of $\mu_{\text{soil}} = \mu_{s_\text{upper}}=1500\frac{Ns}{m^2}$.
The velocities are given in the boundary cells and needs to be calculated for the internal cells $c_1$ and $c_2$.
The velocities at the boundaries are $u_{x_{c_0}} = 0\frac{m}{s}$ and $u_{x_{c_3}} = u_w \neq 0$.
To solve the momentum equation we can make some simplifications.
Assuming a steady state solution, the time derivative can be neglected.
As the flow is parallel to the sediment surface, the convected term will not influence the result and can be neglected, too.
Additionally, we do not consider gravitational forces, also the dynamic pressure gradient is neglected.
The continuity equation is inherently fulfilled.
The momentum equation \eqref{eq_momentum_Bingham} then reduces to the diffusive term
\begin{equation}
 \nabla \cdot \underbrace{\left(\mu_e + \mu_\text{Bingham} \right)}_{\mu} \left(\nabla \mathbf{u} + (\nabla \mathbf{u})^T \right)
 =0
 \hspace{2mm}.
\end{equation}
Integrating over the volume and using the Gauss theorem to convert the volume integrals to surface integrals gives
\begin{equation}
\label{eq_diffusive_allgemein_cell1}
 -\mu_{f_0} \left( \frac{\partial u_x}{\partial z} \right)_{f_0}
 + \mu_{f_1} \left( \frac{\partial u_x}{\partial z} \right)_{f_1}
 = 0
\end{equation}
for cell $c_1$ and
\begin{equation}
\label{eq_diffusive_allgemein_cell2}
 -\mu_{f_1} \left( \frac{\partial u_x}{\partial z} \right)_{f_1}
 + \mu_{f_2} \left( \frac{\partial u_x}{\partial z} \right)_{f_2}
 = 0
\end{equation}
for cell $c_2$, while the subscript $f_i$ marks a value at the $i$th face.
Using 2nd order accurate linear interpolation for the viscosity
and building the velocity gradient directly at the face without special jump treatment lead to the following equation system
\begin{equation}
 -\frac{\mu_{c_0} + \mu_{c_1}}{2} \left( \frac{ u_{x_{c_1}} - u_{x_{c_0}} }{d} \right)
 + \frac{\mu_{c_1} + \mu_{c_2}}{2} \left( \frac{ u_{x_{c_2}} - u_{x_{c_1}} }{d} \right)
 = 0
\end{equation}
\begin{equation}
 -\frac{\mu_{c_1} + \mu_{c_2}}{2} \left( \frac{ u_{x_{c_2}} - u_{x_{c_1}} }{d} \right)
 + \frac{\mu_{c_2} + \mu_{c_3}}{2} \left( \frac{ u_{x_{c_3}} - u_{x_{c_2}} }{d} \right)
 = 0
\end{equation}
with $\mu_0 = \mu_1 = \mu_{s_\text{upper}}$ and $\mu_2 = \mu_3 = \mu_{w}$
and knowing that $\mu_{s_\text{upper}} >> \mu_{w}$
an approximate solution for $u_{x_{c_1}}$ and $u_{x_{c_2}}$ is
\begin{equation}
 u_{x_{c_1}} =
 \frac
 {
  \mu_{s_\text{upper}} ^2 u_{c_0}
  + \mu_{w} \mu_{s_\text{upper}} \left( u_{c_3} - u_{c_0} \right)
  + \mu_{w} ^2 u_{c_3}
 }
 {
  \mu_{s_\text{upper}} ^2
  + \mu_{w} ^2
 }
 \rightarrow
 \sim u_{c_0}
\end{equation}
\begin{equation}
 u_{x_{c_2}} =
 \frac
 {
  \mu_{s_\text{upper}} ^2 u_{c_0}
  + \mu_{w} \mu_{s_\text{upper}} \left( u_{c_0} - u_{c_3} \right)
  + \mu_{w} ^2 u_{c_3}
 }
 {
  \mu_{s_\text{upper}} ^2
  + \mu_{w} ^2
 }
 \rightarrow
 \sim u_{c_0} \hspace{2mm}.
\end{equation}
The results clearly show, that the linear interpolation leads to a wrong velocity for cell $c_2$.
This is a problem for the scour model, as the flow velocity never reaches the water cell next to the sediment.
Therefore no force is acting on the sediment and consequently the sediment viscosity will never be reduced.
In \cite{Volkner2015} it is suggested to use the harmonic mean for the interpolation of the viscosities.
In the following it should be shown that the harmonic mean solves the problem.
Using the harmonic mean the interpolated face value for a equidistant grid is defined as
\begin{equation}
 \label{eq_harmonic_mean}
 \phi_f = \frac{2}{\frac{1}{\phi_N}+\frac{1}{\phi_P}}
\end{equation}

where the subscripts $N$ and $P$ marks the two cells sharing the face $f$.
Interpolating the face values of equations \eqref{eq_diffusive_allgemein_cell1}
and \eqref{eq_diffusive_allgemein_cell2} with the harmonic mean gives the following equation system
\begin{equation}
 -\frac{2}{\frac{1}{\mu_0}+\frac{1}{\mu_1}} \left( \frac{ u_{x_{c_1}} - u_{x_{c_0}} }{d} \right)
 + \frac{2}{\frac{1}{\mu_1}+\frac{1}{\mu_2}} \left( \frac{ u_{x_{c_2}} - u_{x_{c_1}} }{d} \right)
 = 0
\end{equation}
\begin{equation}
 -\frac{2}{\frac{1}{\mu_1}+\frac{1}{\mu_2}} \left( \frac{ u_{x_{c_2}} - u_{x_{c_1}} }{d} \right)
 + \frac{2}{\frac{1}{\mu_2}+\frac{1}{\mu_3}} \left( \frac{ u_{x_{c_3}} - u_{x_{c_2}} }{d} \right)
 = 0
\end{equation}
Solving this equation system under the previous assumptions leads to following approximate solution
\begin{equation}
 u_{c_1} =
 \frac
 {
  3 \mu_{s_\text{upper}}  u_{c_0} + \mu_w \left( u_{c_0} + 2  u_{c_3} \right)
 }
 {
  3 \left( \mu_{s_\text{upper}} + \mu_w \right)
 }
 \rightarrow
 \sim u_{c_0}
\end{equation}
\begin{equation}
 u_{c_2} =
 \frac
 {
  3 \mu_w  u_{c_3} + \mu_{s_\text{upper}} \left( 2 u_{c_0} + u_{c_3} \right)
 }
 {
  3 \left( \mu_{s_\text{upper}} + \mu_w \right)
 }
 \rightarrow
 \sim \frac{1}{3}u_{c_3} \hspace{2mm}.
\end{equation}
The results show that a significant amount of the flow velocity is reaching the cell next to the sediment,
while the sediment cells still keep their correct velocity.
To the authors knowledge, it is not allowed to use the harmonic mean for the interpolation
of the effective viscosity $\mu_e$ without additional adjustments.
Therefore only the soil viscosity $\mu_\text{soil}$ is interpolated with the harmonic mean,
which leads to the second diffusive term in equation \eqref{eq_momentum_Bingham_final} for the final implementation in OpenFOAM.
%
\subsubsection{Relative pressure}
\label{Relative pressure} 

In \cite{Manenti2012}, \cite{Ulrich2013} and \cite{Fourtakas2016} the relative pressure used in equation \eqref{eq_tau_f} is calculated through the equation of state for a weakly compressible fluid.
As we do not simulate a compressible fluid we are introducing a new way to estimate the relative pressure.
For our estimation, we assume that the relative pressure is zero at the sediment surface and only has a vertical spatial dependency.
The basic idea is to calculate the relative pressure $p_\text{rel}$ with the following Poisson equation
\begin{equation}
 \label{eq_rel_pressure}
 \nabla \cdot \left(\mathbf{Z} \cdot \nabla p_\text{rel}^* \right)
 =
 \nabla \cdot \left( \tilde{\rho} \mathbf{g} \right)
\end{equation}
with the gravitational acceleration vector $\mathbf{g}$,
the effective density $\tilde{\rho}$ resulting of the sand particles without the poor-water
and the tensor $\mathbf{Z}$ which decouples the pressure from the horizontal environment.
The $^*$ means that this is not the final relative pressure and a final correction will follow.
The effective density is calculated by
\begin{equation}
 \tilde{\rho} =
    \begin{cases} 
    \rho_\text{rock} \left(1 - \phi_p \right) 	& \textrm{ if } \alpha_s \ge \alpha_\text{wall}\\
    0 & \textrm{ if } \alpha_s < \alpha_\text{wall}\\
    \end{cases}
\end{equation}
with $\phi_p$ as the porosity of the sediment, $\rho_\text{rock}$ as the density of the sand grain
and $\alpha_\text{wall} = 0.6$ defining the sediment surface.
The tensor $\mathbf{Z}$ is defined as
\begin{equation}
        \mathbf{Z} = 
        \begin{bmatrix}
	    0 & 0 & 0 \\
	    0 & 0 & 0 \\
	    0 & 0 & 1 \\
	\end{bmatrix}
\end{equation}
Using this tensor is the main trick to calculate the effective pressure.
It ensures that the pressure is always zero at the sediment surface
and that the pressure only depends on the sediment depth.
But, using this tensor leads to numerical problems.
First of all it is necessary to have a Dirichlet boundary condition on the top of the domain.
For example, a Dirichlet boundary condition only at the outlet will lead to problems
as the internal cells will not be influenced by this boundary condition due to the decoupling tensor $\mathbf{Z}$.
Nevertheless, the solution of this equation always diverged in the first timesteps.
Our solution is to implement the decoupling approach with the help of the deferred correction method.
That means, the Laplace term containing the tensor $\mathbf{Z}$ is treated explicitly on the right hand side.
Additionally, the same Laplace term, but without the tensor $\mathbf{Z}$ is added on both implicit and explicit sides.
This leads to the following equation for the relative pressure
\begin{equation}
 \label{eq_rel_pressure_DC}
 \nabla \cdot \left(\nabla p_\text{rel}^{*^{q+1}} \right)
 =
 \nabla \cdot \left( \tilde{\rho} \mathbf{g} \right)
 + \nabla \cdot \left(\nabla p_\text{rel}^{*^q} \right)
 - \nabla \cdot \left(\mathbf{Z} \cdot \nabla p_\text{rel}^{*^q} \right)
\end{equation}
where the superscript $q$ marks the already known solution of the last iteration and $q+1$ the new solution solved in the current iteration.
If the equations system is converging the Laplace terms without the tensor $\mathbf{Z}$ are canceling each other out
and the solution is the same as for the equation \eqref{eq_rel_pressure}.
The converging behavior of this equation has been satisfying in all of our simulations.
Still small errors in the results for the relative pressure are possible
and in very few cells containing water a noticeable high relative pressure occurred.
A vertical advancing of this error in the water cells is inherently given
by the zero gradient ($\nabla \tilde{\rho} \mathbf{g}$) inside the water cells.
To avoid this errors, resulting in pressure peaks outside the sediment we do an explicit correction.
\begin{equation}
 p_\text{rel} =
    \begin{cases} 
    p_\text{rel}^* 	& \textrm{ if } \alpha_s \ge 0.99 \alpha_\text{wall}\\
    0 & \textrm{ if } \alpha_s < 0.99 \alpha_\text{wall}\\
    \end{cases}
\end{equation}
This correction suppresses the mentioned problem successfully.
Nevertheless, an explicit correction, which is not included in the equation system may destroy the converged solution.
In our case, this explicit correction leads to a slow convergence
and in a few simulations wrong results for the relative pressure occurred for a few timesteps leading to big errors in the result.
Therefore, the last step was, to include this correction into the equation system.
This can be achieved easily, using an implicit relaxation to a target pressure of zero for the sediment cells.
After discretizing equation \eqref{eq_rel_pressure_DC} the final equation system can be written as
\begin{equation}
        \label{eq_rel_pressure_coeff_matrix}
        a_{\text{d},p_\text{rel}} p_\text{rel}^* + \sum_n a_{\text{n},p_\text{rel}} p_\text{rel}^*
        = s_{\text{p}_\text{rel}}
\end{equation}
with $a_{\text{d},p_\text{rel}}$ as the main diagonal elements of the coefficient matrix,
$a_{\text{n},p_\text{rel}}$ as the neighbor elements of the coefficient matrix and $s_{\text{p}_\text{rel}}$ as the right hand side.
Using the implicit relaxation one gets the following equation
\begin{equation}
        \label{eq_rel_pressure_implicit_relax}
        a_{\text{d},p_\text{rel}} p_\text{rel} + r \sum_n a_{\text{n},p_\text{rel}} p_\text{rel}
        =
        r s_{\text{p}_\text{rel}}
        + \left( 1 -r \right) a_{\text{d},p_\text{rel}} p_\text{rel}^t
\end{equation}
with the relaxation factor $r$, and the target pressure $p_\text{rel}^t$, which is zero.
The relaxation factor $r$ is calculated with the help of the sediment volume fraction
\begin{equation}
 r =
    \begin{cases} 
    1.0 	& \textrm{ if } \alpha_s \ge 0.99 \alpha_\text{wall}\\
    0.0 & \textrm{ if } \alpha_s < 0.99 \alpha_\text{wall}\\
    \end{cases}
    \hspace{2mm}.
\end{equation}
With this approach the convergence of the equation system for the relative pressure is very fast.
And we did not observe any more instabilities.
%
%
%
%
\subsubsection{Sliding model}
\label{Sliding model}
As stated out in \cite{Roulund2005} observations have shown, that on the upstream side of the scour hole the slope might exceed the angle of repose by a few degrees.
As a result, shear failures occur at these locations.
For the simulation, \cite{Roulund2005} suggests a sliding model, which moves the sediment until the bedslope deceeds the angle of repose by two degrees.
The whole sliding process is done in the same time step and not resolved in a real transient way.
As this method is based on a mesh boundary representing the sediment surface, it is not directly applicable to our Volume-of-Fluid approach.
Therefore, we have developed a new sliding model based on a similar idea.
The new model can be concluded as follows:
\begin{itemize}
 \item Mark the sediment surface cells: this can be done by iterating over the internal faces
and comparing the volume fraction of the sediment $\alpha_s$ of the two cells sharing this face.
 \item Calculate the gradient of the sediment volume fraction $\nabla \alpha_s$.
 \item For all sediment surface cells: calculate the angle between the local vector $\nabla \alpha_s$ and the gravity vector $\mathbf{g}$.
 \item If the local angle succeeds the angle of internal friction, set the local sediment viscosity $\mu_s$ to zero.
\end{itemize}
Compared to the sliding models presented in \cite{Roulund2005} and \cite{Stahlmann2013}, our new sliding model does not slide the sediment explicitly but it removes the local withstanding force by reducing $\mu_s$.
Therefore the external forces can act unrestricted.
If the gravitational forces dominates, the sediment will slide down.
If another force dominates the sediment may also be picked up by the flow, for example.
%
\subsubsection{Internal wall function}
\label{Method Internal wall function}
Resolving the boundary layer above the sediment surface correctly is important
as the boundary layer size has a significant influence onto the horseshoe vortex.
Simulation methods based on ``grid deforming'' as described in \cite{Roulund2005} can use the standard well-known wall functions
applied to the domain boundary representing the sediment surface.
However, since we use the VOF-method to represent the sediment, its surface is not linked to a domain boundary and standard wall functions are not applicable.
A potential solution could be to resolve the boundary layer with a very fine grid.
But this will increase the computational costs significantly, especially because the whole region of the scour hole development needs to be resolved with the finest cell size.
Therefore we have implemented a new wall function approach in a way allowing it to act on the domain-internal sediment wall.
To achieve this, we use the same principle as the common wall functions.
In the following we will first describe the high-level principle of OpenFOAMs standard wall function for smooth walls.
Afterward, the new internal wall function implementation will be explained.

Using the k-omega-SST turbulence model in combination with OpenFOAM's \textit{omegaWallFunction}, \textit{nutUWallFunction} and \textit{movingWallVelocity}
the solution process of the turbulence model with an applied wall function can be concluded on the following points:
\begin{itemize}
 \item Explicitly calculate the wall distance $y$.
 \item Explicitly calculate $G$ \eqref{eq_G} and the turbulent production rate $P_k$ \eqref{eq_Pk}.\\
 At walls using a wall function:\\
 for the near wall cells overwrite $G$ with a different explicitly calculated result (requires $y$).
 This also changes $P_k$ as its calculation uses $G$.
 \item Set up the equation for $\omega$ (see equation \eqref{eq_omega}).\\
 At walls using a wall function:\\
 manipulate the coefficient matrix for the near wall cells to force a different, explicitly calculated result for these cells (requires $y$).\\
 At the wall set the previously calculated value of the near wall cell for $\omega$.
 Solve the equation for $\omega$.
 \item Set up the equation for $k$ (requires $P_k$, see equation \eqref{eq_k}).\\
 At the wall a Neumann boundary condition with a zero gradient is applied for $k$.
 Solve this equation for $k$.
 \item Calculate the turbulent kinematic viscosity $\nu_t$ explicitly (see equation\eqref{eq_nut}).\\
 At walls using a wall function:\\
 for the wall faces overwrite $\nu_t$ with a different explicitly calculated result (requires $y$).
\end{itemize}
Summarizing the important steps for a wall function:
It is necessary to have $y$, a different $G$ at the near wall cell, a different $\omega$ at the near wall cell and the face of the wall,
a zero gradient boundary condition for $k$ and a different $\nu_t$ at the face of the wall.

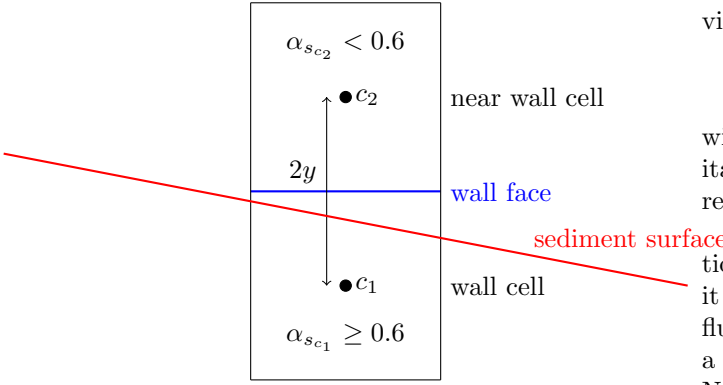
\begin{figure}
 \centering
    \begin{tikzpicture}[scale=2.5]
    
    \draw[black] (0.0,1.0) -- (1.0,1.0);
    \draw[blue, thick] (0.0,0.0) -- (1.0,0.0);  
    \draw[black] (0.0,-1.0) -- (1.0,-1.0);
    
    \draw[black] (0.0,-1.0) -- (0.0,1.0);
    \draw[black] (1.0,-1.0) -- (1.0,1.0);
    
    \draw[black, <->] (0.4,0.5) -- (0.4,-0.5);
    \draw[black, thick] (0.4,0.1) node[left]{$2y$};
    
    \draw[red, thick] (-1.3,0.2) -- (2.3,-0.5); \draw[red, thick] (2.0,-0.35) node[above]{sediment surface};
    
    \node[draw,circle,inner sep=1.5pt,fill] at (0.5,0.5) {}; \draw[black, thick] (0.5,0.5) node[right]{$c_{2}$};
    \node[draw,circle,inner sep=1.5pt,fill] at (0.5,-0.5) {}; \draw[black, thick] (0.5,-0.5) node[right]{$c_{1}$};
    
    \draw[black, thick] (0.5,0.9) node[below]{$\alpha_{s_{c_{2}}} <0.6$};
    \draw[black, thick] (0.5,-0.9)node[above]{$\alpha_{s_{c_{1}}}\ge 0.6$};
    
    \draw[black, thick] (1.0,0.5) node[right]{\text{near wall cell}};
    \draw[blue, thick] (1.0,0.0)node[right]{\text{wall face}};
    \draw[black, thick] (1.0,-0.5)node[right]{\text{wall cell}};
    
    
%
    \end{tikzpicture}
 \caption{Wall cells at sediment-surface}
 \label{Wall cells at sediment-surface}
\end{figure}

For our new, domain internal implementation we use the same equations as the original implementation
and just redefine the wall distance $y$, the near wall cell and the face of the wall as shown in figure \ref{Wall cells at sediment-surface}.
Identifying the wall cells can be done by iterating over all internal faces and comparing the sediment volume fraction of both cells sharing this face.
The sediment wall lies inside these cells, if one cell has a sediment volume fraction smaller than 0.6 and the other cell has a sediment volume fraction greater or equal to 0.6.
As one can see, we use the half distance between the cell centers to estimate the wall distance y.
It might be a more accurate approach to project the distance vector between both cells onto the surface normal of the sediment surface.
But this introduces new problems, as the projected distance might become very small or zero.
Therefore, we do not use such a projection.

The cell with a sediment volume fraction smaller than 0.6 is called the near wall cell,
the cell with a sediment volume fraction greater or equal to 0.6 is called the wall cell.
The face between these two cells is called the wall face.
As a simplification, the wall face is assumed to be the sediment surface in both points,
position and orientation, although this is not the case predominantly.\\

\textbf{Modifications at the near wall cell:}\\
For this cell the modified values for $G$ and $\omega$ are used.
As previously shown the equation system for $\omega$ has to be modified, so that the converged solution contains the modified value for the near wall cells.
In the current implementation the equation system is modified with the help of an implicit relaxation.
For the near wall cells the equation is relaxed to the individual target value with a relaxation factor of $1e-9$.\\

\textbf{Modifications at the wall cell:}\\
The previously modified value for $\omega$ in the near wall cell is copied to the wall cell.
Therefore a linear interpolation will also lead to this value at the wall face.

To achieve a zero gradient like behavior for $k$ at the wall face, we are modifying the coefficient matrix with an implicit relaxation.
For the wall cell the equation is relaxed to the value for $k$ of the near wall cell of the previous iteration.

As mentioned before, the kinematic eddy viscosity $\nu_t$ is modified directly at the face of the wall.
In the code of OpenFOAM there exist different parts where this viscosity or its dynamic equivalent $\mu_t$
or the product of $\mu_t$ and $\left( \nabla \mathbf{u} \right)^T$ are interpolated to the face with the help of a linear interpolation scheme.
A modified version of the linear interpolation scheme which manipulates the values for the wall face is possible but complicated.
Therefore, we decided to manipulate not the face value, but the value of the wall cell for $\nu_t$ in a way,
that a linear interpolation will lead to the desired value at the face.
Then, assuming an equidistant grid, the kinematic eddy viscosity in the wall cell is set to
\begin{equation}
 \nu_{t \text{wallCell}} = \text{max}\left(2\nu_{tf_\text{target}}  - \nu_{t \text{nearWallCell}},0 \right)
\end{equation}
with $\nu_{tf_\text{target}}$ as the target value for the face.
The limitation to a minimal value of zero is done for stability reasons.

In the current implementation we are using the equations for a smooth wall function.
In \cite{Roulund2005} it has been shown that the wall roughness has a small influence onto the result.
Therefore, our implementation of a smooth wall function should be a valid starting point.
Nevertheless, for future work one should implement a rough wall function.
%
%
%
\subsubsection{Special VOF-wall treatment}
\label{Special VOF-wall treatment}
Using the Volume-of-Fluid method to describe the sediment surface, this surface is usually described by at least two cells.
One cell contains a sediment volume fraction greater or equal to $\alpha_s=0.6$, which we will call the sediment-surface-soil-cell (SS-S-Cell),
as it is on the side of the soil.
The other cell contains a sediment volume fraction lesser than $\alpha_s=0.6$ and is called sediment-surface-water-cell (SS-W-Cell),
as it is on the side of the water.
Such surface description leads to two problems:\\

\textbf{(1) Illegal sediment transport:}\\
Using only one velocity field for both phases, as it is the case in our method, one has two problematic possibilities.
(a) Applying the soil viscosity on both sediment-surface cells.
For a sediment acting as a wall, this will lead to a slow or zero velocity in both cells.
Therefore, the water in the SS-W-cell will be too slow and also the internal wall function will not work as desired.
(b) The soil viscosity is only applied for the SS-S-cell and not for the SS-W-cell.
This leads to an almost correct velocity for the water in the SS-W-cell, but to a too high velocity for the soil in this cell.
Therefore, although the soil might act as a solid wall, the sediment in the SS-W-cell is being transported away,
which is an illegal sediment motion.\\

\textbf{(2) Wrong suspension detection:}\\
As given in equation \eqref{eq_blend_mu_susp} the suspension is defined in the interval $[0.01, 0.6]$ for the sediment volume fraction plus a blending region.
With respect to the given VOF-approach, for most SS-W-cells a suspension will be detected,
although the cells contain only such a volume fraction to describe the free surface and not the suspension.\\

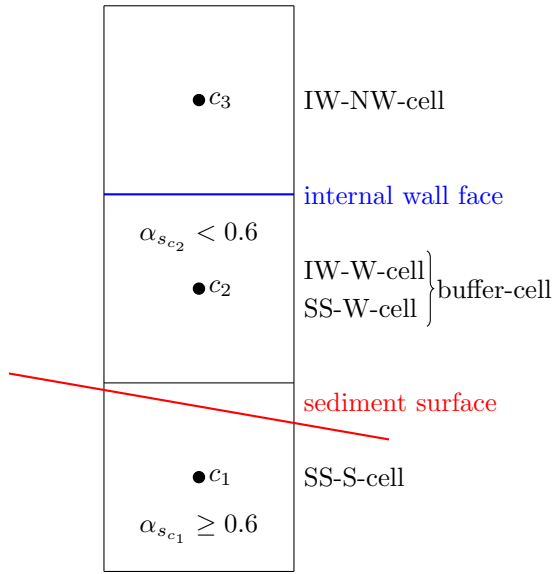
\begin{figure}
 \centering
    \begin{tikzpicture}[scale=2.5, decoration=brace]
    
    \draw[black] (0.0,2.0) -- (1.0,2.0);
    \draw[blue, thick] (0.0,1.0) -- (1.0,1.0);
    \draw[black] (0.0,0.0) -- (1.0,0.0);  
    \draw[black] (0.0,-1.0) -- (1.0,-1.0);
    
    \draw[black] (0.0,-1.0) -- (0.0,2.0);
    \draw[black] (1.0,-1.0) -- (1.0,2.0);
        
    \draw[red, thick] (-0.5,0.05) -- (1.5,-0.3); \draw[red, thick] (1.0,-0.1) node[right]{sediment surface};
    
    \node[draw,circle,inner sep=1.5pt,fill] at (0.5,1.5) {}; \draw[black, thick] (0.5,1.5) node[right]{$c_{3}$};
    \node[draw,circle,inner sep=1.5pt,fill] at (0.5,0.5) {}; \draw[black, thick] (0.5,0.5) node[right]{$c_{2}$};
    \node[draw,circle,inner sep=1.5pt,fill] at (0.5,-0.5) {}; \draw[black, thick] (0.5,-0.5) node[right]{$c_{1}$};

    \draw[black, thick] (0.5,0.9) node[below]{$\alpha_{s_{c_{2}}} <0.6$};
    \draw[black, thick] (0.5,-0.9)node[above]{$\alpha_{s_{c_{1}}}\ge 0.6$};
    
    \draw[blue, thick] (1.0,1.0)node[right]{\text{internal wall face}};
    
    \draw[black, thick] (1.0,1.5) node[right]{\text{IW-NW-cell}};
    
    \draw[black, thick] (1.0,0.6) node[right]{\text{IW-W-cell}};
    \draw[black, thick] (1.0,0.4) node[right]{\text{SS-W-cell}};   
    
    \draw[decorate] (1.7,0.7) -- (1.7,0.3); \draw[black, thick] (1.7,0.5) node[right]{\text{buffer-cell}}; 
    
    \draw[black, thick] (1.0,-0.5)node[right]{\text{SS-S-cell}};
    
    
%
    \end{tikzpicture}
 \caption{Buffer cell approach}
 \label{Buffer cell approach}
\end{figure}

Our solution for the first problem is based on possibility (a), where the soil viscosity is applied on both sediment-surface cells.
Furthermore, we do not apply the internal wall function at the sediment-surface face, but onto the face shifted one cell away from the sediment.
The details are shown in figure \ref{Buffer cell approach}.
The cells $c_1$ and $c_2$ are the SS-S-cell and SS-W-cell as described previously.
The wall face which is used for the internal wall function is shifted one cell away, compare with figure \ref{Wall cells at sediment-surface}.
With respect to the internal wall function, described in \ref{Method Internal wall function} the wall cell is now called the internal-wall wall-cell (IW-W-cell)
and the near wall cell is called the internal-wall nearwall-cell (IW-NW-cell).
As the SS-W-cell and IW-W-cell are the same and as they build some kind of buffer region, one can also call this cell a buffer-cell.
The soil viscosity is calculated in the SS-S-cell and then copied to the buffer-cell.
For a correct calculation of this viscosity the second invariant of the strain rate tensor $j$ is calculated at the buffer-cell and copied to the SS-S-cell,
before the soil viscosity is being updated.

The identification of the SS-S and SS-W cells is straightforward and can be done by comparing the sediment volume fractions of the two cells sharing one face.
After the identification of these cells,
the identification of the IW-W and IW-NW cells can be done
by comparing the sediment volume fractions and SS-cell identification of the two cells sharing one face.

We would like to emphasize, that this approach is only applied for the simulations in section \ref{Scour around a vertical pile}.
Furthermore, it is important that the cells have a uniform size in the region, where this buffer-cell approach is applied.
For future work it might be a better and cleaner solution to use two velocity fields,
one for the water phase and one for the sediment face, instead of the buffer-cell approach.\\
For the solution of the second problem, we stay with the denotation given in figure \ref{Buffer cell approach},
although it is not necessary to apply the buffer-cell approach to solve the second problem.
To check, if the SS-W-cell contains the suspension we do not only use the sediment volume fraction of this cell,
but also the sediment volume fraction of the IW-NW-cell.
The SS-W-cell is than only considered as a suspension cell, if the volume fraction of the IW-NW-cell exceeds a limit of $0.001$.
%
\subsection{Discretization details}
\label{Discretization} 

All terms of the given equations are discretized with 2nd order accurate schemes.
Some local assumptions are possible, where a 1st order scheme is used to fulfill stability criteria as explained later in this chapter.
The major parts of the discretization procedure in OpenFOAM are explained in \cite{Jasak1996}.
In this chapter only the critical parts, which are required for a stable and accurate simulation, are described.

The time derivatives are discretized with a 2nd order scheme. 
OpenFOAM offers two 2nd order schemes, \textit{backward} and \textit{Crank-Nicolson}.
The backward scheme is based on quadratic interpolation using the values of two previous time steps.
The Crank-Nicolson scheme is based on the trapezoidal rule.
Both schemes may lead to oscillating solutions under specific circumstances.
This oscillation tendency is especially large due to a high density-jump at the interface.
Knowing, that future simulations will also contain a water-air interface, this problem is severe.
To avoid oscillations, we have implemented a modified version of the backward scheme.
Under the assumption of a given M-Matrix characteristic as defined by \cite{Hackbusch2017},
the modified scheme locally blends to 1st order Euler, if the right hand side of the unmodified backward scheme has a different sign than the right hand side of the Euler scheme.
This blending is only necessary in very few cells and numerical tests with sea waves have shown,
that the solution still has the accuracy of a full 2nd order solution,
while avoiding any oscillations.

The convective terms, except the one of the Volume-of-Fluid equation \eqref{eq_vof_transport}, are discretized with the 2nd order \textit{Gamma} scheme presented in \cite{Jasak1999}.
The Gamma scheme is based on 2nd order linear interpolation but locally blends to 1st order upwind to fulfill the convective boundedness criterion.
Additionally it redefines the normalized variable approach,
in a way that it does not require the far upwind node addressing, simplifying the implementation of this scheme for unstrucutred grids.

The convective term of the Volume-of-Fluid equation \eqref{eq_vof_transport} is discretized with the
\textit{blended interface capturing scheme with reconstruction}, presented in \cite{Wackers2011}.
This scheme is based on downwind differencing as the downwind scheme has a compressive character leading to a sharp interface.
The scheme locally blends to the Gamma scheme to fulfill the convective boundedness criterion.
Additional local blending is introduced for high Courant-Friedrichs-Lewy-Numbers and for specific angles between the face normal and the normal of the free-surface
to avoid that the free-surface aligns with the grid.
The far upwind node, required for the normalized variable approach, is estimated using a reconstruction of the corresponding value
based on a search path algorithm and a weighted interpolation.
The details of our implementation are described in \cite{Meyer2016}.
It is important to know, that such an high-resolution scheme resolves sharp jumps sufficiently but also converts smooth gradients into a sharp jump.
This inherently suppresses the development of a suspension.
An alternative way, which has been investigated by us, was to use the Gamma-scheme at the sediment surface.
The Gamma scheme resolves smooth gradients sufficiently but smears a sharp jump into a smooth gradient.
The Gamma scheme allows the development of a suspension.
However, the smearing of the sediment interface inherently carries away the sediment,
even if the sediment should act as an solid and a zero sediment velocity is given.
For now, we have decided to use the mentioned high-resolution scheme.

\begin{figure}
 \centering
    \begin{tikzpicture}[xscale=0.013, yscale=2.0]
    %
    \draw[black, thick, domain=0.0:1.0, variable=\y] plot ({-10*\y+20},{\y});
    \draw[black, thick, domain=-1.0:0.0, variable=\y] plot ({-200*\y+20},{\y});
    \draw[black, thick] (20,1.0) node[above]{$p$};
    \draw[black, thick, domain=0.0:1.0, variable=\y] plot ({220},{\y});
    \draw[black, thick, domain=-1.0:-0.0, variable=\y] plot ({260},{\y});
    \draw[black, thick] (240,1.0) node[above]{$\nabla p$};
    \draw[black, thick, domain=0.0:1.0, variable=\y] plot ({320},{\y});
    \draw[black, thick, domain=-1.0:-0.0, variable=\y] plot ({360},{\y});
    \draw[black, thick] (340,1.0) node[above]{$\rho$};
    \draw[black, thick, domain=0.08:1.0, variable=\y] plot ({420},{\y});
    \draw[black, thick, domain=-1.0:-0.08, variable=\y] plot ({420},{\y});
    \draw[black, thick] (420,1.0) node[above]{$\nabla\rho$};
    \draw[black, thick, domain=0.0:1.0, variable=\y] plot ({500},{\y});
    \draw[black, thick, domain=-1.0:0.0, variable=\y] plot ({500},{\y});
    \draw[black, thick] (500,1.0) node[above]{$\frac{\nabla p}{\rho}$};
    \draw[blue, thick] (115.0,0.0) node[above]{$\Gamma$};
    %
    \draw[blue, ->] (-0.25,0) -- (560,0) node[below] {value};
    \draw[<->] (0,-1.25) -- (0,1.25) node[left] {h};
    \end{tikzpicture}
 \caption{Characteristics of pressure and density at the interface $\Gamma$}
 \label{Pressure and Density at the free surface}
\end{figure}
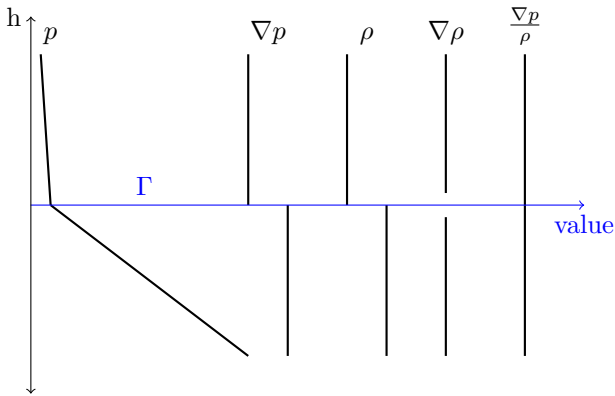

The density jump at the free-surface leads to a kink in the pressure characteristics and a jump in the pressure gradient as shown in figure \ref{Pressure and Density at the free surface}.
Using standard schemes will smear this characteristics.
This leads to errors resulting in overestimated velocities in the cells containing the lighter phase directly above the free-surface.
This error becomes larger if the cell size is reduced, making grid independence studies unpredictable (\cite{Meyer2016}).
To solve this problem the method presented in \cite{Queutey2007} has been implemented.
This method reconstructs the jump induced characteristics using the pressure gradient normalized with the density.
Our implementation is described in \cite{Meyer2016} and it is also shown, that this approach avoids the unphysical high velocities.\\
Besides the calculation of the 2nd invariant of the strain rate tensor $j$, see equation \eqref{eq_j},  some other terms require the explicit calculation of the cell centered gradient of the velocity $\nabla \mathbf{u}$.
In all cases except for the calculation of $j$ the \textit{cellLimited} gradient calculation of OpenFOAM is used.
The limitation works as follows
\begin{itemize}
 \item Calculate the cell centered gradient by converting the volume integrals to surface integrals based on the Gauss theorem
 or calculate the gradient by using a least squares fit.
 \item For each face:
 \begin{itemize}
   \item Extrapolate the cell centered value to the face by using the calculated gradient.
   \item Compare the extrapolated face value with the cell centered value of the neighbour cell also sharing this face.
   \item If necessary limit the cell centered gradient to guarantee that no extrapolated face value exceeds the cell centered value of the neighbour cell.
 \end{itemize}
\end{itemize}
This limitation is used as our standard approach to maintain convergence.
It is important to notice that it is not allowed to use this gradient limitation for the calculation of $j$.
The starting condition for the sediment always includes a zero velocity inside the sediment.
Using the above gradient limitation will inherently reduce $j$ to zero inside the sediment.
As $j$ represents the force acting onto the sediment, the gradient limitation prevents the flow from acting onto the sediment.
%
\subsection{Solution method}
\label{Solution method} 

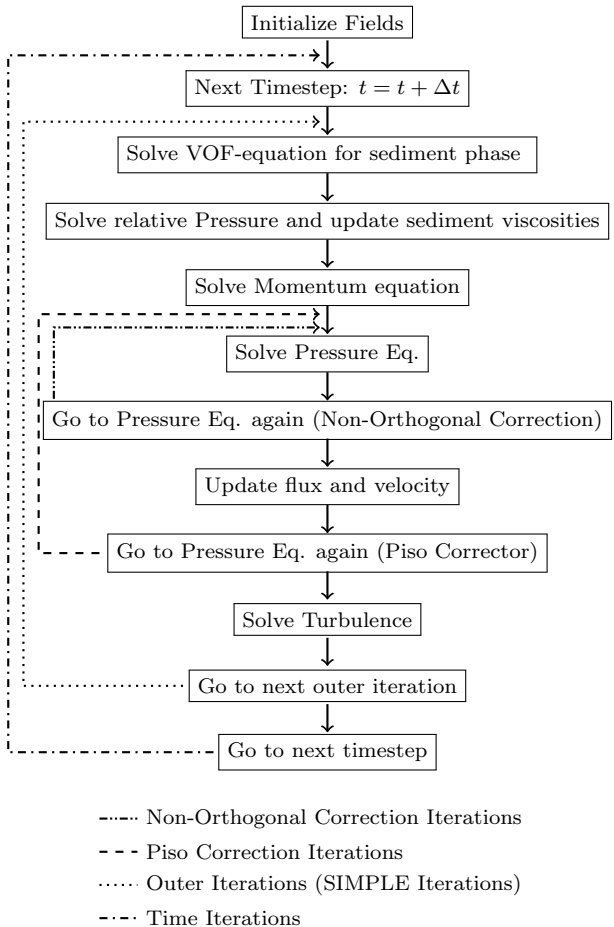
\begin{figure}
 \centering
    \begin{tikzpicture}[xscale=1.0, yscale=2.2]
      \footnotesize

      \draw[black] (0.0,0.0)node[]{$\boxed{\textrm{Initialize Fields}}$};
      \draw[black] (0.0,-0.4*1)node[]{$\boxed{\textrm{Next Timestep: } t = t+ \Delta t}$};
      \draw[black] (0.0,-0.4*2)node[]{$\boxed{\textrm{Solve VOF-equation for sediment phase }}$};
      \draw[black] (0.0,-0.4*3)node[]{$\boxed{\textrm{Solve relative Pressure and update sediment viscosities}}$};
      \draw[black] (0.0,-0.4*4)node[]{$\boxed{\textrm{Solve Momentum equation}}$};
      \draw[black] (0.0,-0.4*5)node[]{$\boxed{\textrm{Solve Pressure Eq.}}$};
      \draw[black] (0.0,-0.4*6)node[]{$\boxed{\textrm{Go to Pressure Eq. again (Non-Orthogonal Correction)}}$};
      \draw[black] (0.0,-0.4*7)node[]{$\boxed{\textrm{Update flux and velocity}}$};
      \draw[black] (0.0,-0.4*8)node[]{$\boxed{\textrm{Go to Pressure Eq. again (Piso Corrector)}}$};
      \draw[black] (0.0,-0.4*9)node[]{$\boxed{\textrm{Solve Turbulence}}$};
      \draw[black] (0.0,-0.4*10)node[]{$\boxed{\textrm{Go to next outer iteration}}$};
      \draw[black] (0.0,-0.4*11)node[]{$\boxed{\textrm{Go to next timestep}}$};

      \draw[black, thick, ->] (0.0,-0.11) -- (0.0,-0.28);
      \draw[black, thick, ->] (0.0,-0.11 -0.4*1) -- (0.0,-0.28 -0.4*1);
      \draw[black, thick, ->] (0.0,-0.11 -0.4*2) -- (0.0,-0.28 -0.4*2);
      \draw[black, thick, ->] (0.0,-0.11 -0.4*3) -- (0.0,-0.28 -0.4*3);
      \draw[black, thick, ->] (0.0,-0.11 -0.4*4) -- (0.0,-0.28 -0.4*4);
      \draw[black, thick, ->] (0.0,-0.11 -0.4*5) -- (0.0,-0.28 -0.4*5);
      \draw[black, thick, ->] (0.0,-0.11 -0.4*6) -- (0.0,-0.28 -0.4*6);
      \draw[black, thick, ->] (0.0,-0.11 -0.4*7) -- (0.0,-0.28 -0.4*7);
      \draw[black, thick, ->] (0.0,-0.11 -0.4*8) -- (0.0,-0.28 -0.4*8);
      \draw[black, thick, ->] (0.0,-0.11 -0.4*9) -- (0.0,-0.28 -0.4*9);
      \draw[black, thick, ->] (0.0,-0.11 -0.4*10) -- (0.0,-0.28 -0.4*10);
      
      \draw[black, thick, densely dashdotdotted, ->] (-3.6,-0.4*6 + 0.13) -- (-3.6,-0.2-0.4*4.1) -- (-0.1,-0.2-0.4*4.1);
      
      \draw[black, thick, dashed, ->] (-3.0,-0.4*8) -- (-3.8,-0.4*8) -- (-3.8,-0.2-0.4*3.9) -- (-0.1,-0.2-0.4*3.9);
      
      \draw[black, thick, dotted, ->] (-1.9,-0.4*10) -- (-4.0,-0.4*10) -- (-4.0,-0.2 -0.4*1) -- (-0.1,-0.2-0.4*1);
      
      \draw[black, thick, dashdotted, ->] (-1.5,-0.4*11) -- (-4.2,-0.4*11) -- (-4.2,-0.2-0.4*0) -- (-0.1,-0.2-0.4*0);
      
      \draw[black, thick, densely dashdotdotted] (-3,-0.4*12.0) -- (-2.5,-0.4*12.0) node[right]{Non-Orthogonal Correction Iterations};
      \draw[black, thick, dashed] (-3,-0.4*12.5) -- (-2.5,-0.4*12.5) node[right]{Piso Correction Iterations};
      \draw[black, thick, dotted] (-3,-0.4*13.0) -- (-2.5,-0.4*13.0) node[right]{Outer Iterations (SIMPLE Iterations)};
      \draw[black, thick, dashdotted] (-3,-0.4*13.5) -- (-2.5,-0.4*13.5) node[right]{Time Iterations};

      \normalsize
    \end{tikzpicture}
 \caption{Solution Algorithm}
 \label{Solution Algorithm}
\end{figure}

The coupled equations are solved with a segregated SIMPLE-like algorithm.
After integrating over the volume, the Gauss Theorem is used to transform the volume integrals to surface integrals.
The linearized, semi-discretized momentum equation \eqref{eq_momentum_Bingham_final} can be written as
\begin{equation}
       \label{semi_disc_momentum_eq_org}
        a_\text{d} \mathbf{u}_\text{d} + \sum_n a_\text{n} \mathbf{u}_\text{n}
        = -\nabla p + \mathbf{s}_\text{w/o p} \hspace{2mm}.
\end{equation}
Here, $a$ represents the elements of the coefficient matrix $\mathbf{A}$ and the subscripts $\text{d}$ and $\text{n}$ mark the main diagonal- and neighbor-elements.
All sources and contributions to the right hand side except the pressure gradient are included in $\mathbf{s}_\text{w/o p}$.
Rearranging \eqref{semi_disc_momentum_eq_org} to $\mathbf{u}_d$ yields the velocity equation:
      \begin{equation}
       \label{velocity_eq_org}
        \mathbf{u}_\text{d}
        = \frac{1}{a_\text{d}}
        \left( 
        -\nabla p + \mathbf{s}_\text{w/o p}
        - \sum_n a_\text{n} \mathbf{u}_\text{n}
        \right) \hspace{2mm}.
       \end{equation}
Substituting and rearranging \eqref{velocity_eq_org} into \eqref{eq_continuity} yields the poisson equation for the pressure:
\begin{equation}
        \nabla \cdot \left( \frac{1}{a_\text{d}}\nabla p \right)
        = \nabla \cdot \frac{1}{a_\text{d}}
        \left(
        \mathbf{s}_\text{w/o p} - \sum_n a_\text{n} \mathbf{u}_\text{n}
        \right) \hspace{2mm}.
\end{equation}
The key points of the solution algorithm are shown in figure \ref{Solution Algorithm}.
The momentum equation is relaxed implicitly with a relaxation factor of 0.7.
The pressure is relaxed explicitly with a relaxation factor of 0.3. 
The pressure relaxation is done after updating the flux and before updating the velocity.
The sediment viscosities are relaxed explicitly with a relaxation factor of 0.1.
For all simulations, the relaxation of the sediment viscosities has been absolutely necessary to achieve convergence.
%
%
%

%% file: chapters/Results.tex
\section{Results}
\label{Results} 
\subsection{Internal wall function}
\label{Internal wall function} 

The internal wall function is verified with a 2D test case.
In this test case, the bottom boundary is treated as a wall.
The flow is initialized with zero velocity and accelerated with a constant horizontal acceleration
so that a boundary layer is being built at the bottom.
Three simulation setups are investigated:
(i) the sediment surface is represented by a fixed no-slip wall boundary condition employing the standard wall function,
(ii) the sediment surface is located inside the domain and represented by the VOF-function without using any special wall treatment and
(iii) our new internal wall function is used additionally.\\
The first simulation is using the standard wall function for smooth walls.
The domain has a length of $18.0 \text{m}$ and a depth of $22.0 \text{m}$.
The cells have a length of $1.0 \text{m}$ and a height of $0.0625 \text{m}$ at the bottom.
The chosen acceleration is $0.00175 \frac{\text{m}}{\text{s}^2}$ and the time step size is $0.02 \text{s}$.\\
For the second simulation the same setup is used, but the bottom domain is extended by $8 \text{m}$.
The extended region is filled with sediment.
No Volume-of-Fluid transport is calculated as the sediment should act like a constant wall.
The sediment viscosity always equals the maximal viscosity of $1500 \frac{\text{Ns}}{\text{m}^2}$ .
For the second simulation no internal wall function approach is applied.\\
The third simulation equals the second one, but with the new internal wall function applied.
The result of the first simulation is used as a reference solution.

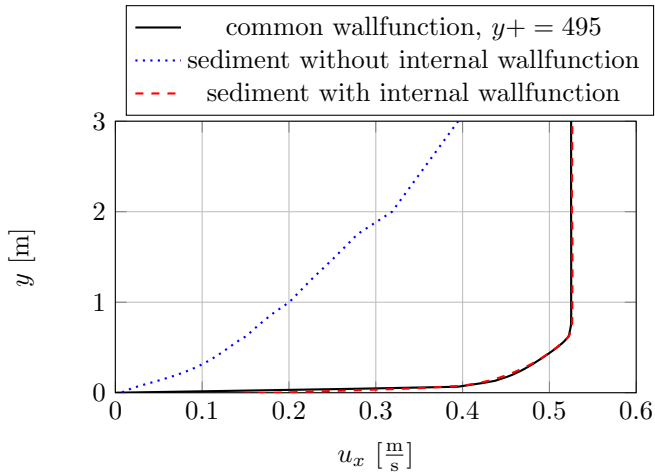
\begin{figure}
 \centering
	\begin{tikzpicture}
		\begin{axis}
			[
			  xlabel={$u_x$ [$\frac{\text{m}}{\text{s}}$]},
			  ylabel={$y $ [m]},			  
			  xmin=0.0,
			  xmax=0.6,
			  ymin=0.0,
			  ymax=3.0,
			  y=1.2cm,
			  grid=both,
			  legend style ={ at={(0.02,1.02)},
			  anchor=south west, draw=black}
			]
			\addplot[black, thick] table[y expr={\thisrowno{0} + 22}, x expr={\thisrowno{1}}]
				{./WallFunctionData/01_line_Ux_WithoutSediment_WithWallFunction.xy};
			\addlegendentry{common wallfunction, $y+=495$};		
			\addplot[blue, dotted, thick] table[y expr={\thisrowno{0} + 22}, x expr={\thisrowno{1}}]
				{./WallFunctionData/02_line_Ux_WithSediment_WithoutInternalWallFunction.xy};
			\addlegendentry{sediment without internal wallfunction};
			\addplot[red, dashed, thick] table[y expr={\thisrowno{0} + 22}, x expr={\thisrowno{1}}]
				{./WallFunctionData/line_Ux.xy};
			\addlegendentry{sediment with internal wallfunction};	
		\end{axis}
	\end{tikzpicture}
\caption{Boundary layer profiles of the horizontal velocity $u_x$ for the three simulations}
\label{wallfunction_ux} 
\end{figure}

Figure \ref{wallfunction_ux} shows the velocity profiles for the three simulations at $t=300\text{s}$.
The results show, that the velocity profile of the second case is clearly underestimated.
Therefore, in a real scour simulation, the forces acting on the sediment are too small.
Additionally, anticipating the simulation of subsection \ref{Scour around a vertical pile}, the horse shoe vortex will not have the correct shape,
as this vortex is influenced by the size of the boundary layer.
The velocity profile of the third test case using the new internal wall function is significantly better.
Figures \ref{wallfunction_k} and \ref{wallfunction_omega} show the profile for the kinetic energy $k$ and the specific dissipation rate $\omega$.
Again, the new internal wall function improves the result significantly.
Only, very close to the wall, the values differ from the reference solution and only have about the half magnitude.
However, with respect to equation \eqref{eq_nut}, both errors are canceling each other out,
which is also noticeable in figure \ref{wallfunction_nut} which shows the profile for the kinematic eddy viscosity $\nu_t$.
This figure also shows, that the kinematic eddy viscosity is significantly overestimated, without using the internal wall function.
Such a high viscosity will influence the flow substantially.
For example the vortices behind a circular pile,
which are an important detail of the simulation in subsection \ref{Scour around a vertical pile},
will be completely different.
The test cases show that the new internal wall function is improving the boundary layer flow significantly.
%
 \begin{figure}
 \centering
	\begin{tikzpicture}
		\begin{axis}
			[			
			  xlabel={$k$ [$\frac{\text{m}^2}{\text{s}^2}$]},
			  ylabel={$y $ [m]},
			  xmax=0.02,
			  ymin=0.0,
			  ymax=3.0,
			  y=1.2cm,
			  grid=both,
			  legend style ={ at={(0.02,1.02)},
			  anchor=south west, draw=black}
			]
			\addplot[black, thick] table[y expr={\thisrowno{0} + 22}, x expr={\thisrowno{1}}]
				{./WallFunctionData/01_line_k_WithoutSediment_WithWallFunction.xy};
			\addlegendentry{common wallfunction, $y+=495$};			
			\addplot[blue, dotted, thick] table[y expr={\thisrowno{0} + 22}, x expr={\thisrowno{1}}]
				{./WallFunctionData/02_line_k_WithSediment_WithoutInternalWallFunction.xy};
			\addlegendentry{sediment without internal wallfunction};
			\addplot[red, dashed, thick] table[y expr={\thisrowno{0} + 22}, x expr={\thisrowno{1}}]
				{./WallFunctionData/line_k.xy};
			\addlegendentry{sediment with internal wallfunction};	
		\end{axis}
	\end{tikzpicture}
\caption{Boundary layer profiles of the kinetic energy $k$ for the three simulations}
\label{wallfunction_k} 
\end{figure}
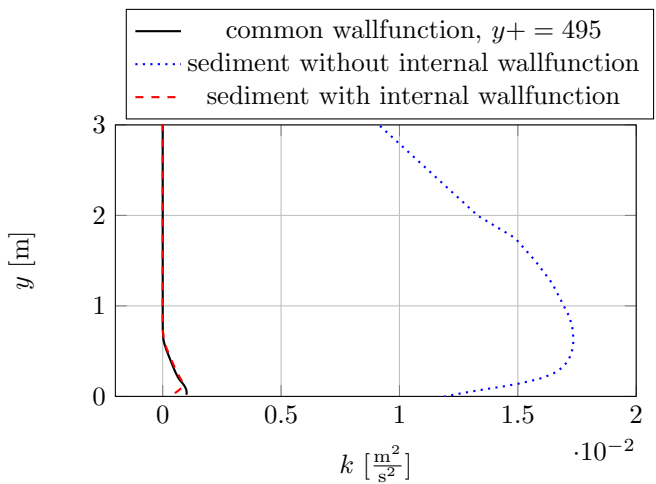
 \begin{figure}
 \centering
	\begin{tikzpicture}
		\begin{axis}
			[			
			  xlabel={$\omega$ [$\frac{1}{\text{s}}$]},
			  ylabel={$y $ [m]},
			  xmin=0.0,
			  ymin=0.0,
			  ymax=3.0,
			  y=1.2cm,
			  grid=both,
			  legend style ={ at={(0.02,1.02)},
			  anchor=south west, draw=black}
			]
			\addplot[black, thick] table[y expr={\thisrowno{0} + 22}, x expr={\thisrowno{1}}]
				{./WallFunctionData/01_line_omega_WithoutSediment_WithWallFunction.xy};
			\addlegendentry{common wallfunction, $y+=495$};			
			\addplot[blue, dotted, thick] table[y expr={\thisrowno{0} + 22}, x expr={\thisrowno{1}}]
				{./WallFunctionData/02_line_omega_WithSediment_WithoutInternalWallFunction.xy};
			\addlegendentry{sediment without internal wallfunction};
			\addplot[red, dashed, thick] table[y expr={\thisrowno{0} + 22}, x expr={\thisrowno{1}}]
				{./WallFunctionData/line_omega.xy};
			\addlegendentry{sediment with internal wallfunction};	
		\end{axis}
	\end{tikzpicture}
\caption{Boundary layer profiles of the specific dissipation rate $\omega$ for the three simulations}
\label{wallfunction_omega} 
\end{figure}
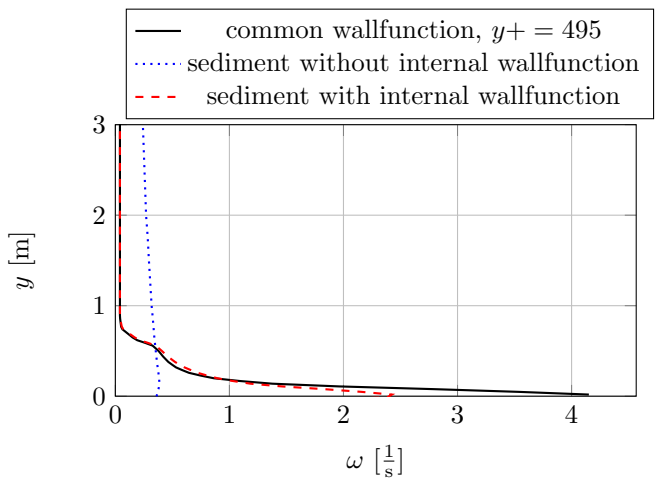
 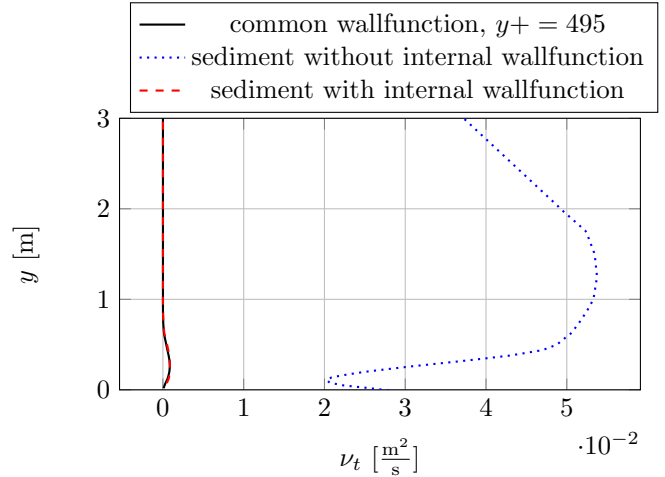
\begin{figure}
 \centering
	\begin{tikzpicture}
		\begin{axis}
			[			
			  xlabel={$\nu_t$ [$\frac{\text{m}^2}{\text{s}}$]},
			  ylabel={$y $ [m]},
			  ymin=0.0,
			  ymax=3.0,
			  y=1.2cm,
			  grid=both,
			  legend style ={ at={(0.02,1.02)},
			  anchor=south west, draw=black}
			]
			\addplot[black, thick] table[y expr={\thisrowno{0} + 22}, x expr={\thisrowno{1}}]
				{./WallFunctionData/01_line_nut_WithoutSediment_WithWallFunction.xy};
			\addlegendentry{common wallfunction, $y+=495$};			
			\addplot[blue, dotted, thick] table[y expr={\thisrowno{0} + 22}, x expr={\thisrowno{1}}]
				{./WallFunctionData/02_line_nut_WithSediment_WithoutInternalWallFunction.xy};
			\addlegendentry{sediment without internal wallfunction};
			\addplot[red, dashed, thick] table[y expr={\thisrowno{0} + 22}, x expr={\thisrowno{1}}]
				{./WallFunctionData/line_nut.xy};
			\addlegendentry{sediment with internal wallfunction};
		\end{axis}
	\end{tikzpicture}
\caption{Boundary layer profiles of the kinematic eddy viscosity $\nu_t$ for the three simulations}
\label{wallfunction_nut} 
\end{figure}
\subsection{Scour downstream of an apron}
\label{Scour downstream of an apron}
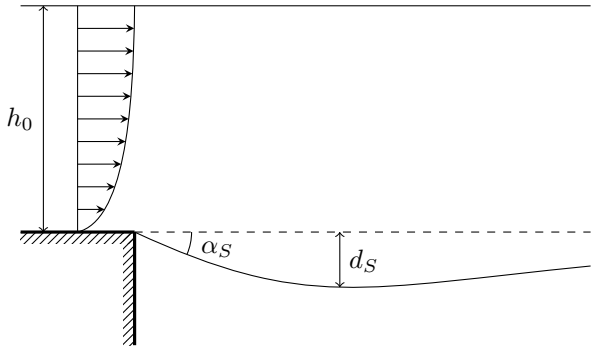
\begin{figure}
 \centering
    \begin{tikzpicture}[scale=1.5]
    
    \draw[black, very thick] (0.0,1.0) -- (1.0,1.0);
    \draw[black, very thick] (1.0,1.0) -- (1.0,0.0);
    
    \draw[black, dashed] (1.0,1.0) -- (5.0,1.0);
    
    \draw[black] (0.0,3.0) -- (5.0,3.0);
    
    \fill[pattern=north east lines, pattern color=black] (0,1) rectangle (1,0.9);
    \fill[pattern=north east lines, pattern color=black] (1,1) rectangle (0.9,0);
    
    \draw[name path=SedimentCurve] (1,1)..controls(2.5,0.3) and(3.0,0.5)..(5.0,0.7);
    
    \path[name path=hSMeasureCurve] (2.8,1.0) -- (2.8,0);
    \draw[-stealth,<->,name intersections={of=SedimentCurve and hSMeasureCurve}] (2.8,1.0) -- (intersection-1);
    \node[right] at (2.8,0.75) {$d_S$};

    \draw[<->](0.2,1)--(0.2,3);
    \node[left] at (0.2,2) {$h_0$};
    
    \draw (1.5,1) arc (0:-23:0.5);
    \node[right] at (1.5,0.85) {$\alpha_S$};
    
      \draw[](0.5,1)--(0.5,3);
      \draw[name path=VelocityCurve] (0.5,1)..controls(0.9,1.1) and(0.99,2.0) ..(1.0,3);
      \foreach \y in {1.2,1.4,...,2.8}
      {
	\path[name path=horizontal] (0.5,\y) -- + (0.5,0);
	\draw[-stealth,name intersections={of=VelocityCurve and horizontal}] (0.5,\y) -- (intersection-1);
      }
    \end{tikzpicture}
 \caption{Testcase: scour downstream of an apron}
 \label{Testcase: scour downstream of an apron}
\end{figure}

The process of scouring of non-cohesive materials behind an apron was investigated experimentally by \cite{Breusers1965}.
First equations describing this process are given which allow evaluating the conformity of the scour hole in model and prototype.
Here the results are used to validate the simulation method.
Figure \ref{Testcase: scour downstream of an apron} shows the setup of the test case.
At the left side a velocity with a suitable boundary layer is given.
The sediment is initialized as a flat bed.
It is shown, that the scouring depth $d_S$ increases exponentially
according to equation \eqref{eq_apron_deltaS}, being valid for a wide range of velocities, water depth and materials \citep{Breusers1965}.
\begin{equation}
 \label{eq_apron_deltaS}
 \frac{d_S}{h_0} = \left(\frac{t}{T_S} \right)^{n_S}
\end{equation}
Here, $h_0$ is the water depth at the end of the bottom protection, $T_S$ is the characteristic timescale of the scouring process
and $n_S$ is the exponent which characterizes the speed of the scour depth growth at the initial stage \citep{Cheng2017}.
Additionally, \cite{Breusers1967} described the behavior of the scour angle $\alpha_S$ with
\begin{equation}
 \label{eq_apron_alphaS}
 \alpha_S = \alpha_{S0}\left(1 - e ^{-\frac{t}{T_{\alpha S}}} \right)
\end{equation}
where $\alpha_{S0}$ is the equilibrium scour angle and $T_{\alpha S}$ is the equilibrium timescale of the scour angle.
For fine sand \cite{Breusers1965} gives a range of $13.4^\circ \le \alpha_{S0} \le 14.95^\circ$ for the equilibrium scour angle.
As mentioned by \cite{Amoudry2009} the experiments of \cite{Breusers1965}, \cite{Breusers1967} and \cite{Dietz1969} show that the shape of the scour hole
is almost independent of the flow velocity and the bed grain size if the flow velocity is significantly larger,
than the critical velocity required for sediment motion.\\

\textbf{Simulationsetup:}\\
The test case is using a two dimensional quadratic domain, with the velocity inlet directly at the edge of the apron.
The domain is $1\text{m}$ long and  $0.2 \text{m}$ height and has a water depth $h_0$ of $0.15 \text{m}$.
The grid cells height is $0.12\text{mm}$ and their length is $0.85\text{mm}$ directly at the edge.
To reduce the total number of cells, the cells are stretched with increasing distance to the edge.
The grid has 300 cells in horizontal direction and 155 cells in vertical direction
with a cell height of $10\text{mm}$ at the top and a cell length of $8.0\text{mm}$ at the outlet on the right side.

The part above the edge of the left side acts as the inlet with Dirichlet boundary conditions for the velocity, volume fractions,
turbulent kinetic energy $k$ and specific rate of dissipation $\omega$.
The part below the edge and the bottom of the domain are treated as walls.
The right side acts as an outlet with Neumann boundary conditions for all variables.
At the top of the domain a Neumann boundary condition is applied for all variables except the pressure which uses a Dirichlet boundary condition.
Specifying the pressure at the top instead of the right side allows that the sediment changes its height directly at the outlet.
We are not using so called Open Boundary Conditions at the outlet
although this has been recommended by \cite{Amoudry2009} reporting small advantages.\\

\textbf{Initial values:}\\
\cite{Cheng2017} imposed a rough-wall log-law velocity profile at the inlet
and an one-dimensional simulation was used to get values like the bed concentration.
Here, we are using a different approach based on an initial simulation to get the initial values
including the velocity and the two turbulence variables $k$ and $\omega$ for the inlet.
This simulation uses the same mesh, but with a shortened domain using only 10 cells in the horizontal direction.
Furthermore, Neumann boundary conditions are given for all variables at the inlet,
proper values for the affected variables are still dictated as a Dirichlet boundary condition at the walls.
The sediment is fixed and acts as a fixed wall using the maximal Bingham viscosity.
Therefore it is not necessary to solve the Volume-of-Fluid transport equation.
On the other hand, this means, that no initial suspension layer is produced.
The flow is initialized with a velocity of zero, the turbulent variables are initialized with a constant initial guess.
Finally a constant acceleration is applied in the whole domain during the whole initial simulation.
The simulation is executed until the changes in the velocity and turbulent profiles are negligible. 
In this study three initial simulations are done with accelerations of
$3.5 \cdot 10^{-3}\frac{\text{m}}{\text{s}^2}$, $7.0 \cdot 10^{-3} \frac{\text{m}}{\text{s}^2}$ and $14.0 \cdot 10^{-3} \frac{\text{m}}{\text{s}^2}$.
In all three cases the changes of the results are being seen as negligible after 2000s.
These initial simulations ensures that the profiles of the affected variables fits perfectly to the given wall function approach.\\

\textbf{Scour simulations:}\\
\nopagebreak
The parameters of the sediment model were adjusted to represent fine sand.
Therefore the rock density is $2650\frac{\text{kg}}{\text{m}^3}$,
the saturation is $0.7$,
the internal friction angle is $25^\circ$,
the cohesion is $0 \frac{\text{N}}{\text{m}^2}$,
and the minimal and maximal Bingham viscosities are set to $0\frac{\text{Ns}}{\text{m}^2}$ and $1500\frac{\text{Ns}}{\text{m}^2}$, respectively.
Furthermore, the internal wall function approach is activated.

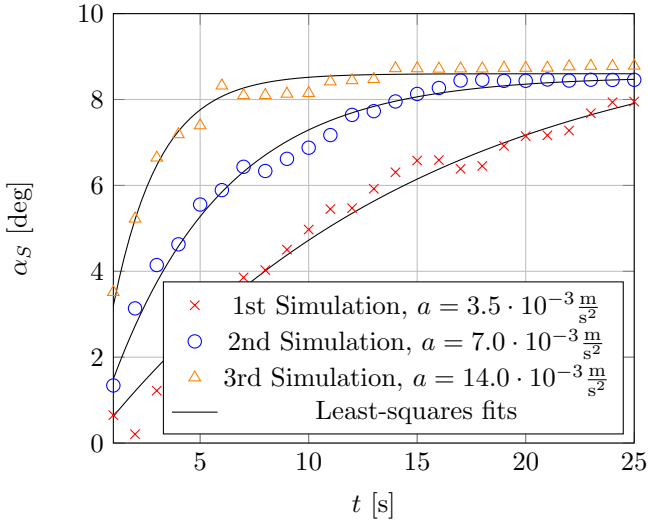
\begin{figure}
 {
	\begin{tikzpicture}
		\begin{axis}
			[			
			  xlabel={$t\textrm{ [s]}$},
			  ylabel={$\alpha_{S}\textrm{ [deg]}$},
			  xmin=1.0,
			  xmax=25.0,
			  ymin=0.0,
			  ymax=10,
			  grid=both,
			  legend style ={ at={(0.98,0.02)},
			  anchor=south east, draw=black}
			]
			\addplot[only marks, mark=x, mark size = 2.5pt,red] table[x expr={\thisrowno{0}}, y expr={\thisrowno{3}}]
				{./ScourApronData/ResultScourApron_a3_5eMinus3};
			\addlegendentry{1st Simulation, $a=3.5 \cdot 10^{-3}\frac{\text{m}}{\text{s}^2}$};
			\addplot[only marks, mark=o, mark size = 2.5pt,blue] table[x expr={\thisrowno{0}}, y expr={\thisrowno{3}}]
				{./ScourApronData/ResultScourApron_a7eMinus3};
			\addlegendentry{2nd Simulation, $a=7.0 \cdot 10^{-3}\frac{\text{m}}{\text{s}^2}$};
			\addplot[only marks, mark=triangle, mark size = 2.5pt,orange] table[x expr={\thisrowno{0}}, y expr={\thisrowno{3}}]
				{./ScourApronData/ResultScourApron_a14eMinus3};
			\addlegendentry{3rd Simulation, $a=14.0 \cdot 10^{-3}\frac{\text{m}}{\text{s}^2}$};
			\addplot[black, domain=0.0:25, samples=100]  {9.8*(1.0-pow(2.7183,-x/15.2))}; 
			\addlegendentry{Least-squares fits};
			\addplot[black, domain=0.0:25, samples=100]  {8.54*(1.0-pow(2.7183,-x/5.2))}; 
			%
			\addplot[black, domain=0.0:25, samples=100]  {8.6*(1.0-pow(2.7183,-x/2.15))}; 
		\end{axis}
	\end{tikzpicture}
}
\caption{Comparison of simulation results and least-squares fits for the scouring angle}
\label{scour_Apron_alphaS} 
\end{figure}

\begin{figure*}
\centering
\subcaptionbox{$t=2\text{s}$}{\includegraphics[width=0.5\textwidth]{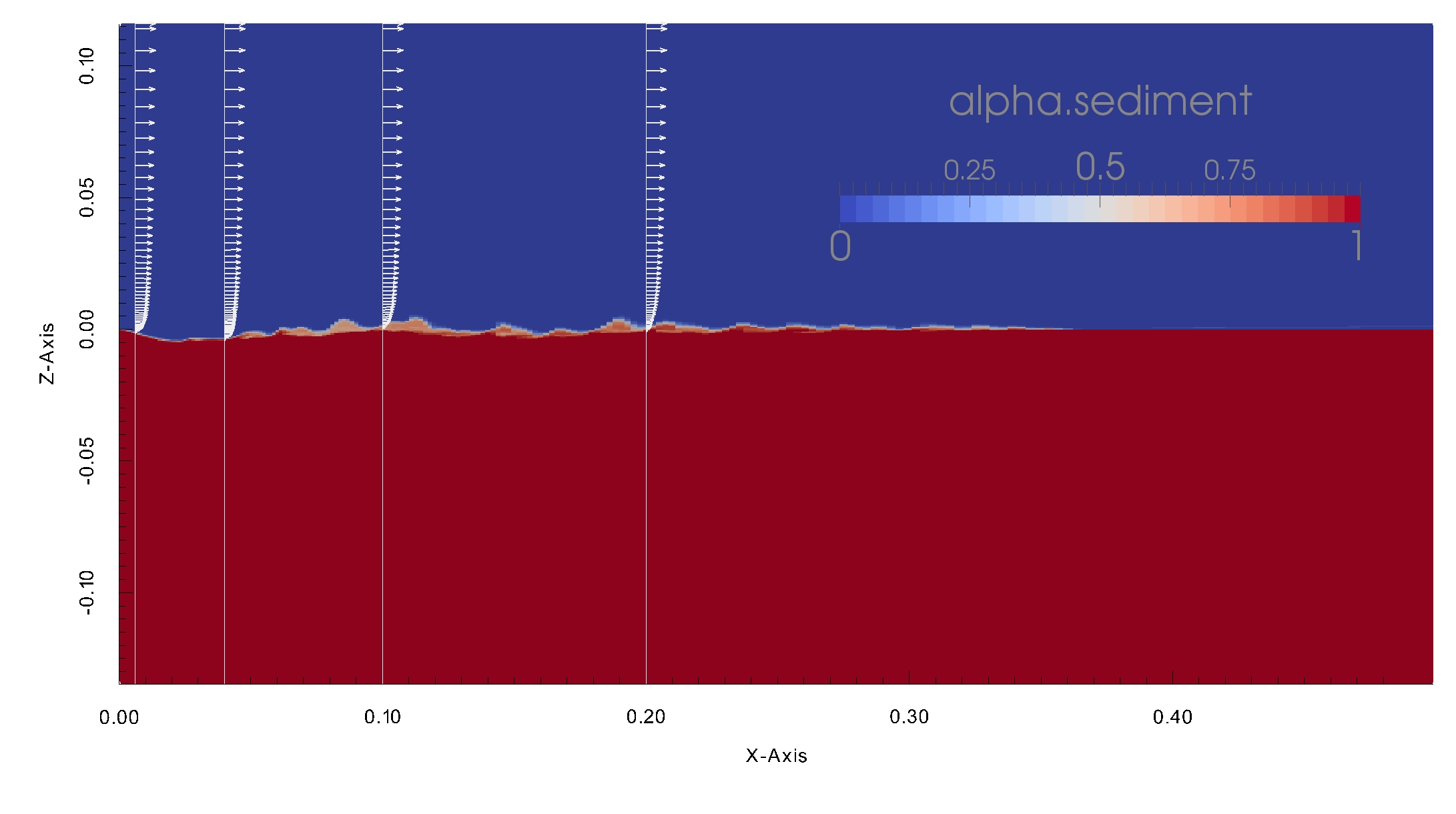}}%
\hfill
\subcaptionbox{$t=4\text{s}$}{\includegraphics[width=0.5\textwidth]{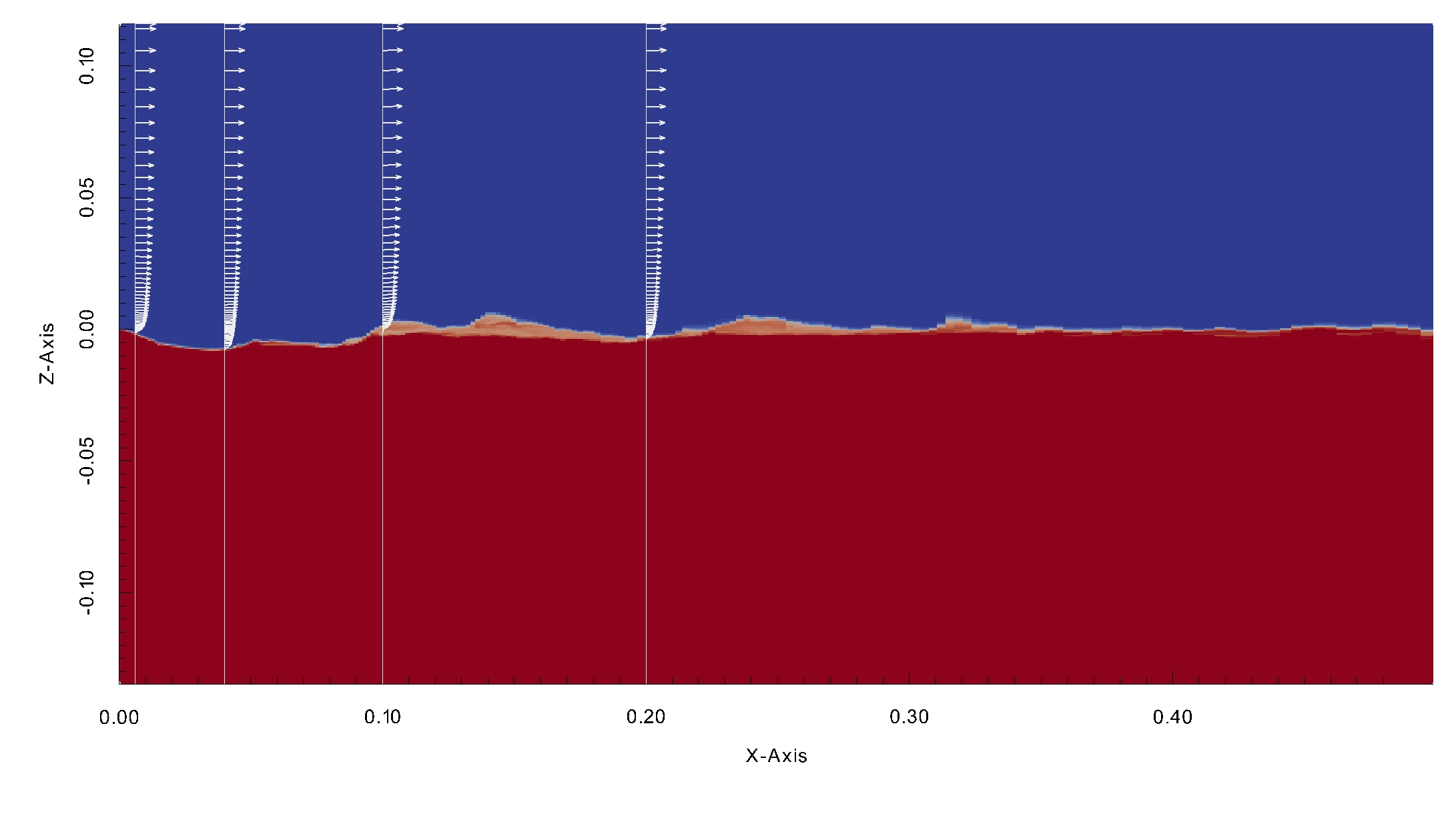}}%
\hfill
\subcaptionbox{$t=25\text{s}$}{\includegraphics[width=0.5\textwidth]{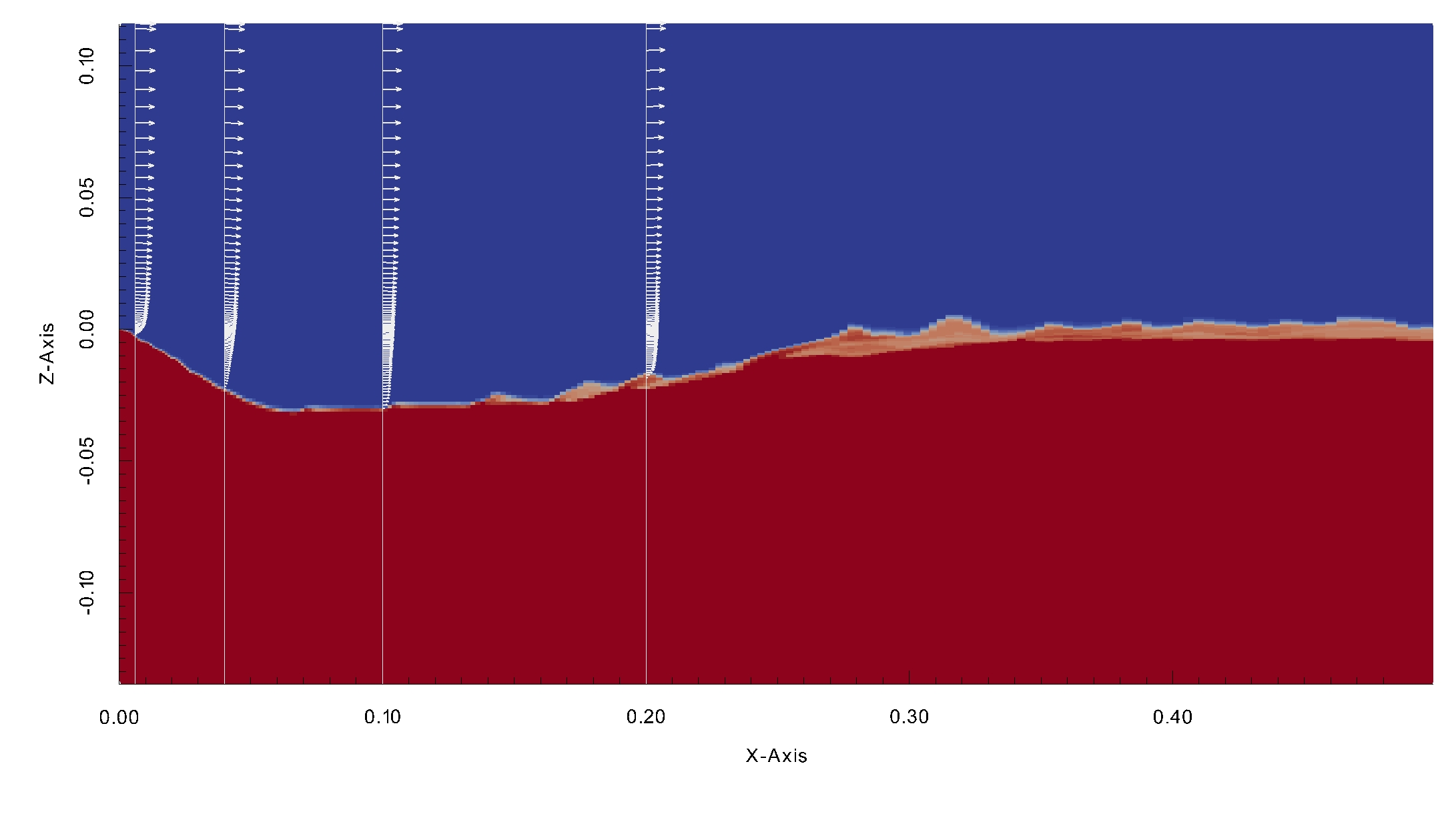}}%
\hfill
\subcaptionbox{$t=200\text{s}$}{\includegraphics[width=0.5\textwidth]{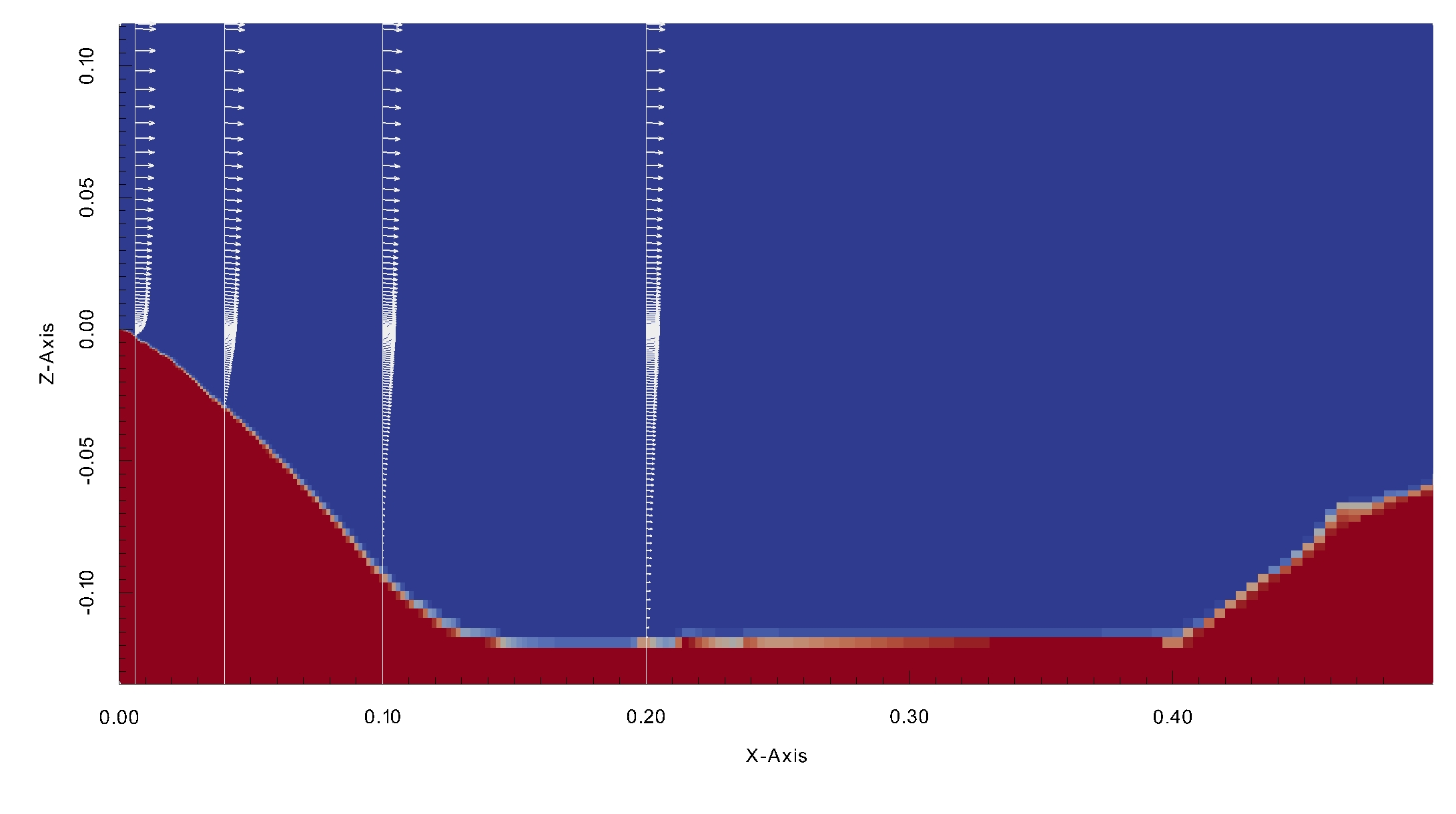}}%
\caption{Sediment volume Fraction and velocity profiles for different time steps}
\label{Sediment volume Fraction and velocity profiles for different time steps}
\end{figure*}

The results are shown in figures
\ref{scour_Apron_deltaS} and \ref{scour_Apron_alphaS}.
Figure \ref{scour_Apron_deltaS} shows the evolution of the scour hole depth $d_S$.
The solid black lines represent the least squares fits to equation \eqref{eq_apron_deltaS}.
The parameters of the optimal fits are given in table \ref{table:Scour Apron Least squares parameters}.
For all three cases a very good fit is achieved,
which shows that the simulation reflects the sediment behavior reported by \cite{Breusers1965}.
The estimated values for $n_S$ are similar to the values of the model of \cite{Amoudry2009} or the model of \cite{Cheng2017}.
As mentioned by \cite{Cheng2017}, referring to \cite{Buchko1988}, this values are at the high range but still reasonable.
\begin{figure}[]
	\begin{tikzpicture}
		\begin{axis}
			[			
			  xlabel={$t\textrm{ [s]}$},
			  ylabel={$d_{S}\textrm{ [m]}$},
			  xmin=0.0,
			  xmax=200.0,
			  ymin=0.0,
			  ymax=0.069,
			  grid=both,
			  legend style ={ at={(0.02,0.98)},
			  anchor=north west, draw=black}
			]
			\addplot[only marks, mark=x, mark size = 2.5pt,red, each nth point=5] table[x expr={\thisrowno{0}}, y expr={\thisrowno{2}}]
				{./ScourApronData/ResultScourApron_a3_5eMinus3};
			\addlegendentry{1st Simulation, $a=3.5 \cdot 10^{-3}\frac{\text{m}}{\text{s}^2}$};
			\addplot[only marks, mark=o, mark size = 2.5pt,blue, each nth point=5] table[x expr={\thisrowno{0}}, y expr={\thisrowno{2}}]
				{./ScourApronData/ResultScourApron_a7eMinus3};
			\addlegendentry{2nd Simulation, $a=7.0 \cdot 10^{-3}\frac{\text{m}}{\text{s}^2}$};
			\addplot[only marks, mark=triangle, mark size = 2.5pt,orange, each nth point=5] table[x expr={\thisrowno{0}}, y expr={\thisrowno{2}}]
				{./ScourApronData/ResultScourApron_a14eMinus3};
			\addlegendentry{3rd Simulation, $a=14.0 \cdot 10^{-3}\frac{\text{m}}{\text{s}^2}$};
			\addplot[black, domain=0.0:200, samples=100]  {0.15*pow(x/4612,0.72)}; 
			\addlegendentry{Least-squares fits};
			\addplot[black, domain=0.0:200, samples=100]  {0.15*pow(x/2387,0.64)}; 
			%
			\addplot[black, domain=0.0:200, samples=100]  {0.15*pow(x/1459,0.56)}; 
		\end{axis}
	\end{tikzpicture}
	\caption{Comparison of simulation results and least-squares fits for the scouring depth}
	\label{scour_Apron_deltaS} 
\end{figure}
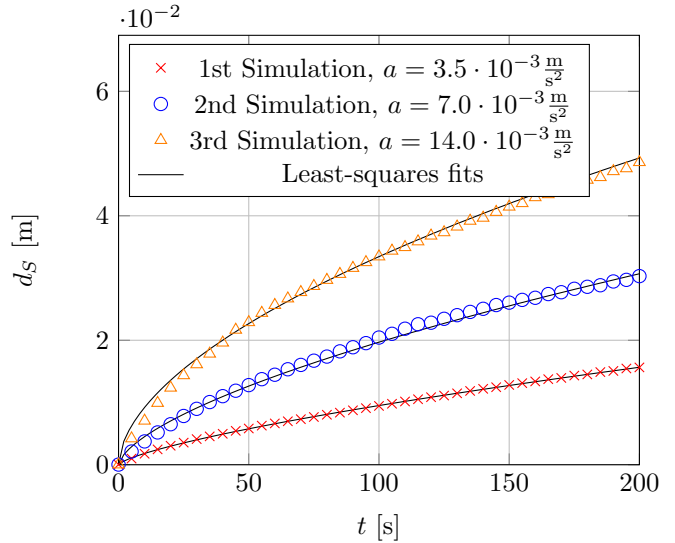
\begin{table}
   \centering
   \caption{Least squares parameters for $d_S$-function
   \eqref{eq_apron_deltaS} estimated over $200\text{s}$
   and $\alpha_S$-function
   \eqref{eq_apron_alphaS} estimated over $25\text{s}$.}
   \label{table:Scour Apron Least squares parameters}
   \begin{tabular}{ccccc}
     $a$ 	& $T_{S}$ & $n_{S}$ & $T_{\alpha S}$ & $\alpha_{S0}$ \\
     $[10^{-3}\frac{\text{m}}{\text{s}^2}]$ &[s]& []& $[\text{s}]$ & $[^\circ]$ \\
     \hline
     3.5	& 4612	& 0.72	& 15.2	& 9.8 \\
     7 		& 2387	& 0.64	& 5.2	& 8.54 \\
     14 	& 1459	& 0.56	& 2.15	& 8.6
   \end{tabular}
\end{table}
Figure \ref{scour_Apron_alphaS} shows the evolution of the scour angle.
Again, the results belong to the empirical equation \eqref{eq_apron_alphaS}.
The equilibrium scour angles are significantly smaller than the given range of the experiments $13.4^\circ \le \alpha_{S0} \le 14.95^\circ$ for fine sand.
The equilibrium scour angle is nearly the same for all three velocities,
which has been expected as the flow velocities are significantly higher than the critical velocity for the motion of the sediment.
%
Figure \ref{Sediment volume Fraction and velocity profiles for different time steps} shows the volume fractions of the sediment and the boundary velocities for different timesteps.
The shape of the sediment is plausible but seems to become too edgy with progressing time.
Simulations with other approaches for the calculation of the relative pressure
have shown that the relative pressure could have a significant influence onto the shape.
Interim results with methods, where the relative pressure is not always zero at the sediment surface, show a much rounder shape.
Therefore, we assume that the here given method for the calculation of the relative pressure is responsible for the too edgy shape.

Furthermore, one can see that no sediment is brought into suspension.
Tests with a given initial suspension layer have shown that the suspension can have a huge influence onto the sediment transport.
Especially the equilibrium scour angle increases, which allows to achieve angles in the correct range.
Nevertheless at the current development state of the scour solver using such a manually given suspension layer
can only be interpreted as \textit{guessing a suspension} or \textit{manipulating the results}.
Therefore we are not showing such simulations.
But, we assume that a better suspension treatment may lead to better results especially for the equilibrium scour angle.

\begin{figure*}
\centering
\subcaptionbox{relative pressure}{\includegraphics[width=0.5\textwidth]{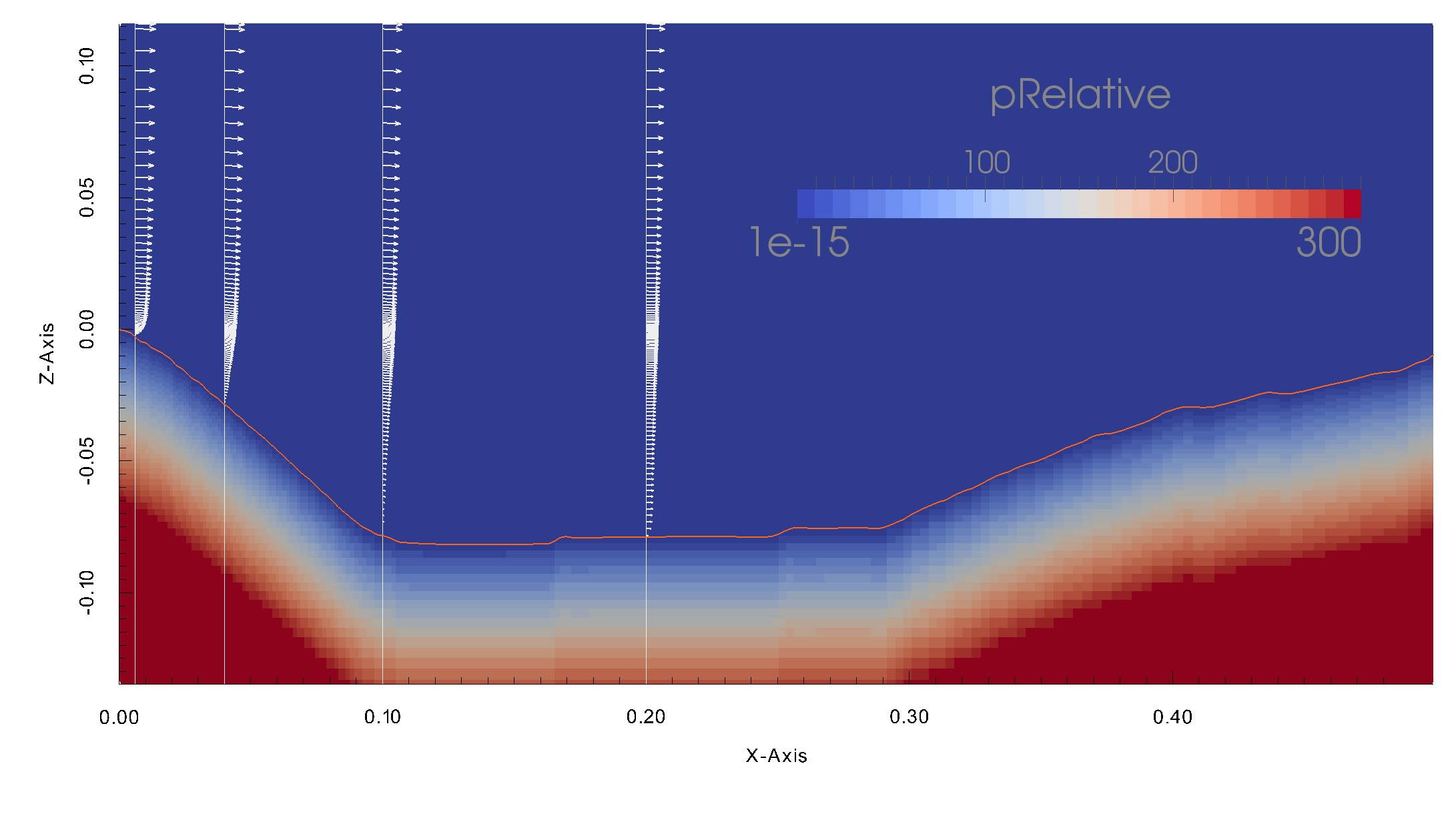}}%
\hfill
\subcaptionbox{soil viscosity}{\includegraphics[width=0.5\textwidth]{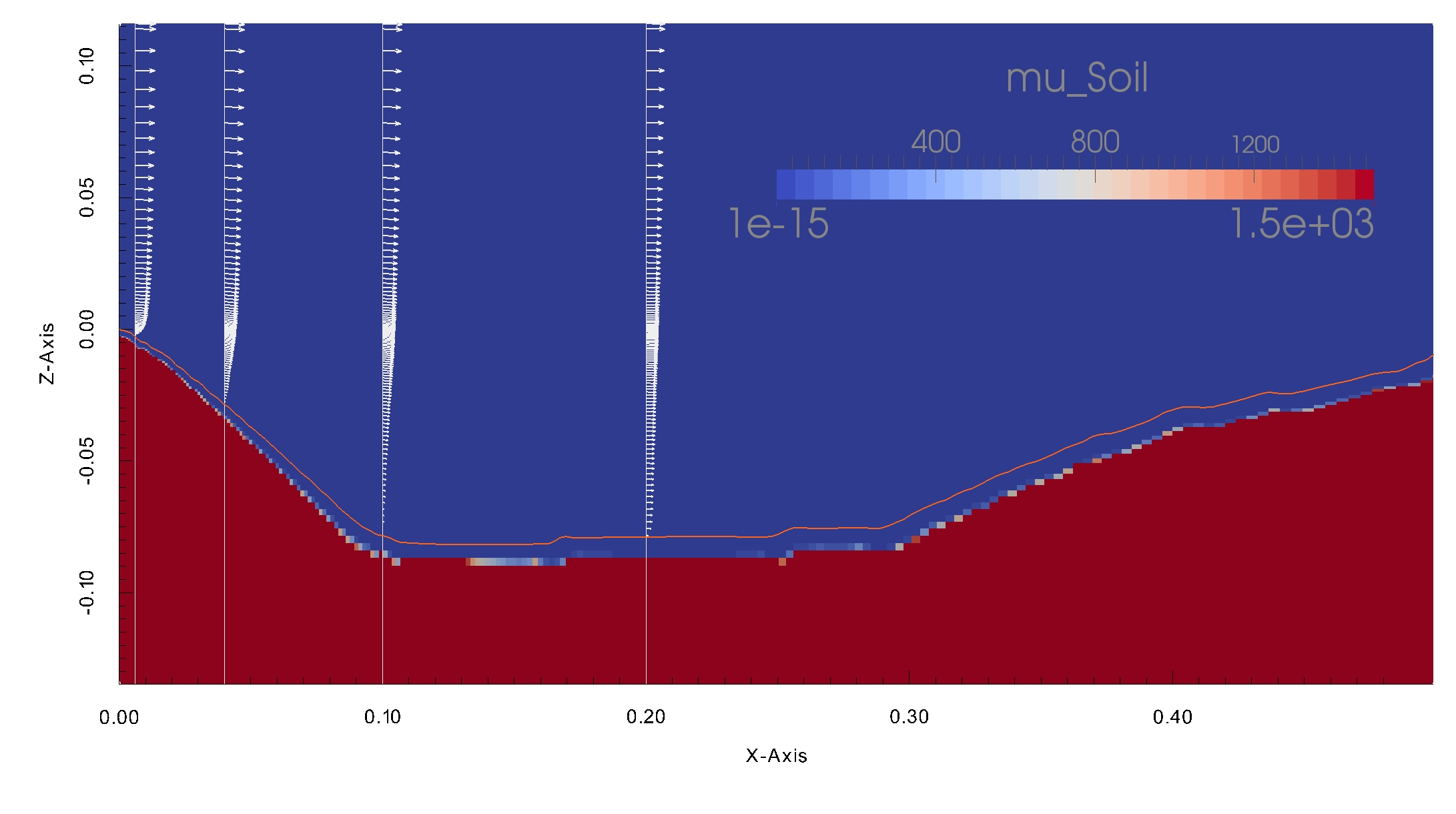}}%
\caption{Relative pressure and soil viscosity at $t=100\text{s}$.
The solid, orange line represents the sediment surface.}
\label{Relative pressure and soil viscosity}
\end{figure*}

Figure \ref{Relative pressure and soil viscosity} (a) shows the relative pressure of the simulation at $t=100\text{s}$.
The solution for the relative pressure is very clean without any disturbances, as desired.
Figure \ref{Relative pressure and soil viscosity} (b) shows the soil viscosity.
It shows that the critical velocity for sediment motion is significantly exceeded.
%
%
%
%
%
%
\subsection{Scour around a vertical pile}
\label{Scour around a vertical pile}

The flow around a vertical circular pile exposed to a steady current is studied numerically and experimentally by \cite{Roulund2005}.
In the following the main phenomena of this test case should be summarized.
The steady current forms a boundary layer at the bottom surface.
This boundary layer forces the flow to build a down-flow on the upwind side of the pile.
This down-flow leads to a horseshoe vortex in front and around the pile, which then trails off downstream.
Additionally, a lee-wake vortex is built downwind of the pile.
Furthermore, the streamlines contract at the sides of the pile.
All three phenomena increase the sediment transport leading to local scour around the pile.

As shown by \cite{Roulund2005} the time required to build the horseshoe vortex depends
on the boundary-layer-thickness to pile-diameter ratio.
The smaller this ratio, the longer the delay until the horseshoe vortex is built.
For very small ratios it is possible, that no separation is formed.
Furthermore, the size of the horse-vortex depends on this ratio and a smaller ratio leads to a smaller vortex.\\

\begin{figure*}
\centering
\includegraphics[width=0.9\textwidth]{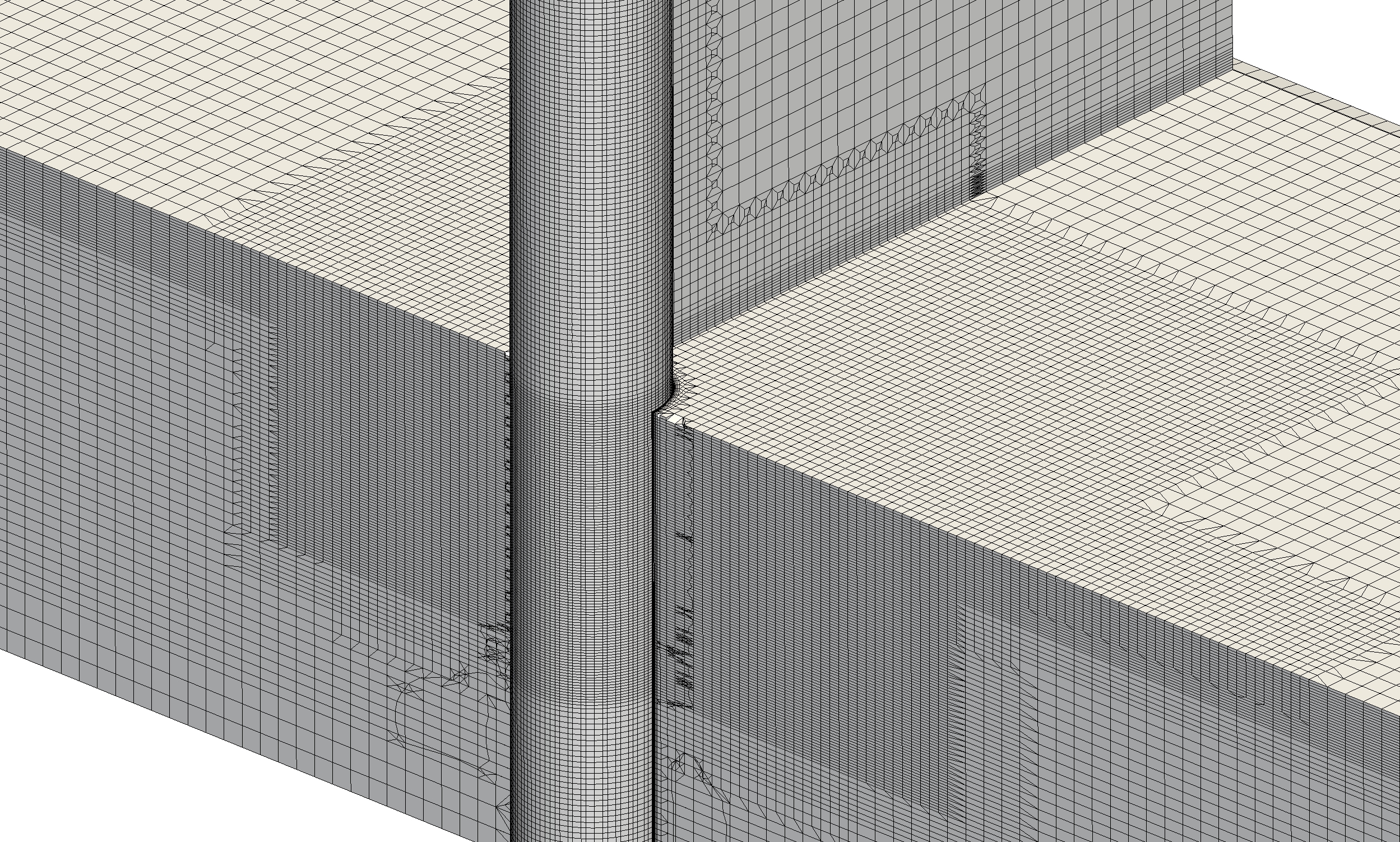}%
\caption{vertical pile in steady current: grid}
\label{vertical pile in steady current: grid}
\end{figure*}

\textbf{Simulation Setup:}\\
The pile has a diameter of $0.1\text{m}$.
The domain has a length of $2.1\text{m}$, a width of $1.2\text{m}$ and a height of $0.75\text{m}$.
It is initialized with sediment up to $0.345\text{m}$.
The domain is discretized with an unstructured hexadominant grid with anisotropic mesh refinement for the sediment surface
as shown in figure \ref{vertical pile in steady current: grid}.
In the region of the sediment surface, the mesh is refined anisotropically in the vertical direction.
The cells have a height of $1.875E-3\text{m}$ and a length and width of $7.5E-3\text{m}$ at the sediment surface.
The final mesh consist of $2.0E6$ cells.

The velocity, the volume fraction and the turbulence values are given
as Dirichlet boundary conditions at the inlet on the left side.
The pressure is given as a Dirichlet boundary condition on the right side.
The initial values for these Dirichlet boundary conditions were generated using a 2D simulation
based on the same principle as in subsection \ref{Scour downstream of an apron}.
The acceleration used to generate the velocity profile was $2.0E-3\frac{\text{m}}{\text{s}^2}$.
A Free-Slip wall boundary condition is applied on the top.
It was not possible to achieve a stable simulation using an open lid boundary condition at the top as recommended by \cite{Roulund2005} and \cite{Baykal2015}.
Symmetry boundary conditions are applied at the sides.

The parameters of the sediment model were adjusted to represent the material of the experiment.
The rock density is $2650\frac{\text{kg}}{\text{m}^3}$,
the saturation is $0.7$,
the internal friction angle is $32^\circ$,
the cohesion is $0 \frac{\text{N}}{\text{m}^2}$.
The maximal Bingham viscosity is set to $1500\frac{\text{Ns}}{\text{m}^2}$.
The simulation was run with different values for the minimal Bingham viscosities.
The investigated minimal viscosities are
$1.0\frac{\text{Ns}}{\text{m}^2}$,
$12.5\frac{\text{Ns}}{\text{m}^2}$
and $25\frac{\text{Ns}}{\text{m}^2}$.
The internal wall function and the buffer cell approach are activated.\\

\begin{figure*}
\centering
\subcaptionbox{$\mu_{S_{\text{lower}}}=1\frac{\text{Ns}}{\text{m}^2}$, $t=20\text{s}$}
{\includegraphics[width=0.33\textwidth]{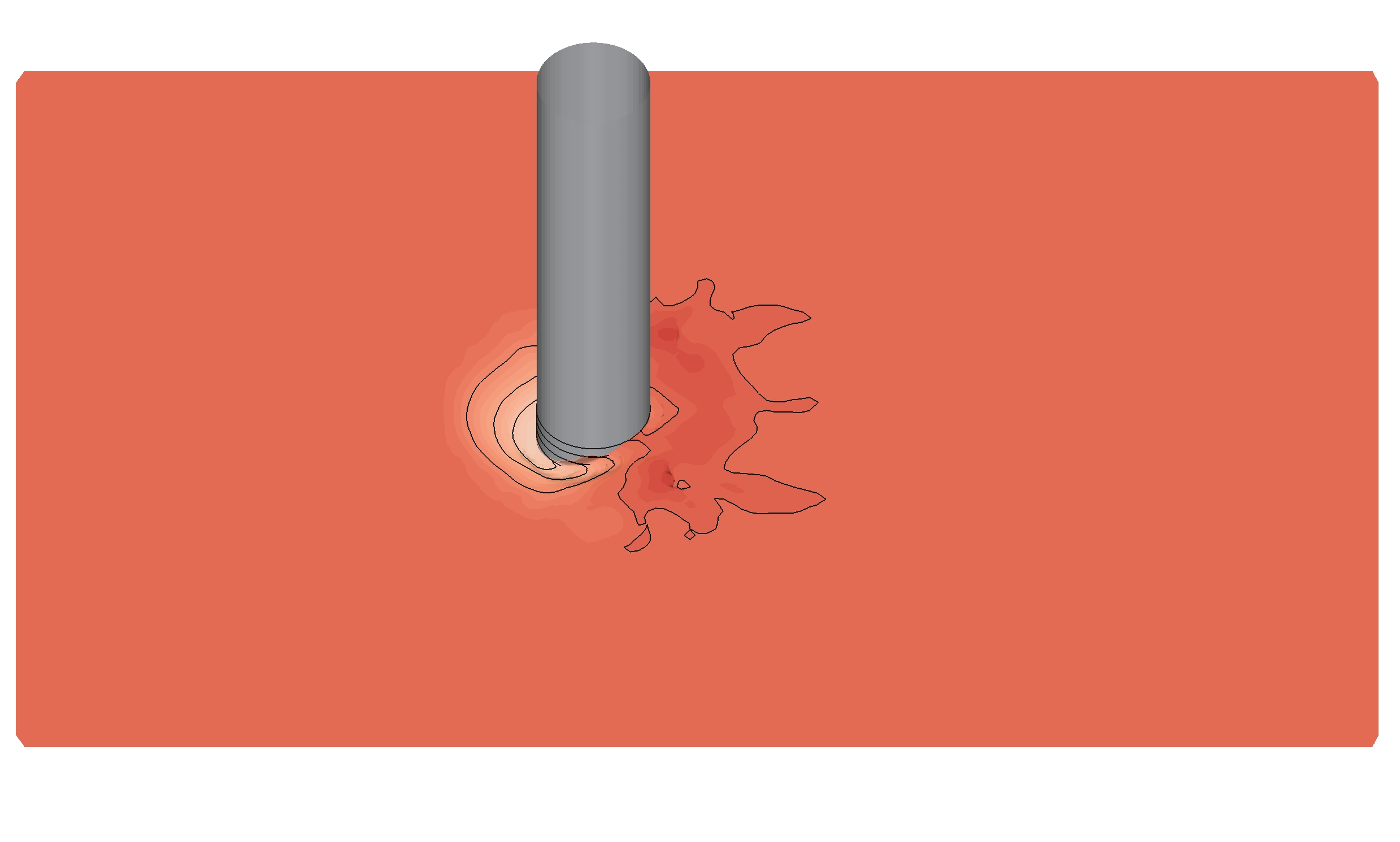}}%
\hfill
\subcaptionbox{$\mu_{S_{\text{lower}}}=12.5\frac{\text{Ns}}{\text{m}^2}$, $t=20\text{s}$}
{\includegraphics[width=0.33\textwidth]{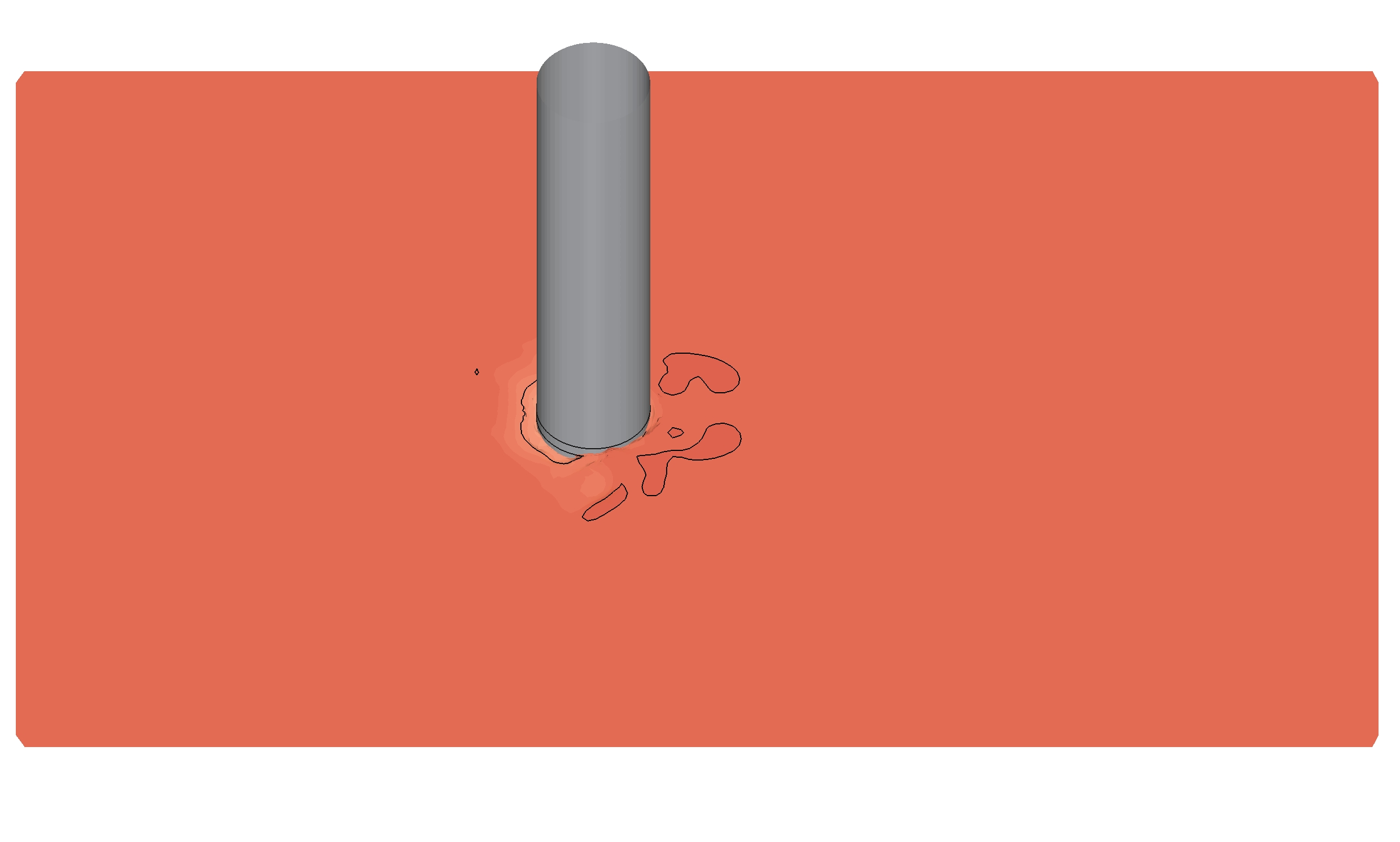}}%
\hfill
\subcaptionbox{$\mu_{S_{\text{lower}}}=25\frac{\text{Ns}}{\text{m}^2}$, $t=20\text{s}$}
{\includegraphics[width=0.33\textwidth]{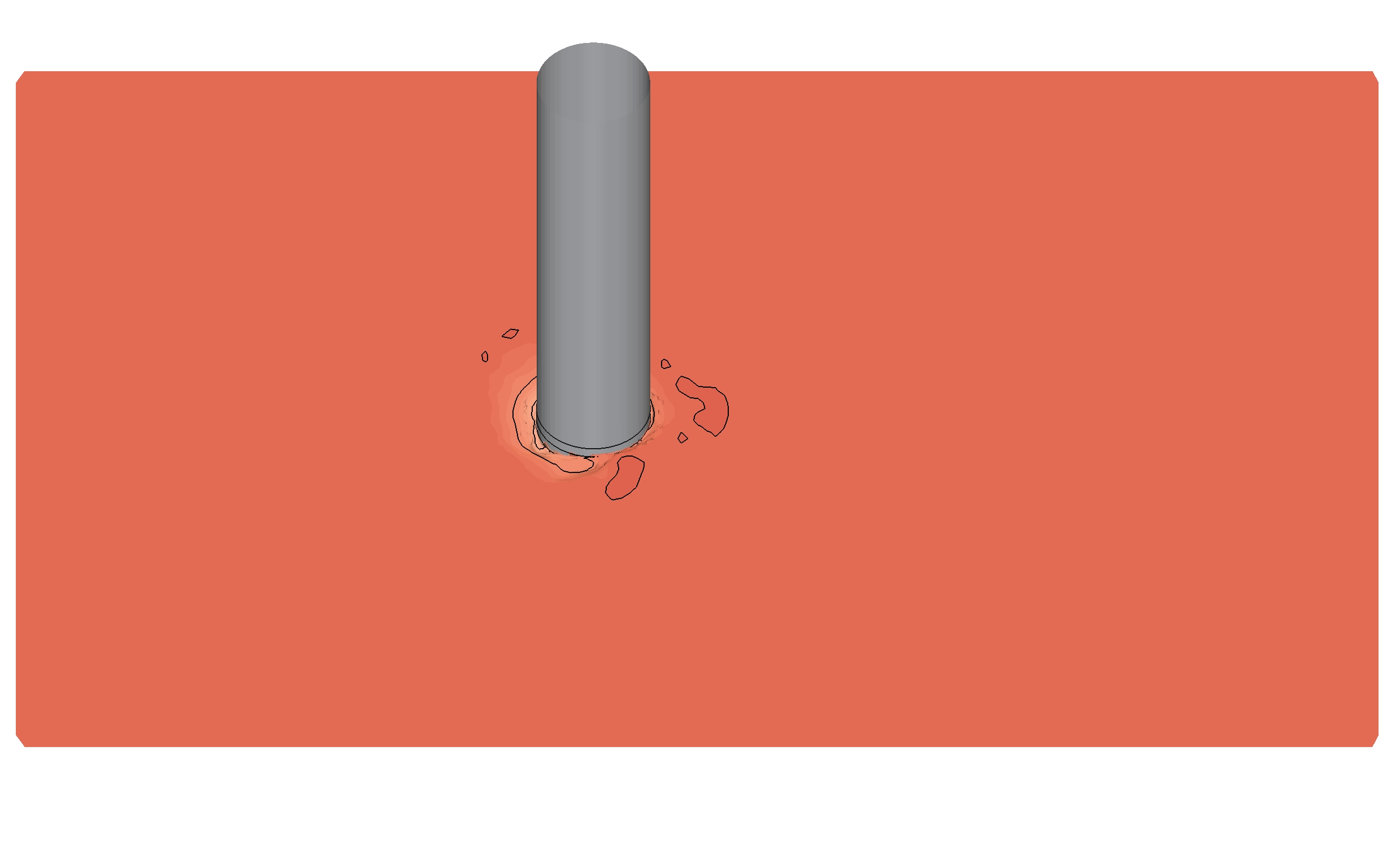}}%
\hfill
\subcaptionbox{$\mu_{S_{\text{lower}}}=1\frac{\text{Ns}}{\text{m}^2}$, $t=60\text{s}$}
{\includegraphics[width=0.33\textwidth]{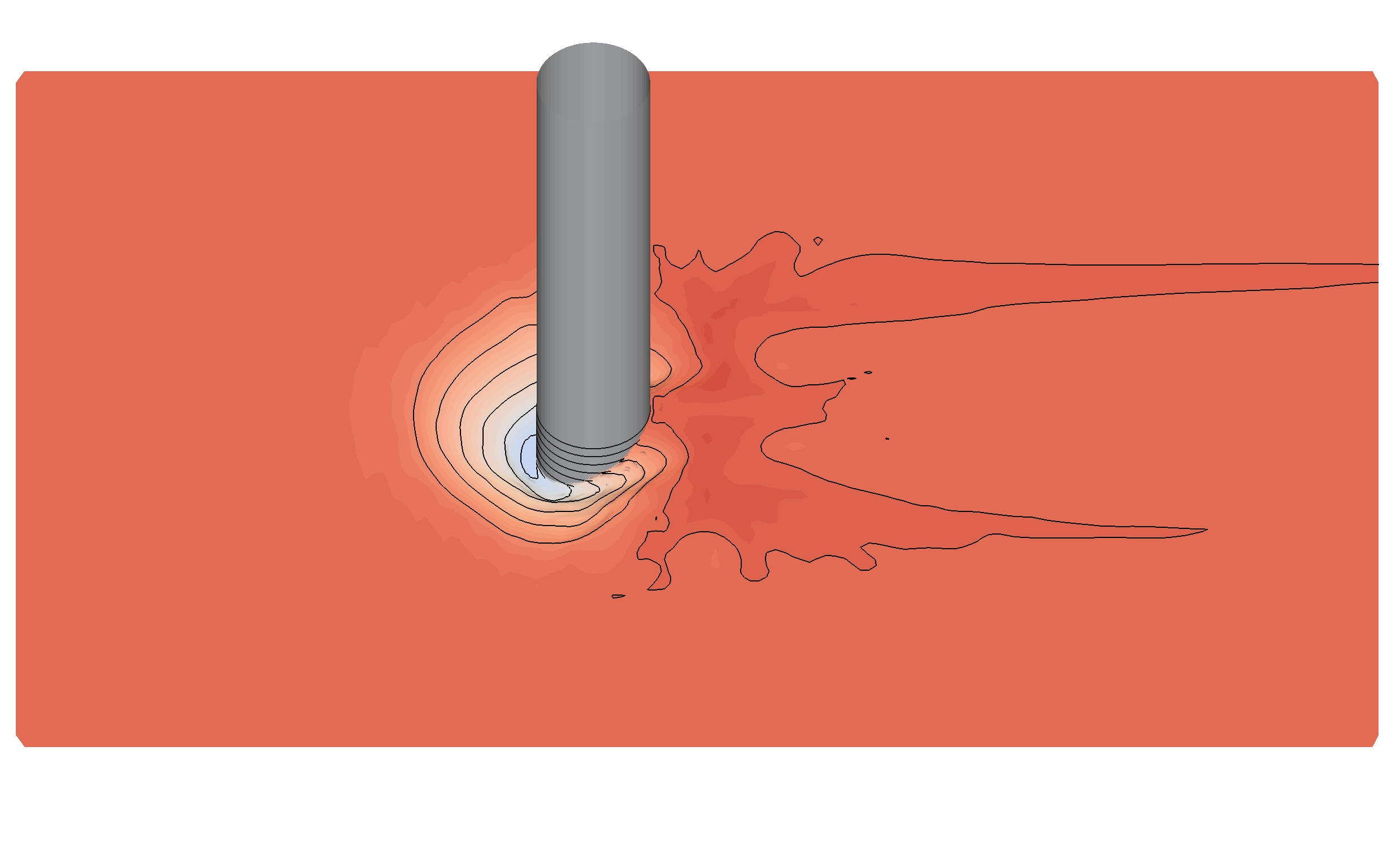}}%
\hfill
\subcaptionbox{$\mu_{S_{\text{lower}}}=12.5\frac{\text{Ns}}{\text{m}^2}$, $t=60\text{s}$}
{\includegraphics[width=0.33\textwidth]{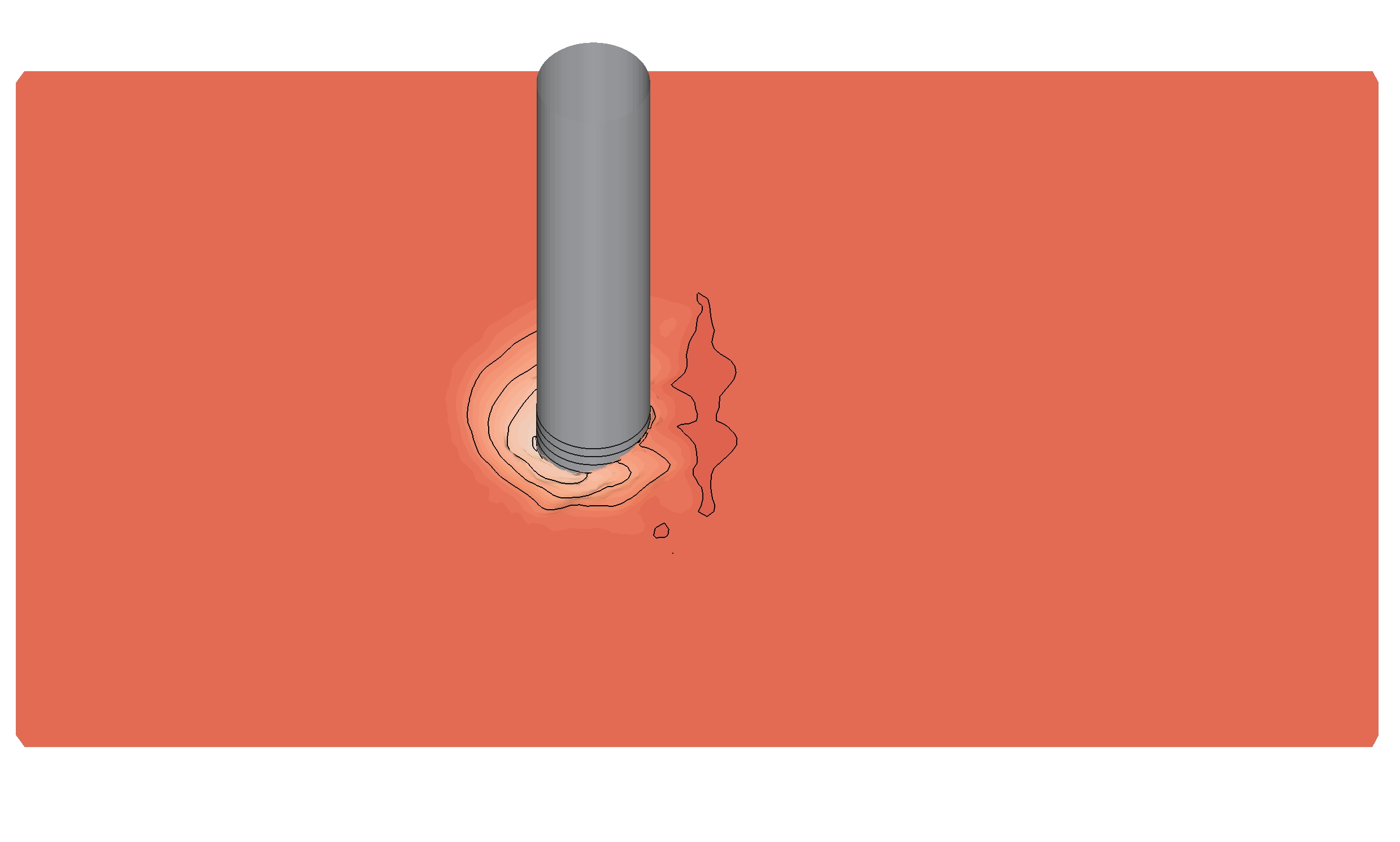}}%
\hfill
\subcaptionbox{$\mu_{S_{\text{lower}}}=25\frac{\text{Ns}}{\text{m}^2}$, $t=60\text{s}$}
{\includegraphics[width=0.33\textwidth]{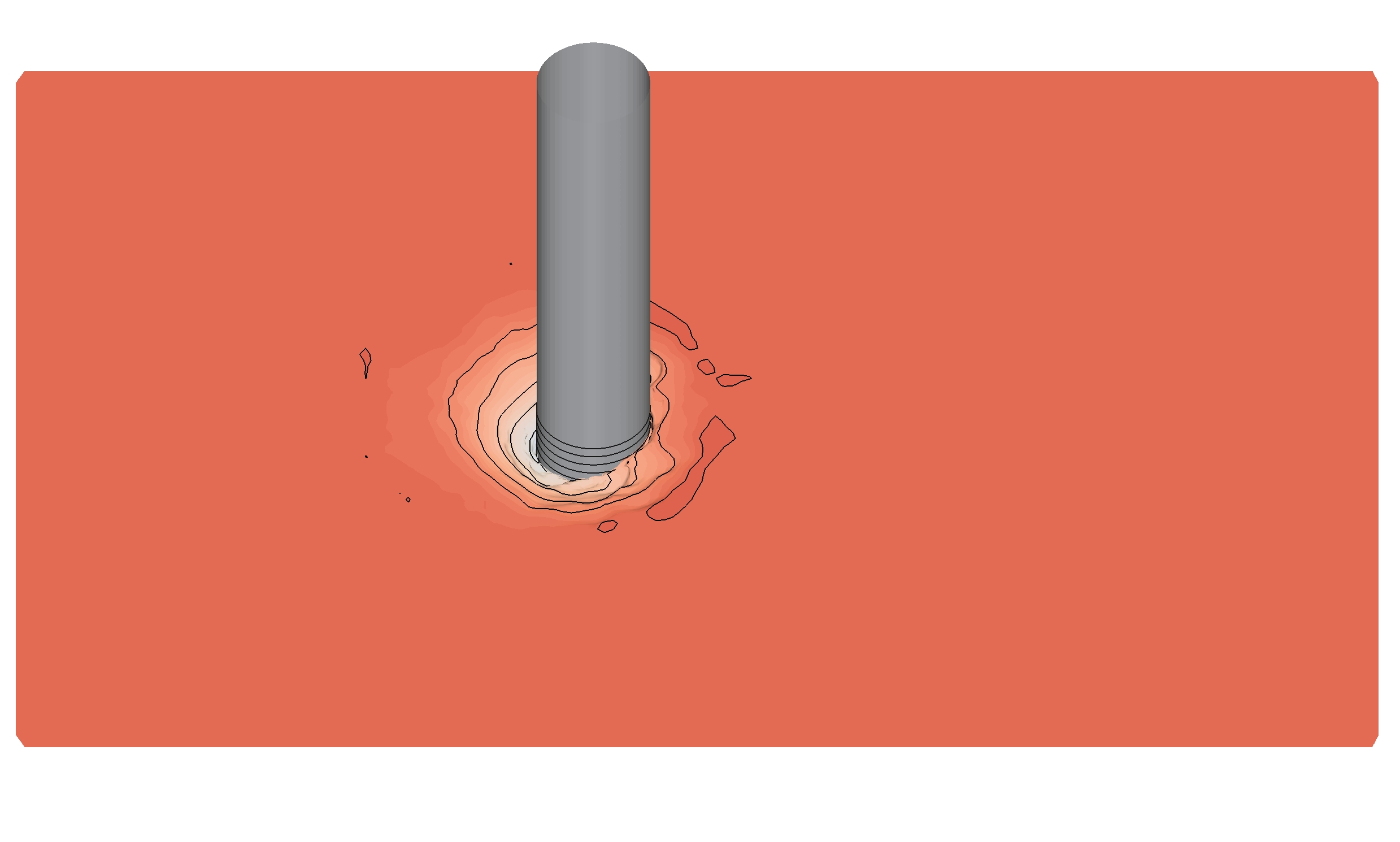}}%
\hfill
\subcaptionbox{$\mu_{S_{\text{lower}}}=1\frac{\text{Ns}}{\text{m}^2}$, $t=120\text{s}$}
{\includegraphics[width=0.33\textwidth]{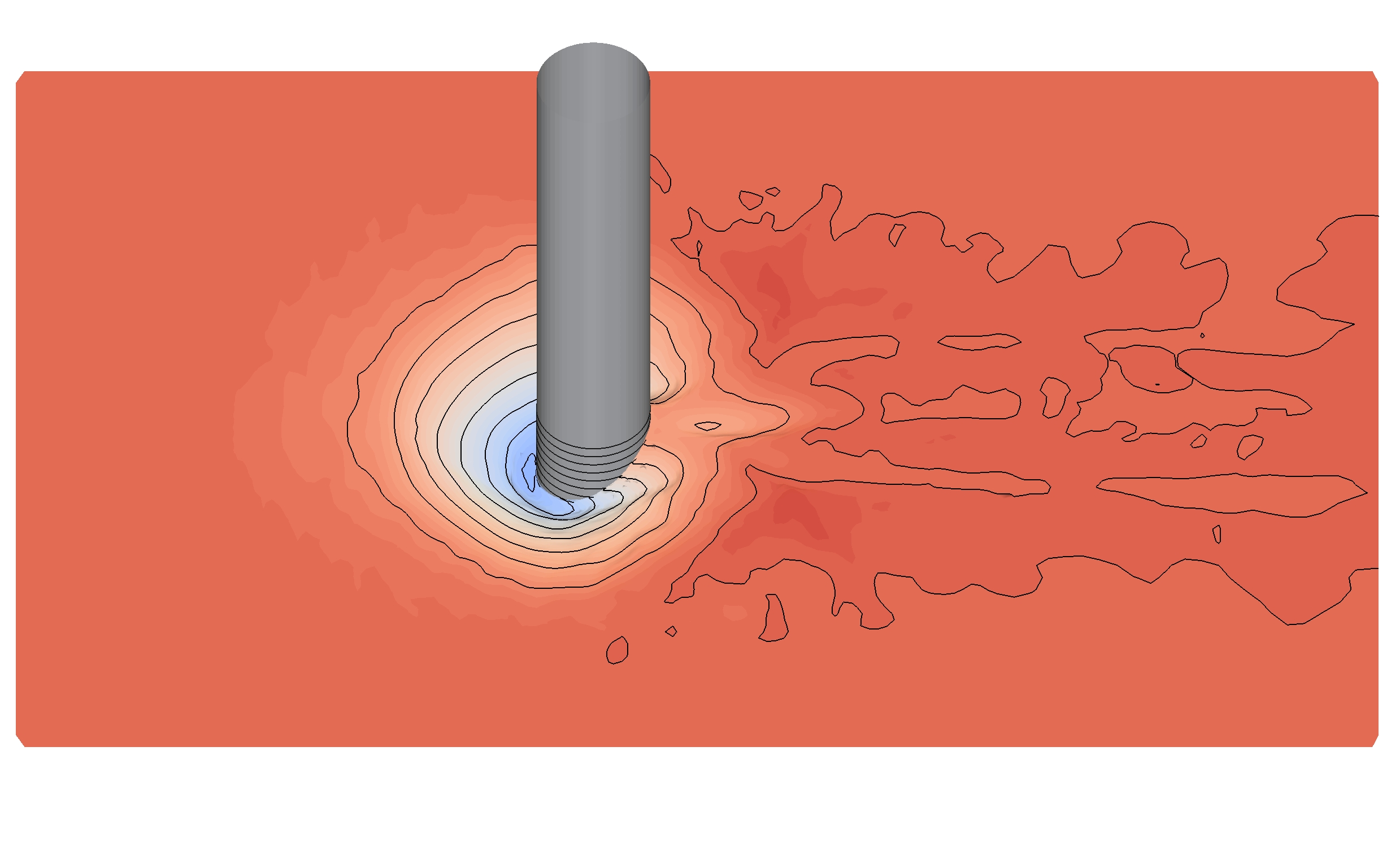}}%
\hfill
\subcaptionbox{$\mu_{S_{\text{lower}}}=12.5\frac{\text{Ns}}{\text{m}^2}$, $t=120\text{s}$}
{\includegraphics[width=0.33\textwidth]{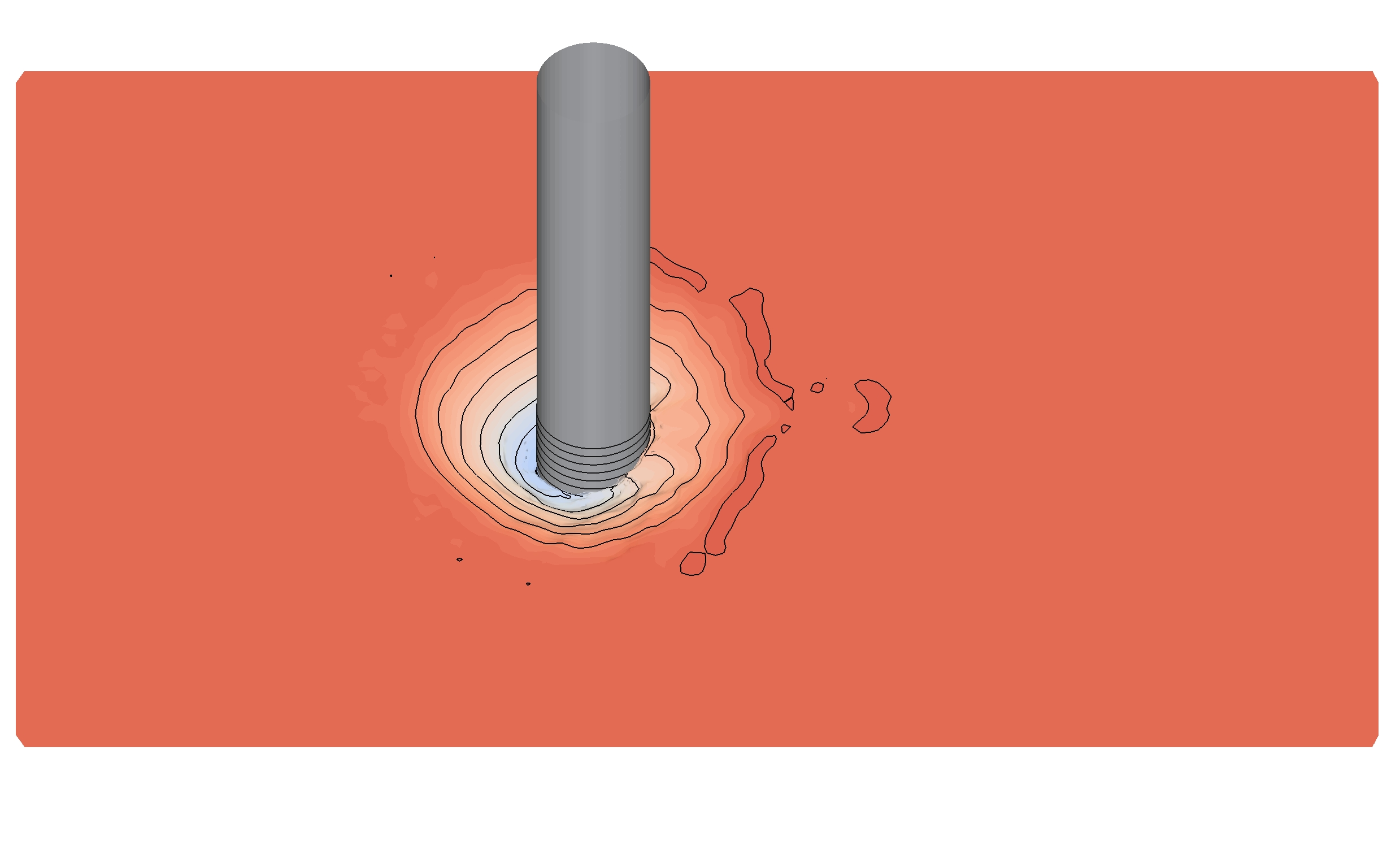}}%
\hfill
\subcaptionbox{$\mu_{S_{\text{lower}}}=25\frac{\text{Ns}}{\text{m}^2}$, $t=120\text{s}$}
{\includegraphics[width=0.33\textwidth]{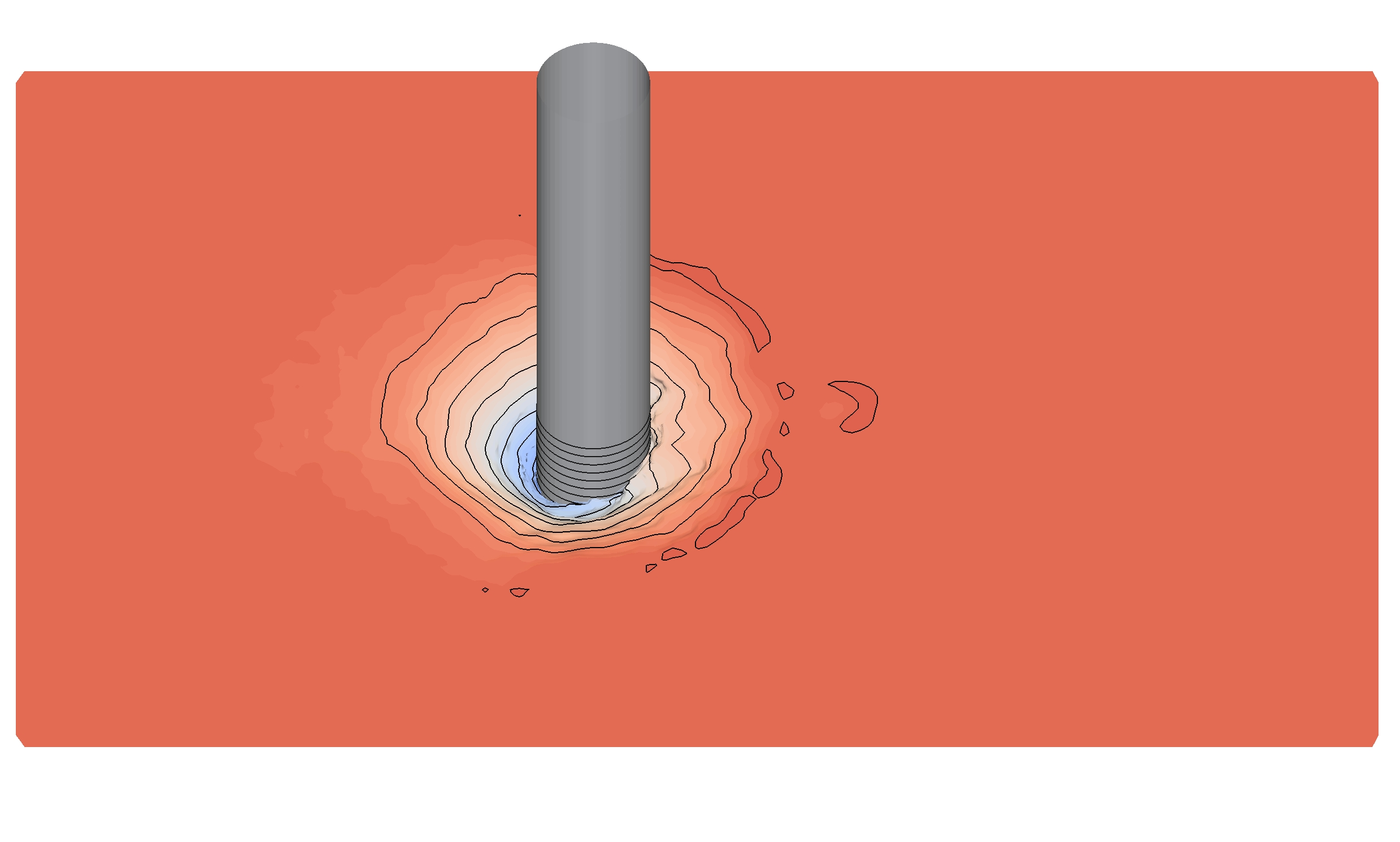}}%
\hfill
\subcaptionbox{$\mu_{S_{\text{lower}}}=1\frac{\text{Ns}}{\text{m}^2}$, $t=240\text{s}$}
{\includegraphics[width=0.33\textwidth]{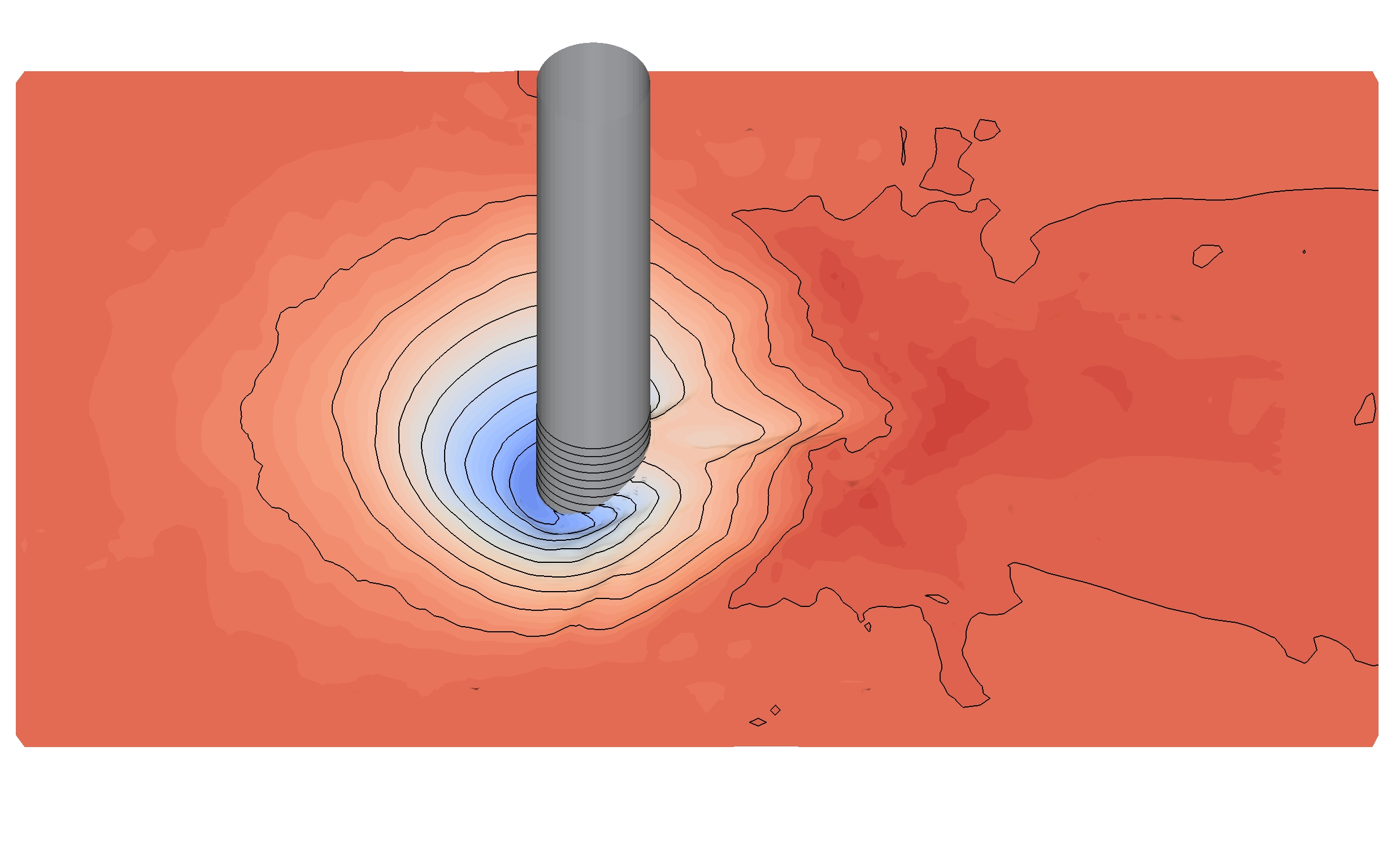}}%
\hfill
\subcaptionbox{$\mu_{S_{\text{lower}}}=12.5\frac{\text{Ns}}{\text{m}^2}$, $t=240\text{s}$}
{\includegraphics[width=0.33\textwidth]{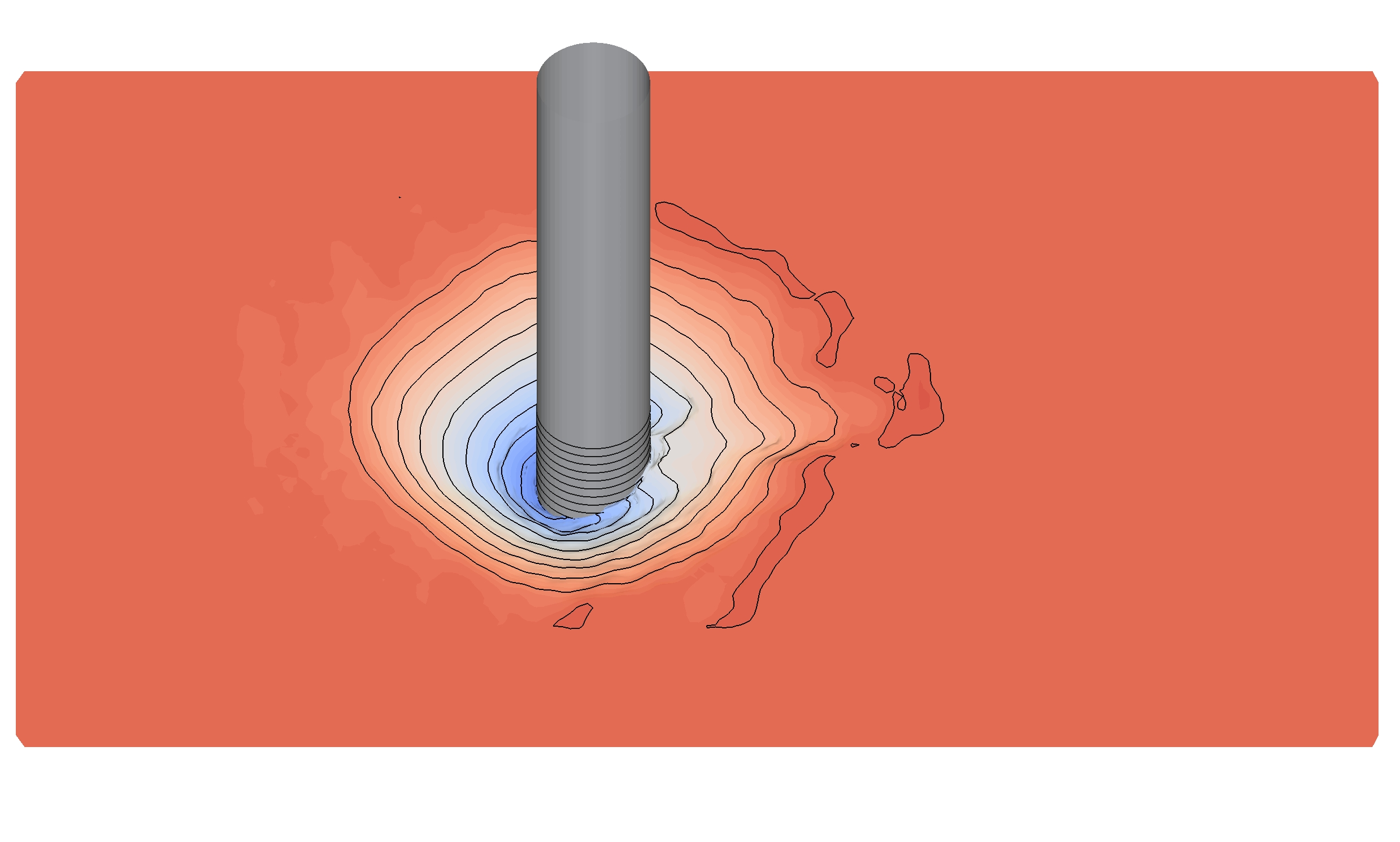}}%
\hfill
\subcaptionbox{$\mu_{S_{\text{lower}}}=25\frac{\text{Ns}}{\text{m}^2}$, $t=240\text{s}$}
{\includegraphics[width=0.33\textwidth]{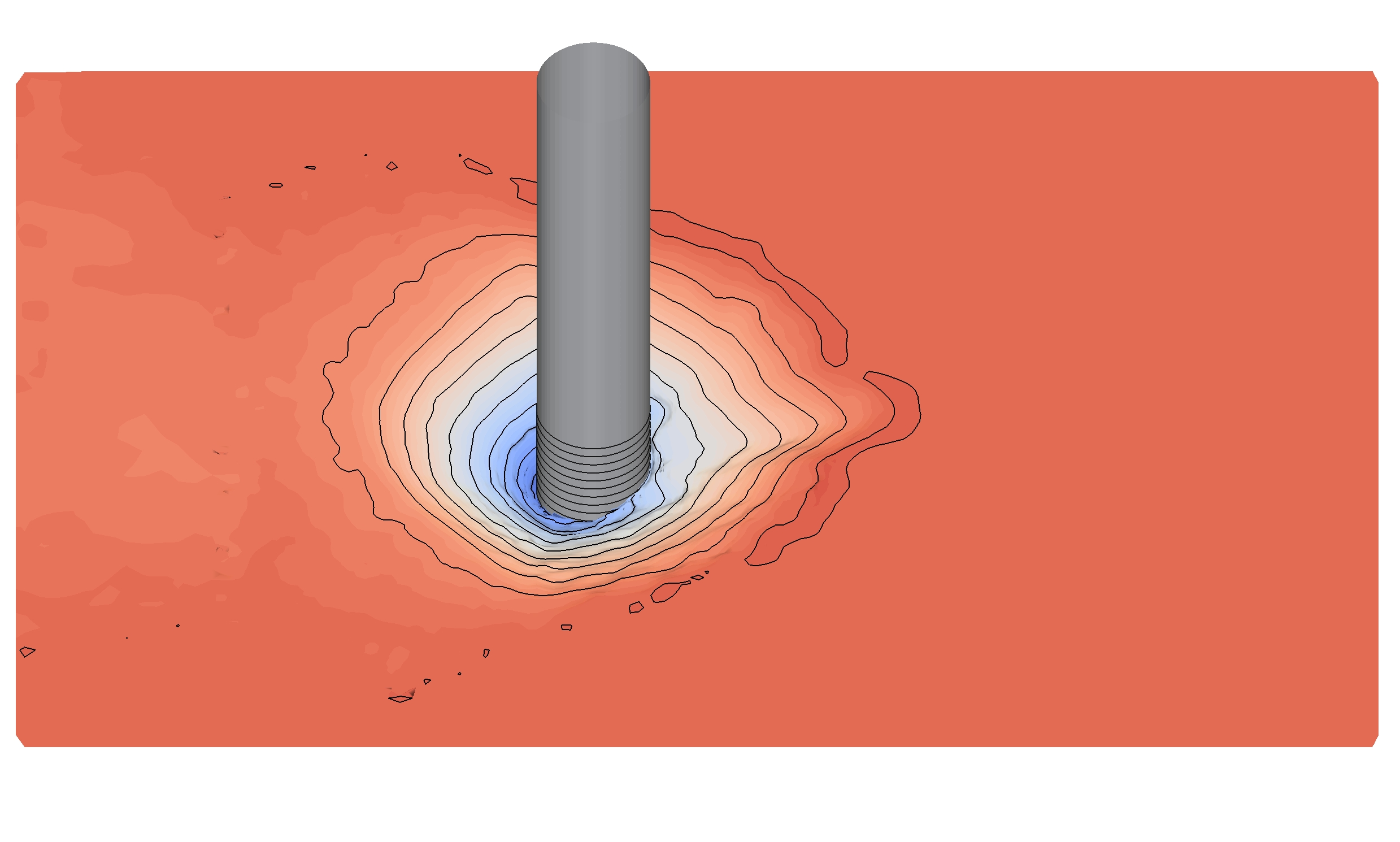}}%
\hfill
\subcaptionbox{$\mu_{S_{\text{lower}}}=1\frac{\text{Ns}}{\text{m}^2}$, $t=500\text{s}$}
{\includegraphics[width=0.33\textwidth]{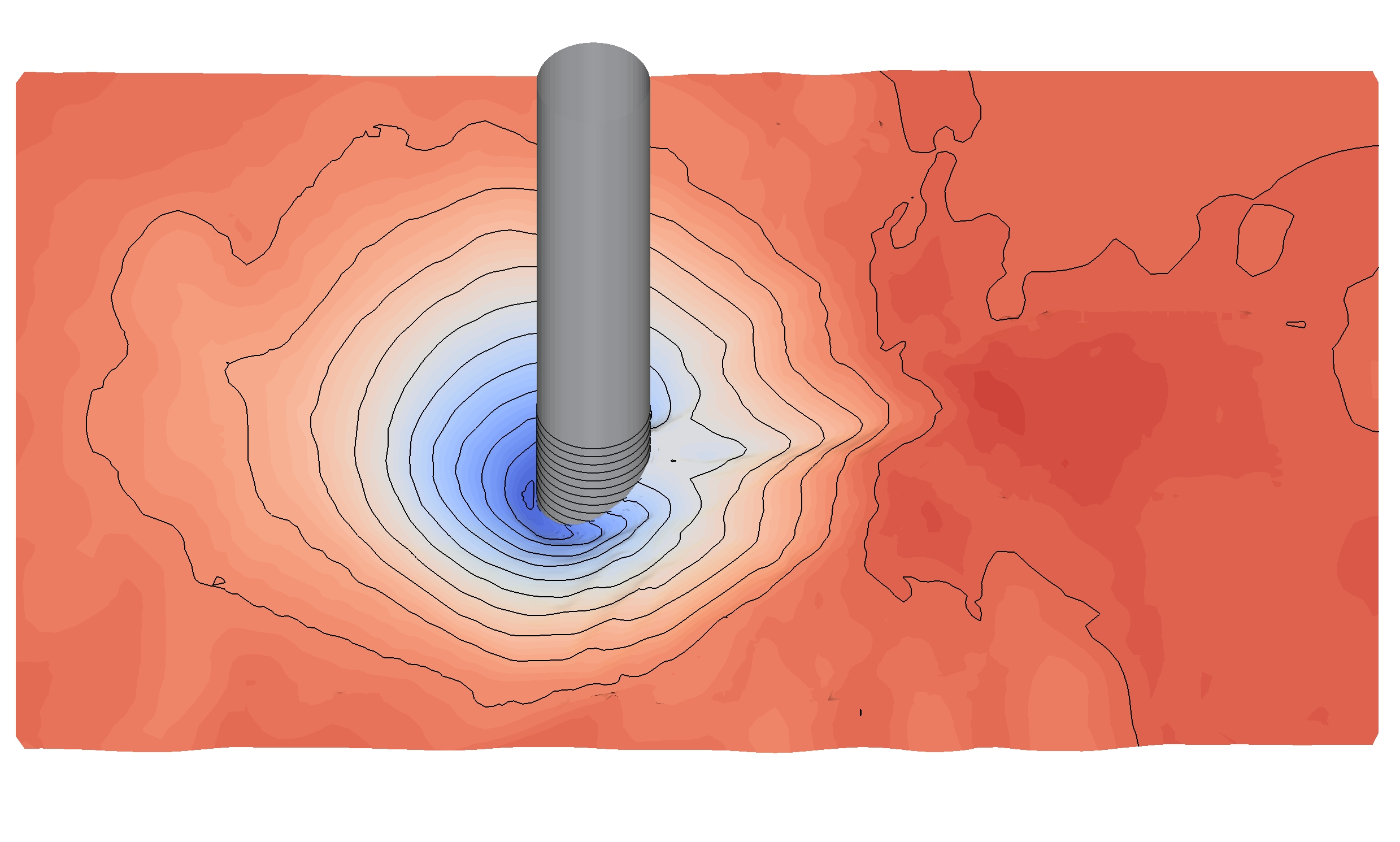}}%
\hfill
\subcaptionbox{$\mu_{S_{\text{lower}}}=12.5\frac{\text{Ns}}{\text{m}^2}$, $t=500\text{s}$}
{\includegraphics[width=0.33\textwidth]{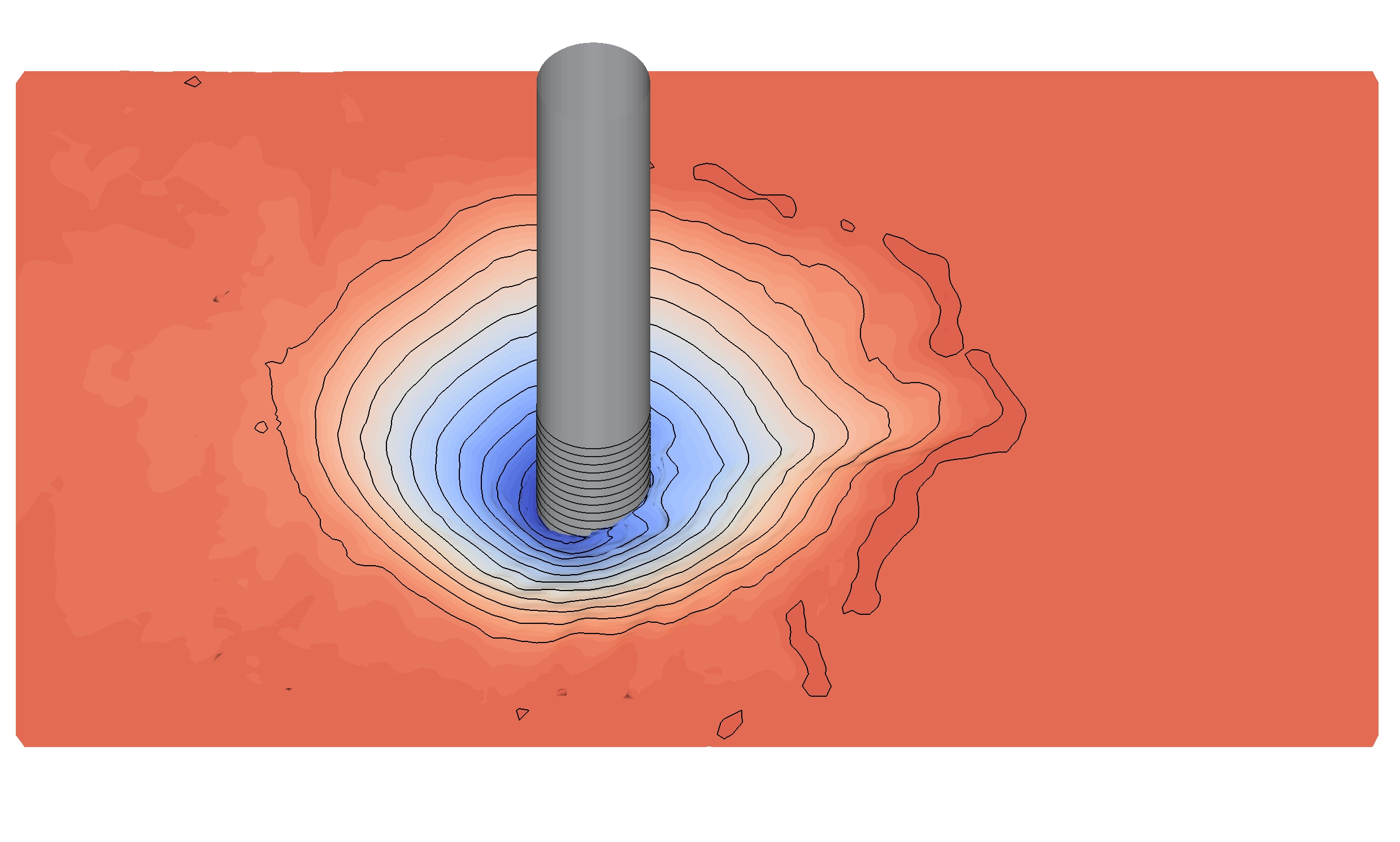}}%
\hfill
\subcaptionbox{$\mu_{S_{\text{lower}}}=25\frac{\text{Ns}}{\text{m}^2}$, $t=500\text{s}$}
{\includegraphics[width=0.33\textwidth]{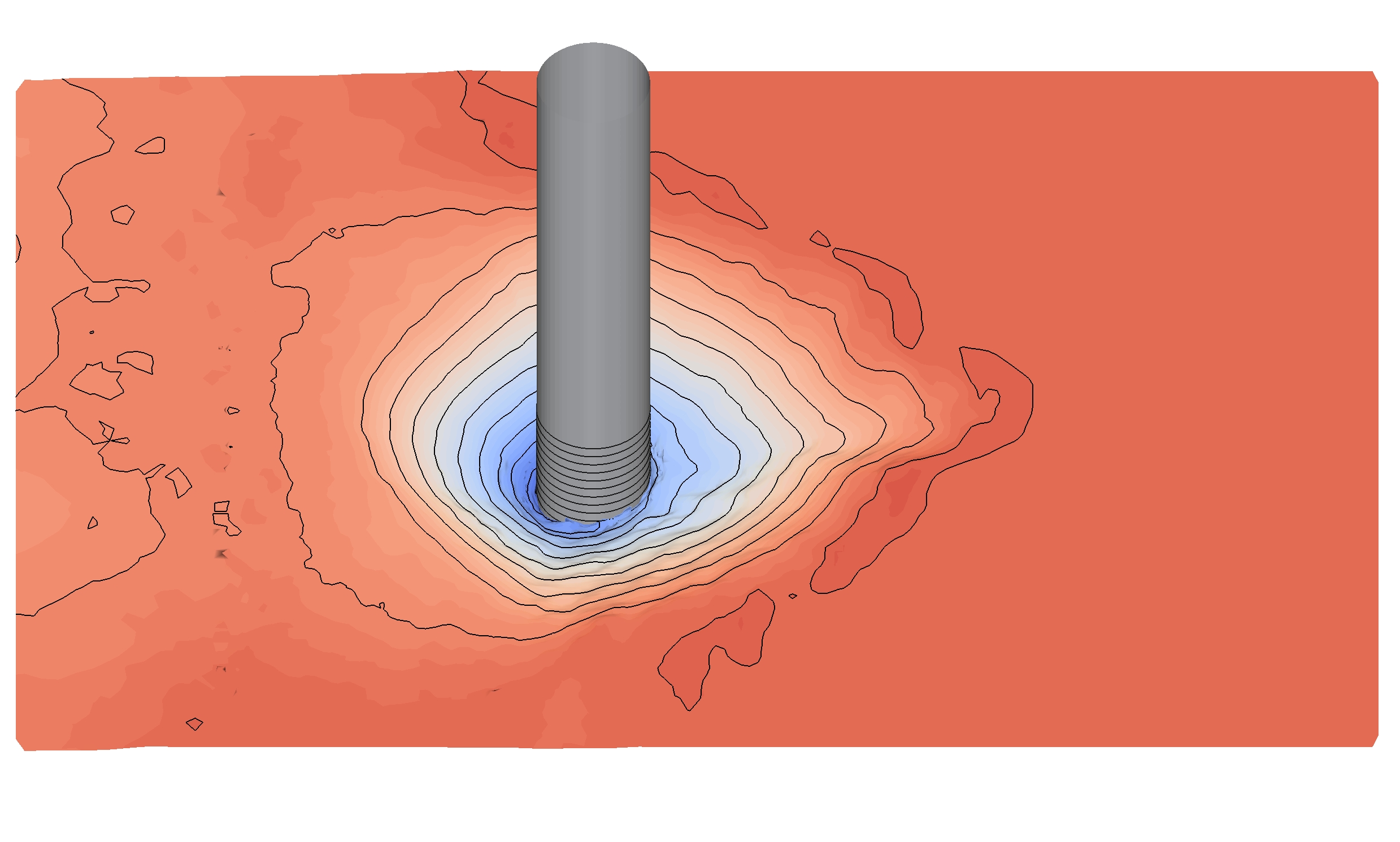}}%
\hfill
{\includegraphics[width=0.4\textwidth]{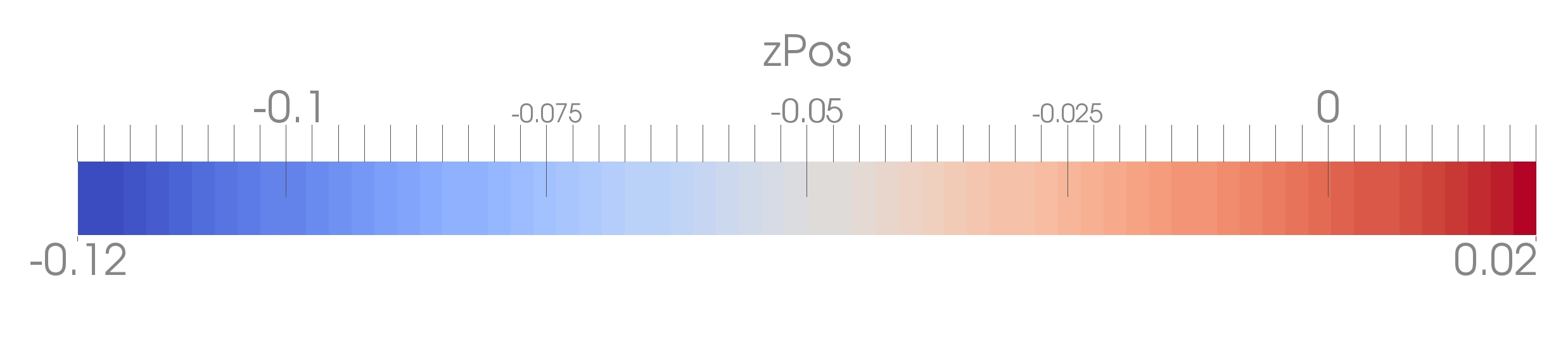}}%
\caption{Vertical pile in steady current: sediment evolution for different minimal soil viscosities.}
\label{vertical pile in steady current: sediment evolution for different minimal soil viscosities}
\end{figure*}

\begin{figure}
\centering
\includegraphics[width=0.5\textwidth]{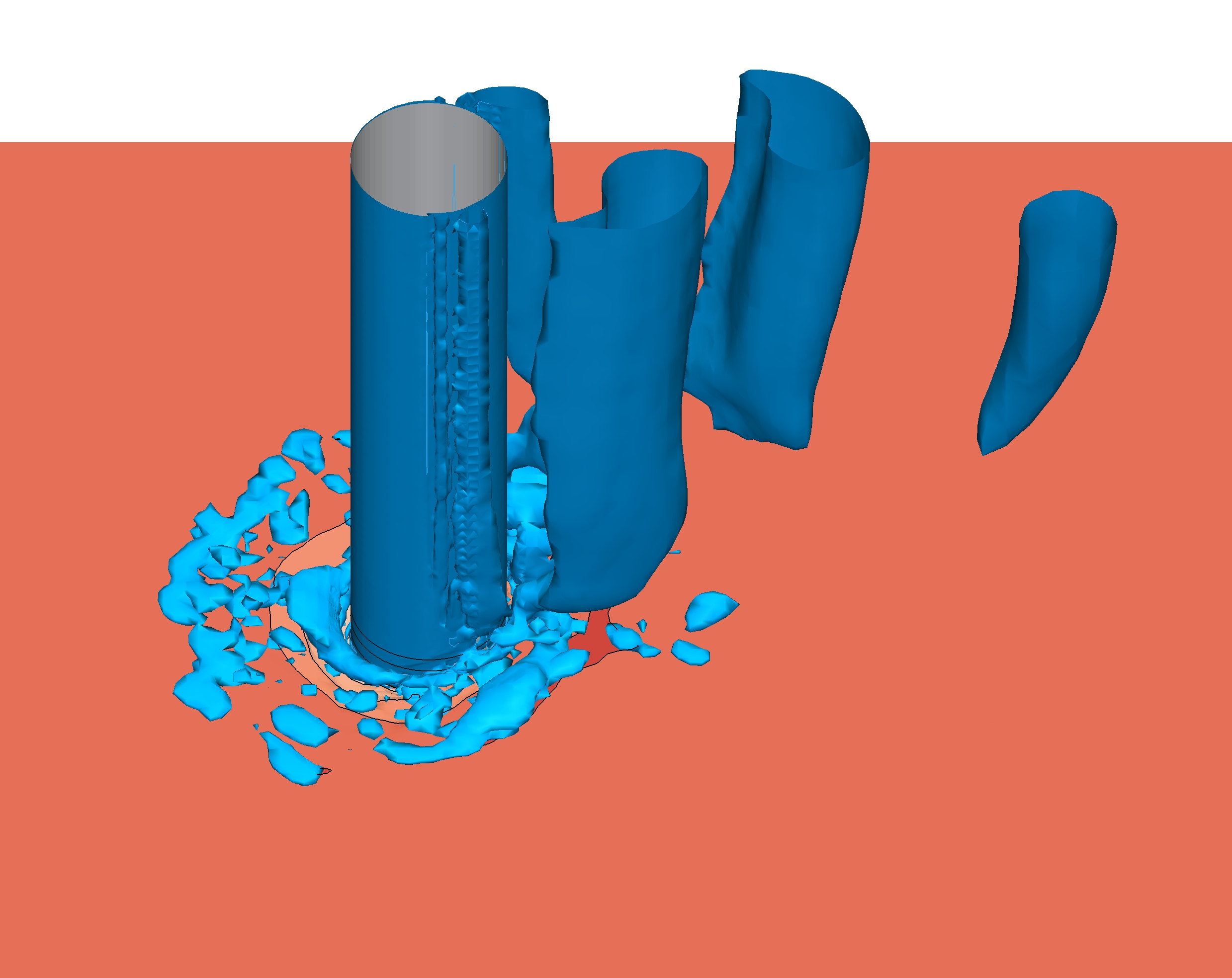}%
\caption{Horse shoe vortex and leewake vortices at $t=40\text{s}$ for $\mu_{S_{\text{lower}}}=25\frac{\text{Ns}}{\text{m}^2}$}
\label{roulund vortices}
\end{figure}

\textbf{Results:}\\
The simulation was executed on a compute cluster with five nodes interconnected via InfiniBand.
Each node holds two 6-core CPUs (Intel Xeon E5-2643 v4 @ 3.40GHz)
and the calculation time averaged 21 hours for one simulation.
Figure \ref{vertical pile in steady current: sediment evolution for different minimal soil viscosities} shows the sediment surface at different timesteps for the four investigated viscosities.
In all cases one can see that the building of the scour begins with the horseshoe vortex on the upwind side of the pile.
Subsequently the hole growth around the pile.
After some time the lee wake vortex is build and also transports the sediment away from the downstream side of the pile.
To emphasize the influence of the different vortices a detailed view is given in figure \ref{roulund vortices}.
The vortices are visualized with the help of the Q-Criterion, which is the 2nd invariant of the velocity gradient tensor.
In comparison of the simulation results one can see,
that the higher the minimal viscosity the steeper the scouring angle on the upstream part.
At $t=240\text{s}$ the slope has an angle of $27^\circ$, $29^\circ$ and $30^\circ$ for the three different minimal viscosities.
The experiment from \cite{Roulund2005} shows, that this angle should equal the angle of repose, which is not reached completely by our simulations.

In all three simulations, the scouring angle at the downstream side is significantly lower than at the upstream side.
At $t=240\text{s}$ the slope has an angle of $16^\circ$, $18.5^\circ$ and $20^\circ$ at the downstream side.
A smaller angle is in accordance with the experiment.
Furthermore, it has been pointed out, that the scouring depth at the upstream side is higher than on the downstream side.
This behavior is resolved in all three simulations till $t=240\text{s}$.
Later at $t=500\text{s}$ only the simulation with the smallest minimal viscosity is able to hold this characteristic.
From comparison with the pictures of the experiment \citep{Roulund2005}
we think that the simulation with the smallest minimal viscosity represents the scouring depth very well.

In all simulations one can observe an erosion in front of the scouring hole.
We assume, that the buffer cell approach, presented in subsection \ref{Special VOF-wall treatment}, is not sufficient to protect the sediment surface from this erosion.
Furthermore, we would like to emphasize that the suspension may have a significant influence onto the result.
The suspension increases the erosion at the downstream side of the pile
which is in accordance with the observations represented by \cite{Baykal2015},
reporting a decrease of 50\% if the suspension is neglected.
For the simulation with the highest minimal viscosity
we observed that the suspension protects the sediment in front of the scouring hole.
Without the suspension the erosion would have been much higher in that region.

Due to the sensible reaction to the suspension
in combination with an insufficient representation of the suspension generation
it is currently not possible to calibrate the minimal soil viscosity and give a final value.
Nevertheless, for the presented results all characteristics of the experiments are represented in all three simulations
and a value of $1.0\frac{\text{Ns}}{\text{m}^2}$ seems to be the best compromise.

\begin{figure}
\centering
\subcaptionbox{$\mu_{S_{\text{lower}}}=1\frac{\text{Ns}}{\text{m}^2}$, $t=20\text{s}$}
{\includegraphics[width=0.33\textwidth]{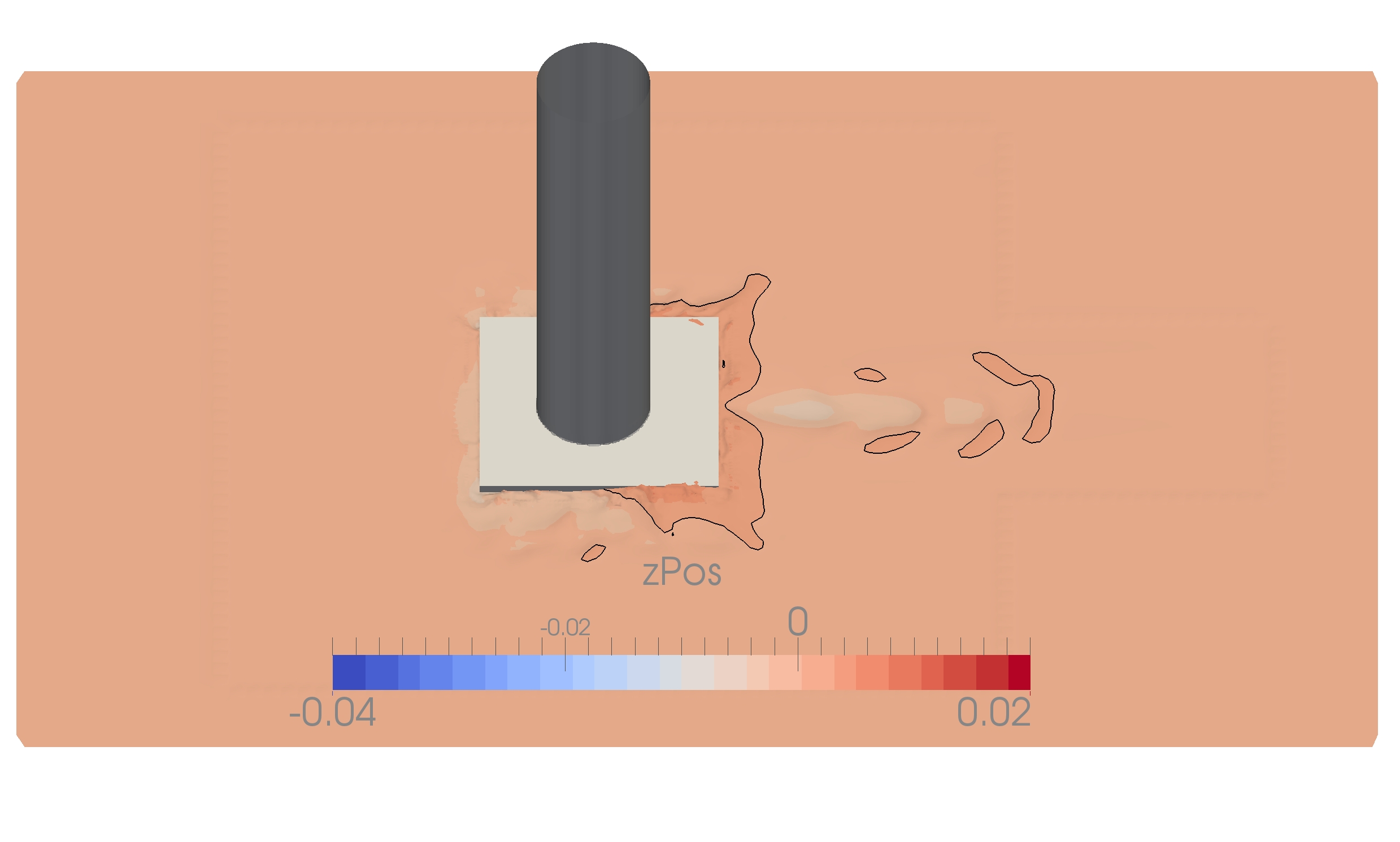}}%
\hfill
\subcaptionbox{$\mu_{S_{\text{lower}}}=1\frac{\text{Ns}}{\text{m}^2}$, $t=60\text{s}$}
{\includegraphics[width=0.33\textwidth]{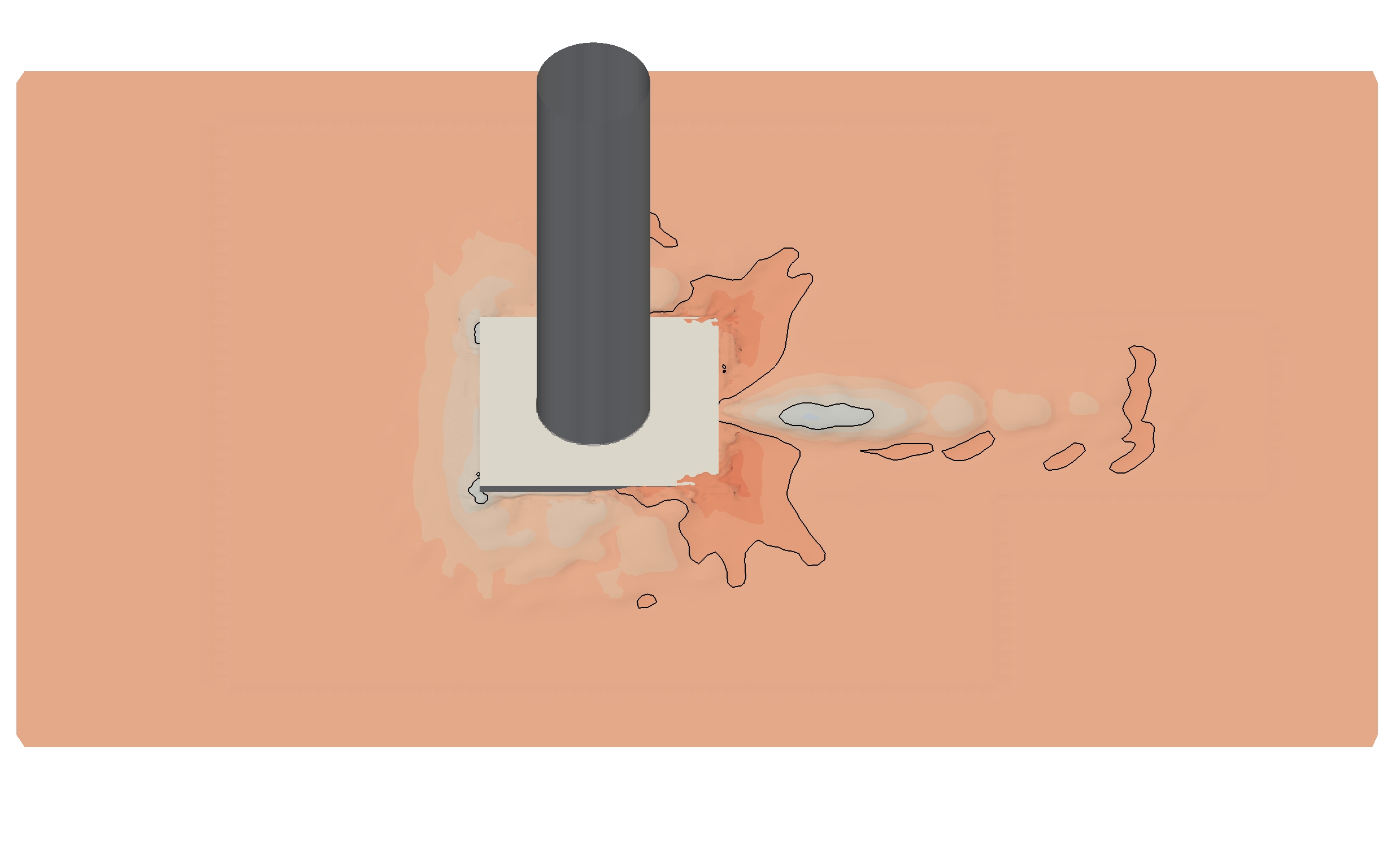}}%
\hfill
\subcaptionbox{$\mu_{S_{\text{lower}}}=1\frac{\text{Ns}}{\text{m}^2}$, $t=120\text{s}$}
{\includegraphics[width=0.33\textwidth]{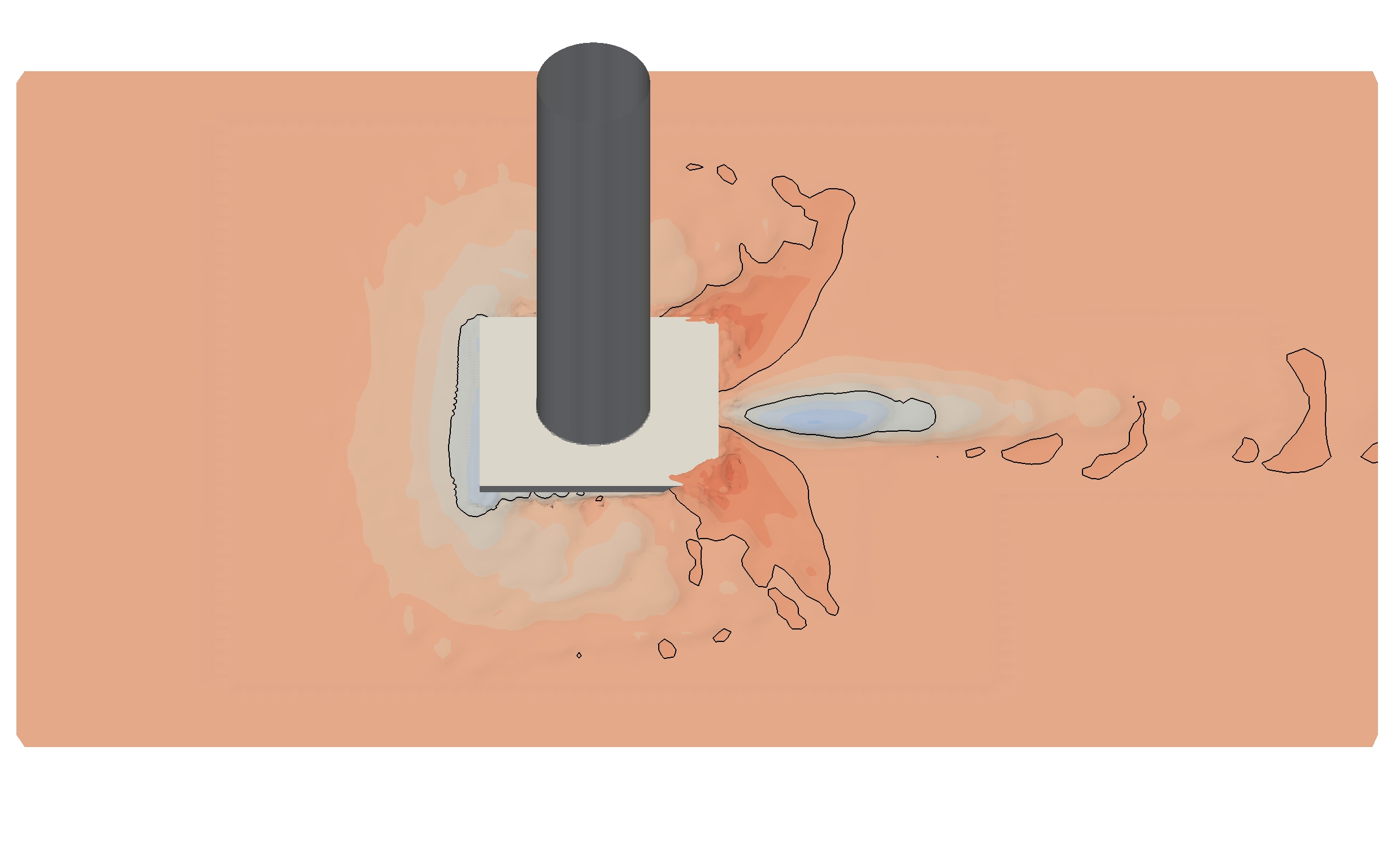}}%
\hfill
\subcaptionbox{$\mu_{S_{\text{lower}}}=1\frac{\text{Ns}}{\text{m}^2}$, $t=240\text{s}$}
{\includegraphics[width=0.33\textwidth]{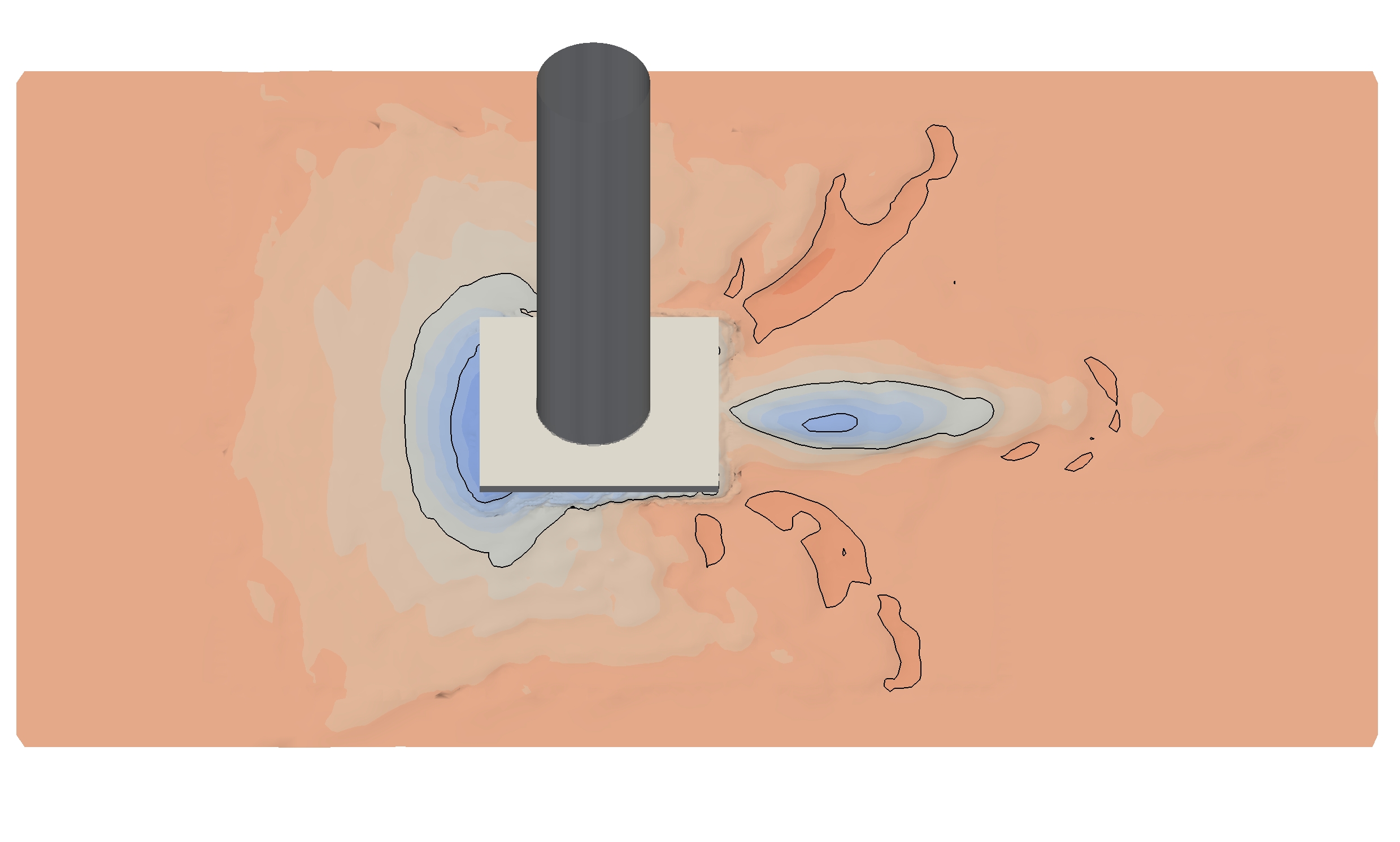}}%
\hfill
\subcaptionbox{$\mu_{S_{\text{lower}}}=1\frac{\text{Ns}}{\text{m}^2}$, $t=500\text{s}$}
{\includegraphics[width=0.33\textwidth]{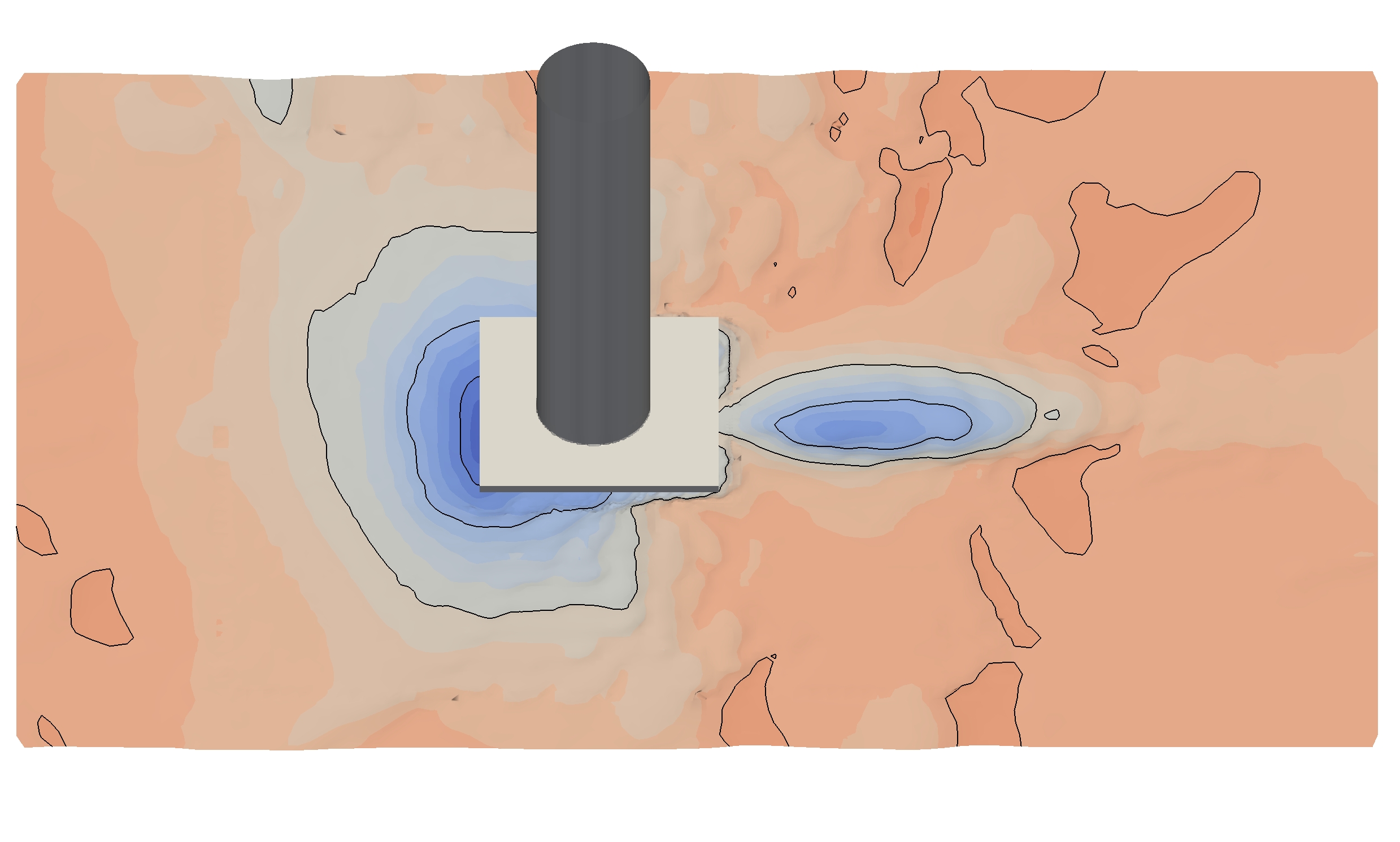}}%
\caption{Scour development around a vertical pile with a mudplate in steady current}
\label{vertical pile with mudplate}
\end{figure}

\begin{figure}
\centering
\subcaptionbox{$\mu_{S_{\text{lower}}}=1\frac{\text{Ns}}{\text{m}^2}$, $t=20\text{s}$}
{\includegraphics[width=0.33\textwidth]{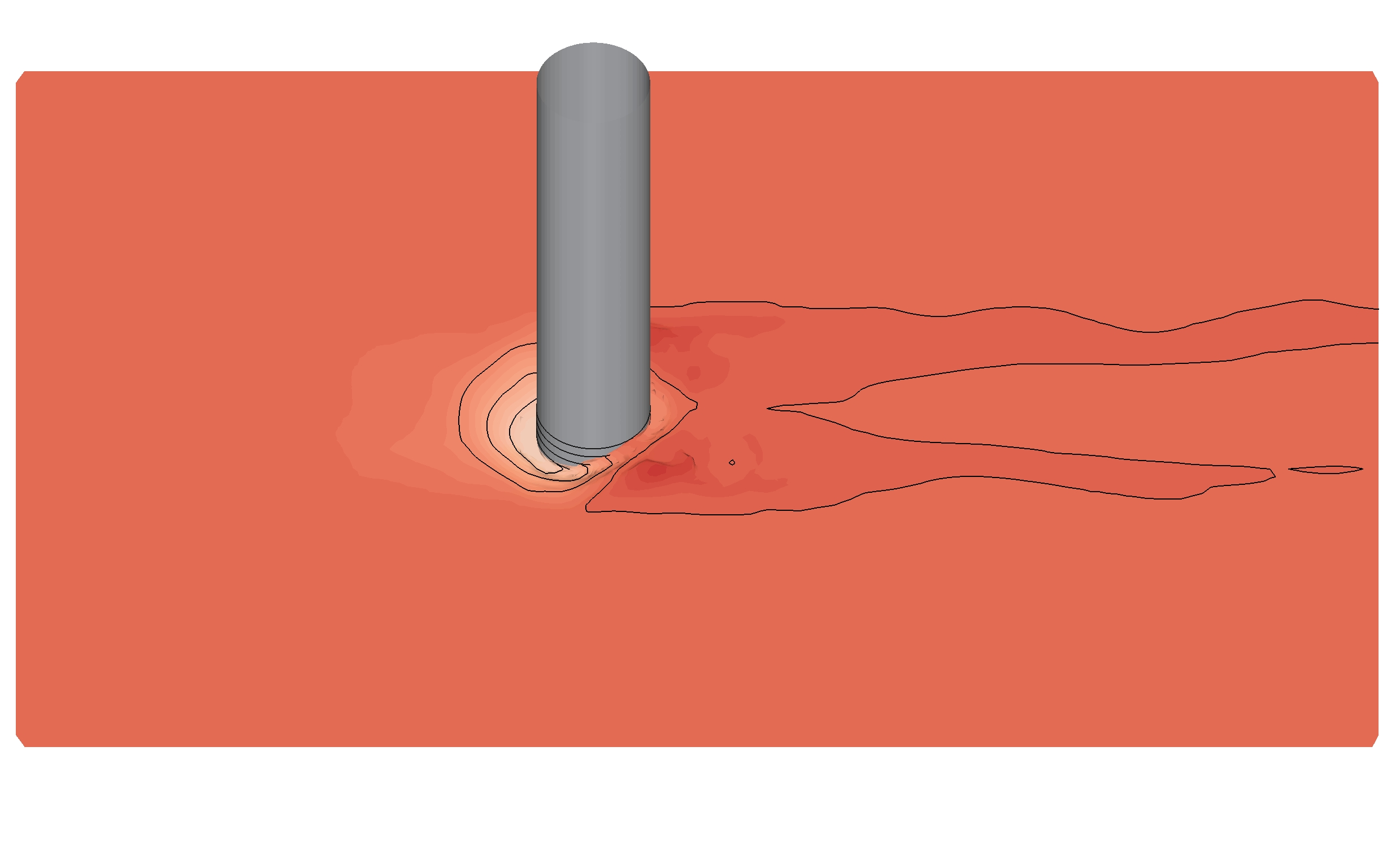}}%
\hfill
\subcaptionbox{$\mu_{S_{\text{lower}}}=1\frac{\text{Ns}}{\text{m}^2}$, $t=60\text{s}$}
{\includegraphics[width=0.33\textwidth]{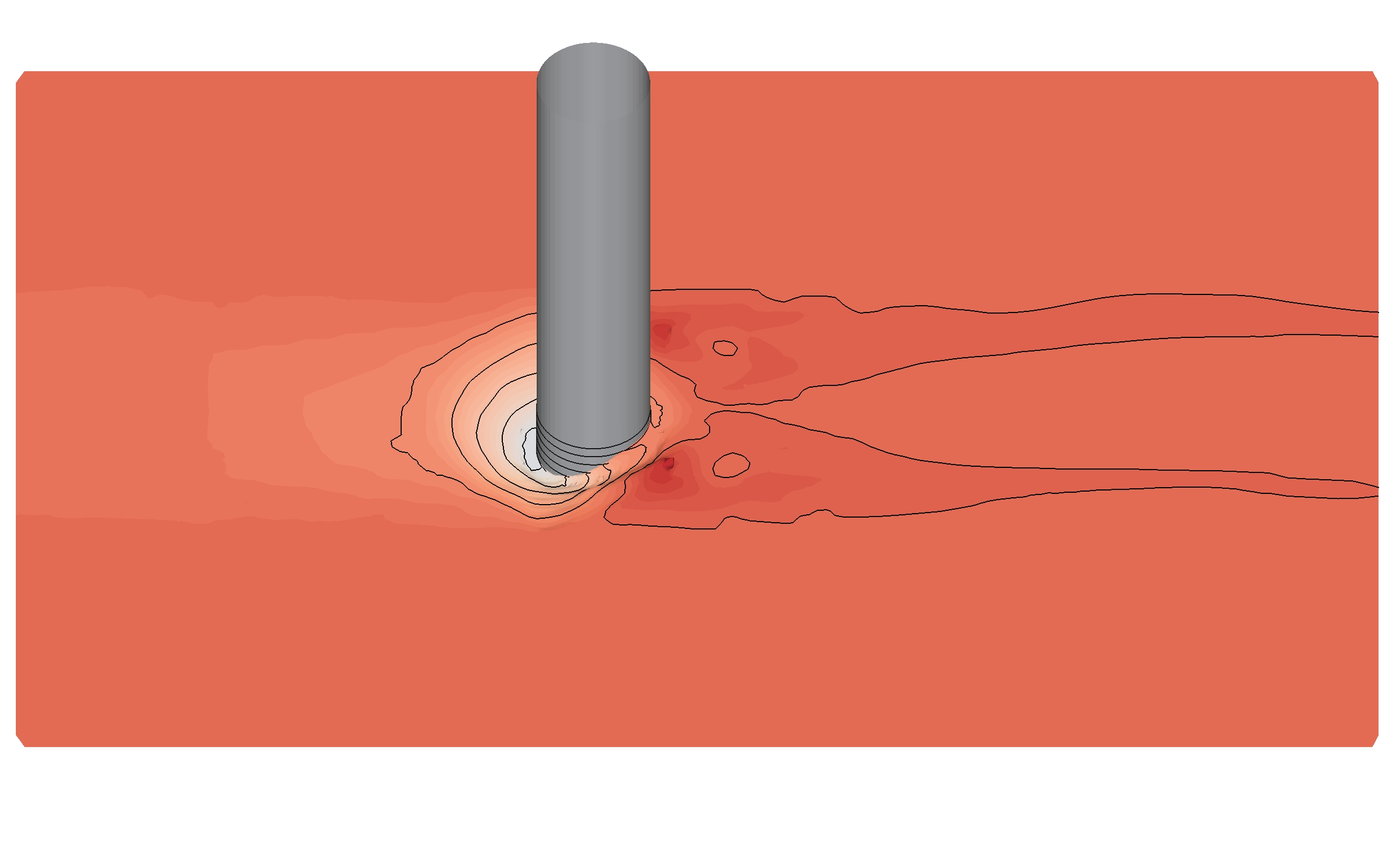}}%
\hfill
\subcaptionbox{$\mu_{S_{\text{lower}}}=1\frac{\text{Ns}}{\text{m}^2}$, $t=120\text{s}$}
{\includegraphics[width=0.33\textwidth]{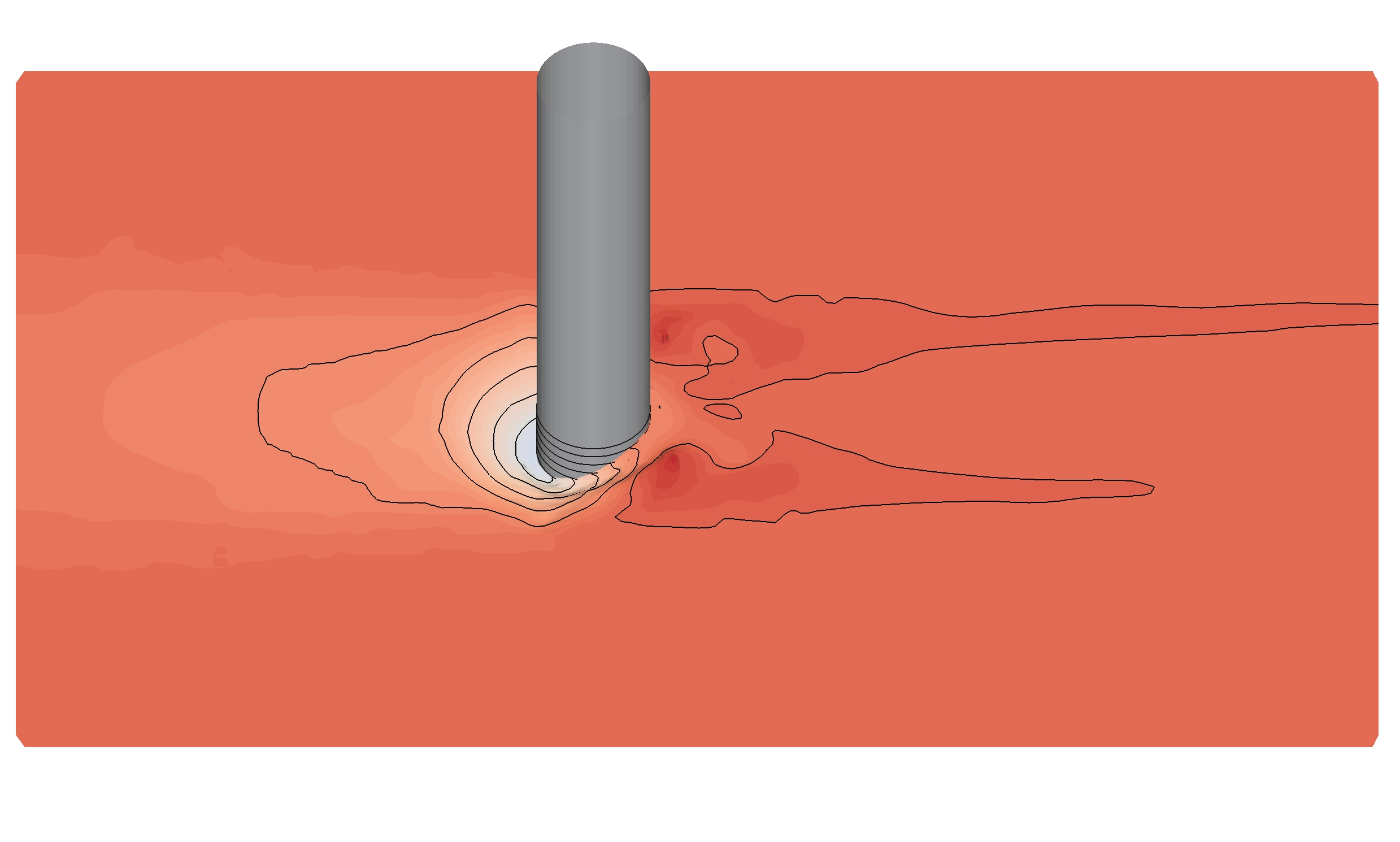}}%
\hfill
\subcaptionbox{$\mu_{S_{\text{lower}}}=1\frac{\text{Ns}}{\text{m}^2}$, $t=240\text{s}$}
{\includegraphics[width=0.33\textwidth]{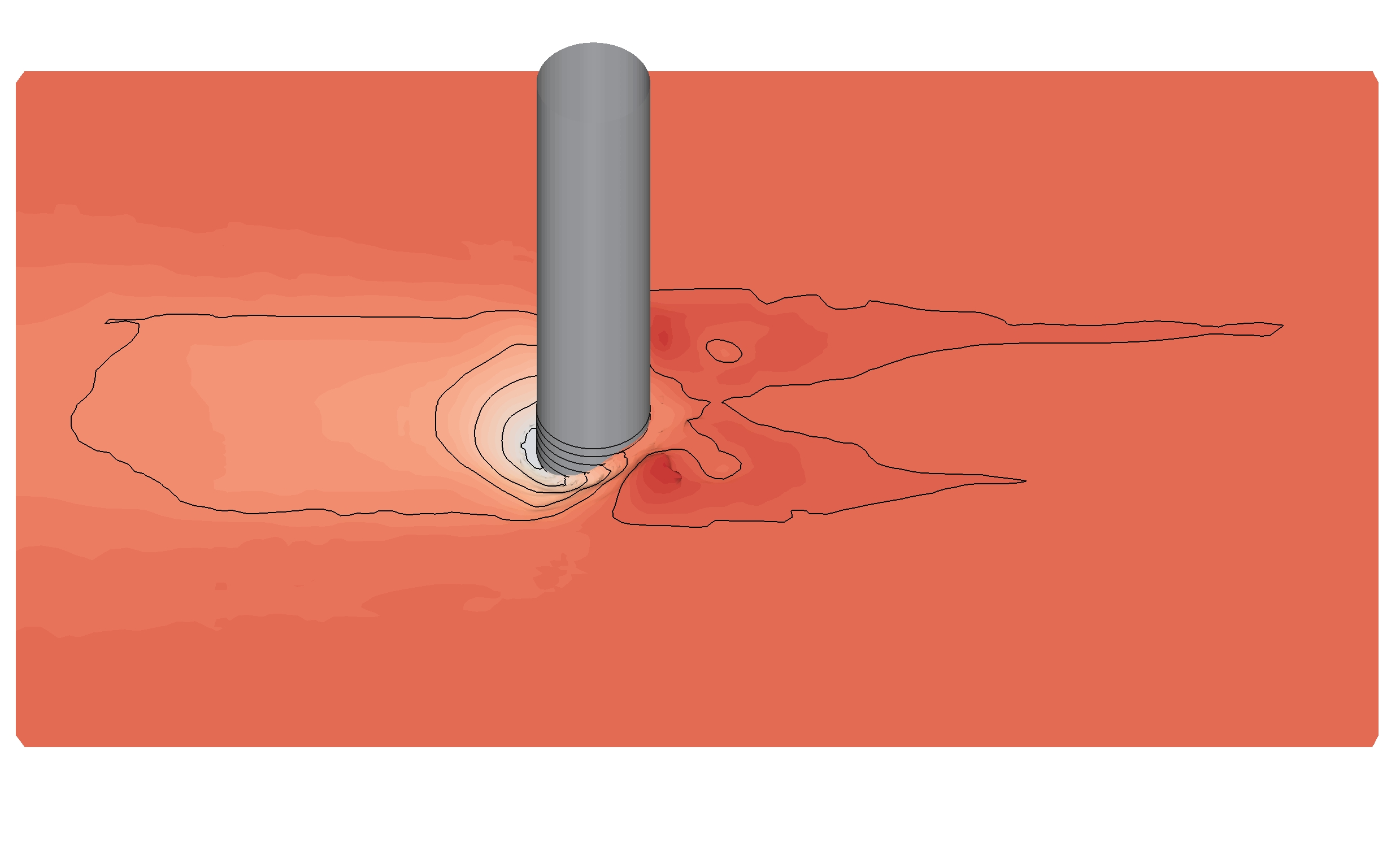}}%
\hfill
\subcaptionbox{$\mu_{S_{\text{lower}}}=1\frac{\text{Ns}}{\text{m}^2}$, $t=500\text{s}$}
{\includegraphics[width=0.33\textwidth]{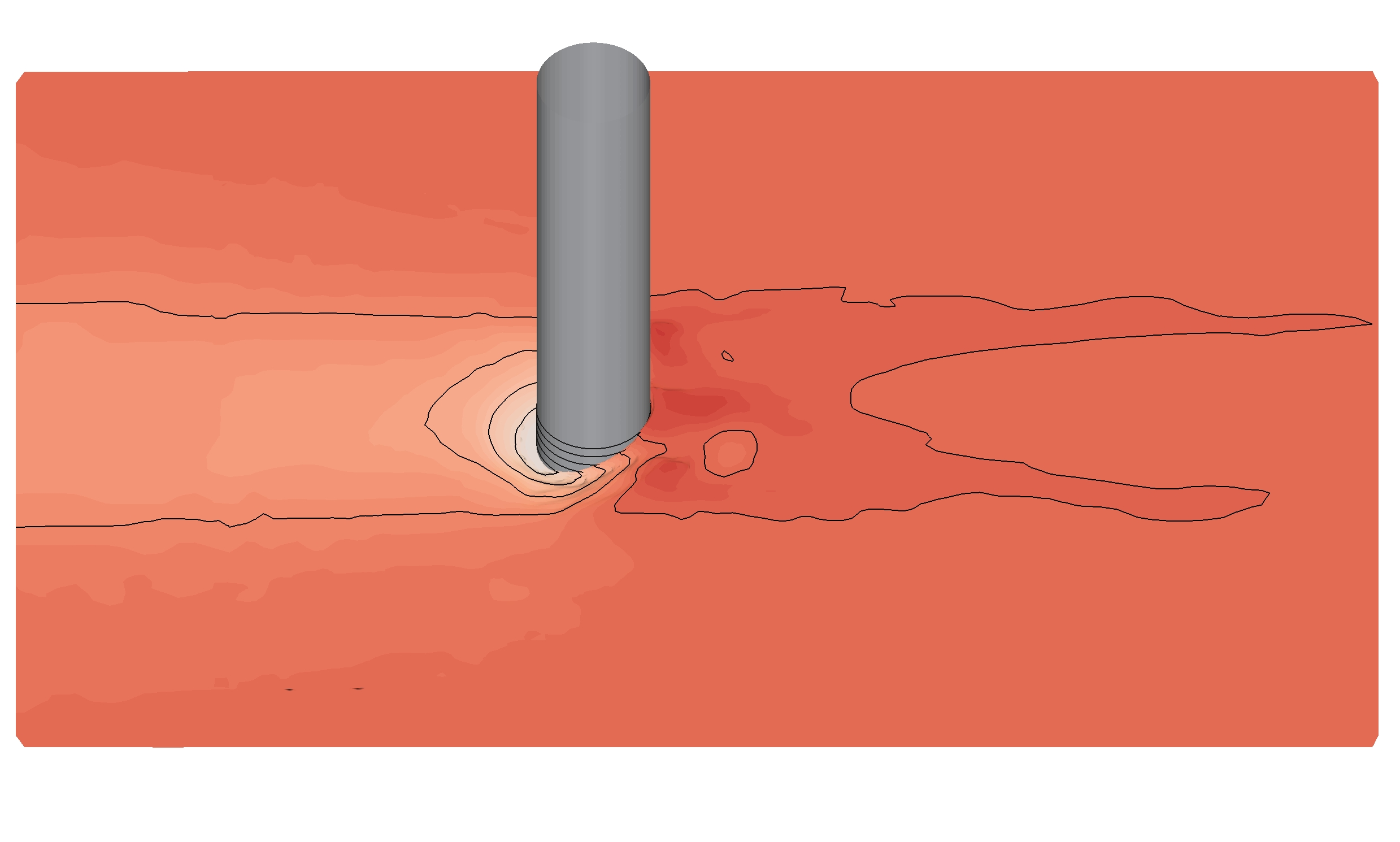}}%
\caption{Vertical pile in steady current: sediment evolution without using the buffer cell approach.}
\label{vertical pile in steady current: sediment evolution for different minimal soil viscosities without buffer cell}
\end{figure}

To demonstrate the ability to simulate arbitrary structures, we have done the same simulation with an additional mudplate.
Usually such a mudplate prevents scouring as the horse shoe vortex can not act onto the sediment.
On the other hand, if scouring occurs, the mudplate can have a negative influence,
as the flow is accelerated between the plate and the sediment.
Figure \ref{vertical pile with mudplate} shows the simulation results at different timesteps.
One can see that the influence of the horse shoe vortex is suppressed and only the leewake vortex can act onto the sediment.
At $t=60\text{s}$ an additional erosion begins at the edges on the upstream side of the plate.
As one can see, this erosion is growing until the plate gets underwashed.
At this point the whole scouring process is initiated and one can assume a significant reduction of the stability, after some additional time.\\
Figure \ref{vertical pile in steady current: sediment evolution for different minimal soil viscosities without buffer cell}
shows the simulation results for the single pile but without the buffer cell approach applied.
The pictures clearly show the importance of such an approach.
The wrong erosion in front of the pile completely suppresses the erosion directly at the pile
which finally leads to totally different and wrong results.

%% file: chapters/conclusion.tex
\section{Conclusions}
\label{Conclusions} 
In this study a new numerical code for the simulation of scour around offshore structures has been presented.
The method is based on a Bingham model for the soil and an additional model for the suspension.
The sediment shape is described using the Volume-of-Fluid method.
It has been shown, that the method resolves complex problems like the scour development around a circular pile in current.
The results are in a good agreement with experiments.
Nevertheless, the treatment of the suspension requires further improvements,
especially for the modeling of its generation.
As the influence of the suspension is high
a final calibration of the minimal Bingham viscosity can not be given at this development state.

It has been shown, that the solver is applicable to complex structures, which can not be simulated by other methods based on grid morphing.
Furthermore, the calculation times are small enough to allow an industrial application.
For the simulation of the scour around a vertical pile in current the solver requires approximately 25 CPU hours for 10 seconds of simulation.
This is a big improvement compared to the method investigated by \cite{Nagel2020}, where 6000 CPU hours are reported.
However, we would like to emphasize again, that the method used by \cite{Nagel2020} is modeling the sediment in more detail.

To achieve a wall behavior at the sediment surface, while using only one velocity field for all phases,
a new approach based on buffer-cells has been presented.
Its application demonstrates the importance of such an method, and the wall behavior was resolved sufficient for first simulations.
For future work, the authors recommend to replace this approach, by an approach based on an additional velocity field for the sediment.

To improve the wall behavior even more, the typical wall functions have been transferred successfully to the domain internal sediment surface.
The results show a very good agreement with a standard smooth wall function for domain boundaries.
Further improvement should be achieved using a rough wall function instead of a smooth wall function.
%
%
%
%
%
%
%
%
%
%

%% file: chapters/acknowledgements.tex
\section{Acknowledgements}
\label{Acknowledgements} 

This work was supported by the \textit{Ministerium f\"{u}r Energiewende, Landwirtschaft, Umwelt und l\"{a}ndliche R\"{a}ume des Landes Schleswig-Holstein}.